\RequirePackage{amsthm,amsmath, amssymb}
\RequirePackage{amsfonts,amsopn,color,mathtools,bm}
\documentclass{article}
\usepackage{a4}
\usepackage{epsfig}
\usepackage{booktabs}
\usepackage{amsmath,amsthm,amssymb,amsfonts,amsopn,color,mathtools}
\usepackage[latin3]{inputenc}
\def\bbr{{\mathbb R}}
\def\bbe{{\mathbb E}}
\def\bbp{{\mathbb P}}

\def\ep{{\varepsilon}}
 \newcommand{\eps}{\varepsilon}
  \newcommand{\epst}{\varepsilon^\star}

    \def\bbr{{\mathbb R}}
    \def\bbe{{\mathbb E}}
    
    \def\bbp{{\mathbb P}}

\newcommand{\R}{\mathbb{R}}

\newcommand{\ri}{\to}
\newcommand{\toas}{\stackrel{a.s.}{\to}}
\newcommand{\toP}{\stackrel{P}{\to}}

\newcommand{\n}{\noindent}

\newcommand{\DX}{\Delta_{i} X}

\newcommand{\DXq}{(\Delta_{i} X)^2}
\newcommand{\DXz}{\Delta_{i} X_0}
\newcommand{\DiJ}{\Delta_{i} J}
\newcommand{\DuJ}{\Delta_{1} J}
\newcommand{\DjX}{\Delta_j X}
\newcommand{\sumi}{\sum_{i=1}^n}
\newcommand{\sumjni}{\sum_{j\neq i}}

\newcommand{\IDXqleqr}{ I_{\{ \DXq\leq \ep^2\}} }
\newcommand{\hatIV}{\hat{IV}\!\!_n}
\newcommand{\IDXqleqri}{ I_{\{ \DXq\leq r(\sigma_{t_{i-1}}, h_i)\}} }
\newcommand{\sqrh}{\sqrt{r(\sigma, h)}}
\newcommand{\sqri}{\sqrt{r_i(h)}}
\newcommand{\IDXqleqrieta}{ I_{\{ \DXq\leq (1+\eta)r_i(h)\}} }
\newcommand{\DN}{\Delta_{i} N}
\newcommand{\IDNeqZ}{ I_{\{ \DN=0\}} }
\newcommand{\IDNneqZ}{ I_{\{ \DN\neq 0\}} }
\newcommand{\intIi}{\int_{t_{i-1}}^{t_i}}
\newcommand{\Ii}{]t_{i-1},t_i]}
\newcommand{\beq}{\begin{equation}}
\newcommand{\eeq}{\end{equation}}
\newcommand{\beqlab}{\begin{equation} \label}

\newcommand{\vp}{\vspace{0.3cm}}
\newcommand{\hot}{h.o.t.}

\newtheorem{thm}{Theorem}
\newtheorem{cor}{Corollary}

\theoremstyle{definition}
\newtheorem{lemma}{Lemma}
\newtheorem{prop}{Proposition}
\newtheorem{rem}{Remark}
\theoremstyle{remark}

\definecolor{Red}{rgb}{1, 0.0, 0.0}

\definecolor{DRed}{rgb}{0,0,0}%{0.4, 0.0, 0.0}

\definecolor{Green}{rgb}{0,0.5,0}%{0.0, 0.6, 0.0}%{0,0,0}%

\definecolor{Black}{rgb}{0, 0, 0}

\definecolor{Orange}{rgb}{0,0,0}%{1, 0.4, 0}

\definecolor{Brown}{rgb}{0,0,0}%{0.59, 0.29, 0}

\definecolor{Pink}{rgb}{1, 0.2, 1}

\begin{document}

\title{Optimum thresholding {using mean and conditional mean square error}}
\author{Jos\'e E. Figueroa-L\'opez\footnote{Department of Mathematics, Washington University in St. Louis, MO, 63130, USA ({\tt figueroa@math.wustl.edu})}\,  and Cecilia Mancini\footnote{Department of Management and Economics, University of Florence, via delle Pandette 9, 50127 ({\tt cecilia.mancini@unifi.it})}}

\date{\today} \maketitle

\begin{abstract}
We consider a univariate semimartingale model for (the logarithm of)
an asset price, containing jumps having possibly infinite activity (IA).
The nonparametric threshold estimator $\hat{IV}_n$ of the integrated
variance $IV:=\int_0^T\sigma^2_sds$ proposed in \cite{Man09} is
constructed using observations on a discrete time grid, and
precisely it sums up the squared increments of the process when they
are below a {\it threshold}, a deterministic function of the
observation step and possibly of the coefficients of $X$. All the
threshold functions satisfying given conditions allow asymptotically
consistent estimates of $IV$, however the finite sample properties
of $\hat{IV}_n$ can depend on the specific choice of the threshold.
We aim here at optimally selecting the threshold by minimizing either the estimation mean square
error (MSE) or the conditional mean square error (cMSE). The last criterion allows to reach
a threshold which is optimal not in mean but for the specific  volatility {and jumps paths} at hand.

A parsimonious characterization of the optimum is established, which turns out to
be asymptotically proportional to the Lévy's modulus of continuity of the underlying Brownian motion.
Moreover, minimizing the cMSE enables us to  propose a novel
implementation scheme for approximating the optimal threshold. Monte Carlo simulations
%and an empirical application (in
%progress)
illustrate the superior performance of the proposed
method.
\end{abstract}



\section{Introduction}

The importance of including jump components in assets prices models
has been extensively highlighted. For instance Huang and Tauchen (in \cite{HuaTau05}) documented  empirically that jumps account for 7\% of the S\&P500 market price variance, and many different tests for the presence of jumps in asset prices have been proposed and applied in the literature (see \cite{ManCal12}, Sec.~17.3, for a review of the most used tests). From an economic point of view,
jumps {may reflect, for instance,} reactions of the market to important announcements or events. Thus semimartingale models with jumps are  broadly used in a variety of financial applications, for example for derivative pricing, and also infinite activity jump components have  been considered % for modeling assets prices
(see e.g. \cite{ConTan04}, ch.15).

Separately identifying the contribution of the Brownian part (through the {\em Integrated Variance} IV) and the one of the jumps to the asset price variations when we can observe prices discretely is crucial in many respects, for instance, for model assessing and for improving volatility forecasting: e.g. in \cite{BaNShe06} the proposed test for the presence of jumps is obtained after having filtered out the jump component; in  \cite{AitJac11TestFaIa}, the separation allows to construct two tests for recognizing whether the jumps have finite or infinite variation; in \cite{AndBolDie07} it is  shown that including a separate factor accounting for the jumps in an econometric model for the realized variance substantially improves the out of sample volatility
 forecasts. The correct identification of a model has a significant impact on option pricing and on risk management and thus on assets allocation: for instance Carr and Wu (in \cite{CarrWu03}) show that the asymptotic behavior of the price of an option as the time-to-maturity approaches zero is substantially different depending on whether the model for the underlying contains jumps or not, and whether the jumps have finite or infinite variation; Liu, Longstaff, and Pan (in \cite{LiuLonPan03}) find that incorporating jumps events in the model  dramatically affects the optimal investment strategy.

With discrete (non-noisy) observations, non parametrically disentangling the jumps from integrated variance (IV) has mainly been done by using Multipower Variations (MPVs) and Truncated (or Threshold) Realized Variance (TRV) (see \cite{ManCal12}, {Sec.}~17.2, for a review of also other methods).
MPV relies on the observation that, when the jumps have finite activity, the probability of having jumps
among subsequent sampling intervals is very small, however with infinite activity jumps, this probability is much larger. Hence, MPV %(without any truncation)
may not work well in the general case. In contrast, TRV has been shown to be consistent also in the presence of any infinite activity jumps component (\cite{Man09}). Further, it is efficient  as soon as the jumps  have finite variation. 

However the choice of the truncation level (threshold) has an impact on the estimation performance of IV on finite samples.
The estimation
error is large when either the threshold is too small or when it is
too large. In the first case too many increments are discarded,
included the increments bearing relevant information about the Brownian part, and
TRV underestimates IV. In the second case too many increments are
kept within TRV, included many increments containing jumps, leading
to an overestimation of IV.
Many different data driven choices of the threshold have been proposed  in the literature, for instance Ait-Sahalia and Jacod \cite{AitJac11TestFaIa} (Sec. 4 therein)  chose a truncation level of the form $\alpha h^{0.2}$, where $h$ is the observation step and $\alpha$ is a multiplier of the standard deviations of the continuous martingale
part of the process 
 (other choices are described in \cite{ManCal12}, p.418). However  it is important to control for the estimation error for a given time resolution $h$, and here we look for an endogenous,  theoretically supported, optimal choice.

We consider the  model
\beq\label{Mod} dX_t= \sigma_t
dW_t + dJ_t,\eeq
where W is a standard Brownian motion, $\sigma$ is a
 c\'adl\'ag process, and  $J$ is a  pure jump semimartingale {(SM)} process.
{We assume} that we have at our disposal a record $\{x_0, X_{t_1},.. ,
X_{t_n}\}$ of discrete observations of $X$ spanned on the fixed time
interval $[0,T]$. {We also} define $\Delta_i Z${,} {or $\Delta_i^n Z$}{,} the increment $Z_{t_i}-
Z_{t_{i-1}}$ for any process $Z$, and  a {\it threshold function}
$r(\sigma, h)$ any deterministic non-negative function of the
observation step $h$, and possibly of a summary measure $\sigma$ of the realized volatility
 path of $(\sigma_{t})_{t\geq{}0}$, such that for any value $\sigma\in \R$ the following conditions are satisfied
$$\lim\limits_{h\ri 0}r(\sigma, h)= 0, \quad \lim\limits_{h\ri 0} \frac{r(\sigma, h)}{h \log \frac{1}{h}}=
+\infty.$$ We know that then TRV, given by 
\begin{equation}\label{TRV00}
\hat{IV}\!\!_n := \sum_{i=1}^n \DXq \IDXqleqri,
\end{equation}
where $h_i:=t_i-t_{i-1}$, is a consistent estimator of %the {\em Integrated Variance}
$ IV:= \int_0^T \sigma^2_s ds,$ as $\sup_i h_i\ri 0$, as soon as $(\sigma_{t})_{t\geq{}0}$ is a.s. %cad lag => a.s. bounded su [0,T]
bounded away from zero on $[0,T]$. In the case where the
 jump process $J$ has finite variation (FV) and the  observations are
evenly spaced, the estimator is also  asymptotically
Gaussian and efficient.

For the choice of the threshold (TH) in finite samples, we consider the following  two optimality criteria: minimization of MSE, the expected quadratic error in the estimation of IV; and minimization of cMSE,
the expected quadratic error conditional {on} the realized paths of the jump process $J$ and of the volatility process $(\sigma_s)_{s\geq{}0}$.
Even though, as mentioned above, many different TH selection procedures have been proposed, the literature for \emph{optimal TH selection} is rather scarce. In \cite{FigNis13} the TH that minimizes the expected number of jump misclassifications is considered for a class of additive processes with finite activity (FA) jumps and absolutely continuous characteristics. Even though it is shown therein that the proposed criterion is asymptotically equivalent to the minimization of the MSE in the case of L\'evy processes with FA jumps, the latter optimality criterion was not directly analyzed in \cite{FigNis13}. Here we go beyond and not only investigate the MSE criterion in the presence of FA jumps but also consider infinite activity jumps and further introduce the novel cMSE criterion. The last criterion allows to reach a threshold which is optimal not in mean but for the specific  volatility and jumps paths at hand, so it is particularly appealing in the cases of non-stationary processes, for which,
even if the MSE was feasible, the deviation of each realization from the
 unconditional mean value could be quite large, yielding a poor performance of the unconditional
 criterion. Moreover, minimizing the cMSE is important from a practical point of view, as will be seen in Section \ref{FACVP0}, where we propose a new TH selection method in the presence of FA jump processes.

Assuming evenly spaced observations, it turns out that for any
semimartingale $X,$  for which the volatility and the
jump processes are independent {of} the underlying Brownian motion,
the two quantities MSE and cMSE are explicit functions of the TH and  under each criterion  an optimal TH exists, and is a solution of an explicitly given equation, the equation being different under the two criteria. Under certain specific assumptions we also show uniqueness of the optimal TH: for L\'evy processes $X$, under the first criterion;  for constant volatility processes with general FA jumps, under the second criterion.

The equation characterizing the optimal threshold depends on the {observations' time} step $h$
and so does its solution. The optimal TH has to tend to 0 as h tends
to zero and, under each criterion, an asymptotic expansion with
respect to $h$ is possible for some terms within the equation, which
in turn implies an asymptotic expansion of the optimal TH. Under the
MSE criterion, when $X$ is Lévy  and $J$ has either  finite activity
jumps or  the activity is infinite but $J$ is symmetric strictly
stable, the leading term of the expansion is explicit in $h$, and in
both cases is proportional to the modulus of continuity of the
Brownian motion paths and to the spot volatility of X, the
proportionality constant being $\sqrt{2-Y}$, where $Y$ is the jump
activity index of $X.$ Thus the higher the jump activity is, the
lower the optimal threshold has to be if we want  to discard the {higher}
noise represented by the jumps and to catch information about
$IV$.

The leading term of the optimal TH does not satisfy the classical
assumptions under which the truncation method has been shown in \cite{Man09}
 to consistently estimate $IV$, however, at least in the
finite activity jumps case,  we show herein
that the threshold estimator of  IV constructed with the optimal TH is still consistent.

The assumptions needed for the asymptotic characterization {for the} cMSE criterion {are less} restrictive, and also allow for a drift. We find that, for constant $\sigma$  and {general} FA jumps,
the leading term of the
optimal TH still has to be proportional to the modulus of continuity of
the Brownian motion paths and to $\sigma$. One of the main motivations for considering the cMSE arises from a novel application of this to
tuneup the threshold parameter. The idea consists {in} iteratively updating the optimal TH and {estimates of} the increments of
the continuous and jump components  $X^{c}_{t}=\int_{0}^{t}\sigma_{s}dW_{s}$ and $\{J_{t}\}_{t\geq{}0}$ of $X$. We illustrate this
 method {on simulated data}.
Minimization of cMSE in the presence of infinite activity jumps in $X$ is a further topic of ongoing research.

The constant volatility assumption of some of our results is obviously restrictive. It is possible to allow {for} stochastic volatility and {leverage} but, since the proofs are still ongoing, we only discuss here some ideas and present some simulations experiments that show that also in {such contexts} our methods outperform other popular estimators appearing in the literature.

An outline of the paper is as follows. Section 2 deals with
the MSE: the existence of an optimal threshold $\ep^\star(h)$  is
established for a SM $X$ having volatility and jumps independent on the underlying Brownian motion $W$; for a Lévy process $X$,
uniqueness is also established (Subsection 2.1) and the asymptotic
expansion for the optimal TH is found in Section 2.3, in both the
cases of a { finite jump activity} Lévy $X$ and of an infinite activity symmetric
strictly stable $X$. In Section 3, for any {finite jump activity}  SM $X$,
consistency of $\hatIV$ is verified even when the threshold function
consists of the leading term of the optimal threshold, which does
not satisfy the classical hypothesis. Section 4 deals with the cMSE
in the case where $X$ is a SM with constant volatility and FA jumps:
existence of an optimal TH $\bar\ep(h)$ is established, its
asymptotic expansion is found, then uniqueness is obtained. In
Section 5 the results of Section 4 are used to construct a new
method for iteratively determine the optimal threshold value in
finite samples, and a reliability check is
executed on simulated data. Section 6 presents a Monte Carlo study that shows the superior performance of the new methods over other methods available in the literature under stochastic volatility and leverage. Section 7 concludes and Section 8 contains the proofs of the presented results.\\

%%%%%%%%%%%%%%
%%%%%%%%%%%%%%

{\bf Acknowledgements}. Jos\'e Figueroa-L\'opez's research was
supported in part by the National Science Foundation grants:
DMS-1561141 and DMS-1613016. Cecilia Mancini's work has benefited
from support by GNAMPA (Italian Group for research in Analysis,
Probability and their Applications. It is a subunit of the INdAM
group, the Itaian Group for research in High Mathematics, with site
in Rome) and EIF (Institut Europlace de Finance, subunit  of the
Institut Louis Bachelier in Paris).

%\tableofcontents

%\vspace{0.5cm}
%\n {\bf NOTATION} $\star\star$ means "to be hidden"

\section{MEAN SQUARE ERROR}\label{sec2}
We compute and
optimize the  mean square error (MSE) of $\hatIV$ passing through the
{\it conditional} expectation with respect to the paths of $\sigma$
and $J$:
$$
	 MSE:=E[(\hatIV- IV)^2]=E\left[E\left[(\hatIV- IV)^2| \sigma,
J\right]\right].
$$  Conditioning on $\sigma$,  as well as assuming no drift
in $X$,  is standard in papers where MSE-optimality is
looked for, in the absence of jumps (see e.g. \cite{BanRus05}).
We  also assume evenly spaced observation over a fixed time horizon $[0,T]$, so that $t_{i}=t_{i,n}=ih_{n}$, for any $i=1\dots n$, with $h=h_{n}=T/n$.
Denoted by $\ep$ the square root  $\sqrh$ of a given threshold function, in this work we focus on the performance of the threshold estimator:
\begin{equation}\label{TRV00a}
\hat{IV}\!\!_n(\varepsilon) := \sum_{i=1}^n \DXq I_{\{|\Delta_{i}X|\leq{}\varepsilon\}}.
\end{equation}
We indicate the corresponding MSE by $MSE(\ep)$. Note that for $\ep{\equiv } 0$ we have $\hatIV=0,$ so
$MSE(\ep)=E[IV^2]$; as $\ep$ increases some squared
increments $\DXq$ are included within $\hatIV$, so $\hatIV$ becomes closer to $IV$ and
$MSE(\ep)$ decreases. However, {if $J\not\equiv 0$}, for $\ep\ri+\infty$ the quantity
$MSE(\ep)$ increases again, since $\hatIV$ includes all the squared
increments $\DXq$ and thus $\hatIV$ estimates the global quadratic
variation $IV+\sum_{s\leq T} \Delta X_s^2$ of $X$ at time $T$, and
$MSE(\ep)$ becomes close to $E[(\sum_{s\leq T} \Delta X_s^2)^2]$. We
look for a threshold $\ep^\star$ giving
$$
    {MSE(\ep^{\star})=\min_{\ep\in[0,\infty[ } MSE(\ep)}.
$$
In this section we  analyze the first derivative $MSE'(\ep)$ and we
find that  an optimal threshold exists, in the general  framework
where $X$ is a  semimartingale satisfying {\bf A1} below, {and
we furnish an equation to which $\ep^\star$ is a solution}, while in
Section \ref{sec.Levy}, we find that {$\epst$} is even unique.
The equation has no explicit solution, but $\ep^\star$ is a
function of $h$ and we can explicitly characterize the first order
term of its asymptotic expansion in $h$, for $h\to 0$. Clearly
 we can always find an approximation of the optimal threshold with arbitrary precision making use of numerical methods.

Let us denote
$$ \DX_\star:= \DX \IDXqleqr, {\quad\sigma_{i}^2:= \intIi \sigma_s^2 ds, 
\quad m_i:=\DiJ}.$$
We assume the following\\

\begin{minipage}[c]{0.9\linewidth}
{\bf A1}. A.s. $\sigma^2_s > 0 $ for all $s$; $J\not\equiv 0$;
% otherwise I cannot divide by $\sigma_i$ for all $h$
 and $\sigma$, $J$ are independent on $W$.\\
 %and $m_i\in L^4(\Omega,P)\ \forall i=1..n.$
 \end{minipage}

\n The independence condition is needed to guarantee that $W$ remains a Brownian motion conditionally to $\sigma$ and $J$. We analyze the leverage case in our simulation study of Sec. 6. With the next theorem we compute the first derivative $MSE'$ of the mean square error. The proof is {deferred to the} Appendix.\\

\begin{thm}\label{expressionMSEingen}
Under {\bf A1} and the finiteness of the expectations of the terms below, for fixed $h$ and $\ep>0$, we have that
%we have that $MSE'(\ep) >0$ if and only if  $G(\ep)>0$, where
$MSE'(\ep)=\ep^2 G(\ep)$, where
\begin{equation}\label{EqG}
G(\ep):=\sumi E\Big[a_i(\ep) \Big(\ep^2 + 2  \sum_{ \substack{j=1\\ j\neq i}}^{n}
% \sum_{
%\footnotesize\begin{array}{c}
%j=1{\Red \dots}n\\
%j\neq i
%\end{array}
%}}
b_j (\ep) - 2 IV\Big)\Big],
\end{equation}
 with $a_{i}(\ep)$ and $b_{i}(\ep)$ defined as
\begin{align*}
a_i(\ep)&:= \frac{e^{-\frac{(\ep-m_i)^2}{2\sigma^2_i}} +
e^{-\frac{(\ep+m_i)^2}{2\sigma^2_i}}}{\sigma_i\sqrt{2\pi}},\\
b_i(\ep)&:= E[(\DX_\star)^2| \sigma, J]=-\Big(e^{-\frac{(\ep-m_i)^2}{2\sigma^2_i}}(\ep + m_i)
+e^{-\frac{(\ep+m_i)^2}{2\sigma^2_i}}(\ep -
m_i)\Big)\frac{\sigma_i}{\sqrt{2\pi}}
+ \frac{m_i^2+\sigma_i^2}{\sqrt{2\pi}}
    \int_{\frac{m_i-\ep}{\sigma_i}}^{\frac{m_i+\ep}{\sigma_i}} e^{-\frac{x^2}{2}} dx.
\end{align*}
\end{thm}

\vp\n
It clearly follows that $MSE'(\ep) >0$ if and only if  $G(\ep)>0$ and, thus, to our aim of finding an optimal threshold, it suffices to study the
sign of $G(\ep)$ as $\ep$ varies.

\vp
\n {\bf Notation.}  For brevity we sometimes omit to precise the dependence on $\ep$ of $a_i(\ep)$ and $b_i(\ep)$.\\
For a function $f(\ep)$ we sometimes use $f(+\infty)$ for
$\lim_{\ep\ri+\infty}f(\ep).$\\
For two functions $f(x), g(x)$ of a non-negative variable $x$ which tends  to 0 (respectively to $+\infty$),  by $f\ll g$, we mean
that $f=o(g)$ as $x\to 0$ (respectively  $x\to+\infty$),  by $f\asymp g$ we mean that both $f=O(g)$
and $g=O(f)$ as $x\to 0$ (respectively  $x\to+\infty$), while by $f\sim g$ we mean that $f(x)/g(x) \to 1$ as $x\to 0$ (respectively  $x\to+\infty$).\\
We denote $\phi(x)=\frac{e^{-\frac{x^2}{2}}}{\sqrt{2\pi}}, \quad \bar{\Phi}(x)= \int_x^{+\infty} \phi(s) ds. $\\
{\em h.o.t} means {\em higher order terms}\\

\begin{rem}
Under {\bf A1} and the finiteness of the expectation of the terms in MSE we have
$$MSE(0)=E[IV^2]>0\quad \mbox{ and, for small $h$, } \lim_{\ep\ri
+\infty} MSE(\ep) >0.$$
\end{rem}

\n
{The next Corollary states the existence of an optimal threshold (see the proof in the  Appendix).}

\begin{cor}\label{OptThrExists}
Under the same assumptions of Theorem \ref{expressionMSEingen} an
optimal threshold exists and is solution of  the equation $G(\ep)=0$.
\end{cor}

\n To find an optimal threshold $\ep^{\star}$ to estimate $\sigma$ we need to find the zeroes of $G$, which in turn depends on $\sigma$. Also, $G$ depends on the jump process increments ${\bf m}=(m_{1},\dots,m_{n})$, which we don't know. An analogous problem arises when dealing with the minimization of the conditional MSE introduced in Section \ref{Sect:CMSE}, where the optimal threshold $\bar\ep$ has to satisfy the equation $F(\ep)=0$, with $F(\ep):=a_i(\ep) (\ep^2 + 2  \sum_{ j\neq{}i}^{n}b_j (\ep) - 2 IV)$. However, when we apply our theory to the case of constant $\sigma$ and finite activity jumps, as precisely explained in Section \ref{FACVP0}, we can proceed by estimating $\sigma,$ ${\bf m}$ and $\bar\ep$ iteratively. Another method yet to implement $\ep^{\star}$ is to study the infill asymptotic behavior of $\ep^{\star}$ in a stationary or deterministic state of $\sigma$. In some situations, the leading order terms of $\ep^{\star}$ will only depend on a few summary measures of the stationary distribution or path of $\sigma$, which could be estimated separately or jointly with $IV$.

\begin{rem}
In principle $MSE(\ep)$ could even have many points $\ep$ where the absolute
minimum value $\underline{MSE}$ of MSE on $[0, +\infty)$ is reached; also, MSE could have an infinite number
of local not absolute minima.
\end{rem}

\n To determine the number of solutions to $G(\ep)=0$, we need to study the sign of
$G'(\ep)$ (corresponding {to} the convexity properties of
$MSE(\ep)$), but this is not easy. Define $$g_i(\ep) := \ep^2 + 2
\sum_{j\neq i} b_j - 2 IV,$$ so that
$$ G(\ep)= \sum_i E[a_i(\ep) g_i(\ep)].$$
We can easily study the functions $g_i,$ since we know that $g_i(0)=
-2IV <0,$ $\lim_{\ep \ri +\infty} g_i(\ep)= + \infty$ and $g_i'(\ep)
= 2 \ep  (1 + \ep \sum_{j\neq i} a_i) > 0$ for all $\ep>0$. However
within the joint function $G(\ep)$ the presence of the terms
$a_i(\ep)$ makes it difficult even to know whether $(a_i g_i)'$ is
positive.

\subsection{{When $X$ is L\'evy}}\label{sec.Levy}

Let us assume\\

\begin{minipage}[c]{0.9\linewidth} {\bf A2}. $X$ is a Lévy process.\\
 \end{minipage}

\noindent We now have that $\sigma> 0$ is constant and $\DX_\star$ are i.i.d.,
so the equation characterizing $MSE'(\ep)=0$ is much simpler  to analyze. Indeed, from
(\ref{EqG}), since within $a_i\sum_{j\neq i} b_j$, the
term $m_i$ of $a_i$ is independent on the terms $m_j$ of $b_j,$ we
have
$$MSE'(\ep)=\ep^2 G(\ep)=
\ep^2 n E[{a_1(\varepsilon)}]\Big( \ep^2+ 2(n-1)E[{b_1(\varepsilon)}]- 2 IV\Big).$$
The next  result establishes uniqueness of the optimal threshold under {\bf A2}. The proof is in the Appendix.

\begin{thm}\label{TeoEsUNicOptThLevycase}
If $X$ is Lévy,  equation
\beq\label{eqOptThreshXLevy}\ep^2+ 2(n-1)E[b_1(\ep)]- 2 IV=0\eeq
has {a} unique solution $\epst$ {and, thus,} there exists a unique optimal threshold, which
is $\epst$.
\end{thm}

\vp \n The equation in (\ref{eqOptThreshXLevy}) has no explicit solution, however we can give some
important indications to approximate $\ep^\star$.

\subsection{Asymptotic behavior of $\bbe\left(b_{i}(\varepsilon)\right)$}

{For the rest of Section \ref{sec2}}, {in order to emphasize the dependence of $\epst$ on $h$, we write}  $\ep:=\ep(h)=\ep_h$. %, even when for brevity we {sometimes} omit to indicate the dependence on $h$.
 {We still are under {\bf A2}}, so recall that
% and that $X$ is a L\'evy process $X_{t}=\sigma W_{t}+J_{t}$
%so that
\[
    \bbe\left[ b_{i}(\varepsilon)\right]%=\bbe\left[ b_{i}(\varepsilon,h)\right]
    =\bbe\left[\left|\sigma\Delta_{i}^{n}W+
    \Delta_{i}^{n}J\right|^{2}{\bf 1}_{\{\left|\sigma\Delta_{i}^{n}W+\Delta_{i}^{n}J\right|\leq{}\varepsilon\}}\right],
\]
 is constant in $i$.  Note that $\bbe[ b_{i}(\varepsilon)]$ is finite for any L\'evy process $J$, regardless of whether
 $J$ has bounded first moment or not. We consider two cases: the case where $J$ is a finite jump activity process and the one where {it} is
 a symmetric {strictly} stable process.
 The asymptotic characterization of  $\bbe\left[ b_{i}(\varepsilon)\right]$ will be used in Subsection \ref{subsec.AsimptEpstarMSE} to deduce the
 asymptotic behavior {in $h$} of the optimal threshold $\varepsilon^{\star}$.

We anticipate that in Subsection \ref{subsec.AsimptEpstarMSE} we will also see that an optimal threshold {$\ep^\star$} has to tend to
0 as $h\to 0$ and in such a way that $\frac{\ep^\star}{\sqrt h} \to +\infty.$

\subsubsection{Finite Jump Activity L\'evy {process}}

 \n {The asymptotic characterization of  $\bbe\left[ b_{i}(\varepsilon)\right]$ in the case where $J$ has finite activity jumps is given in the following Theorem. Its proof is in the  Appendix.}

\begin{thm}\label{thrmKA}
    Let $X$ be a finite jump activity L\'evy process with  jump size density $f$ and with jump intensity  $\lambda$. Suppose
    also that the restrictions of $f$ on $(0,\infty)$ and $(-\infty,0)$ admit $C_{1}$ extensions on $[0,\infty)$ and $(-\infty,0]$,
    respectively. Then, for any $\varepsilon=\ep(h)$ such that $\ep\to{}0$ and $\ep\gg \sqrt{h}$, as $h\to{}0$, we have
\[
\bbe\left[b_{1}(\varepsilon)\right]=\sigma^{2}h-\frac{2}{\sqrt{2\pi}}\sigma\varepsilon\sqrt{h}
e^{-\frac{\varepsilon^{2}}{2\sigma^{2}h}}+ \lambda h
\frac{\varepsilon^{3}}{3}C(f)+
O\left(h^{2}\right)+o\left(\varepsilon\sqrt{h}
e^{-\frac{\varepsilon^{2}}{2\sigma^{2}h}}\right)+o\left(h
\varepsilon^{3}\right),
\]
where above $C(f):=f(0^{+})+f(0^{-})$.
\end{thm}

\subsubsection{Strictly stable symmetric L\'evy {Jump} process}
 Let us start by noting that
    \begin{align*}
        \bbe[b_{1}\left(\varepsilon\right)]&=\bbe\left[\left(\sigma W_{h}+J_{h}\right)^{2}{\bf 1}_{\{|\sigma W_{h}+J_{h}|\leq{}\varepsilon\}}\right]\\
        &=\sigma^{2}\bbe\left[W_{h}^{2}{\bf 1}_{\{|\sigma W_{h}+J_{h}|\leq{}\varepsilon\}}\right]+
        2\sigma\bbe\left[W_{h}J_{h}{\bf 1}_{\{|\sigma W_{h}+J_{h}|\leq{}\varepsilon\}}\right]+
        \bbe\left[J_{h}^{2}{\bf 1}_{\{|\sigma W_{h}+J_{h}|\leq{}\varepsilon\}}\right]\\
        &=:C_{h}(\varepsilon)+D_{h}(\varepsilon)+E_{h}(\varepsilon).
    \end{align*}
The first term above can be written as
\begin{align*}
    C_{h}(\varepsilon)=\sigma^{2}h-\sigma^{2}\bbe\left[W_{h}^{2}{\bf 1}_{\{|\sigma W_{h}+J_{h}| >\varepsilon\}}\right]=
    \sigma^{2}h-\sigma^{2}h\left(C^{+}_{h}(\varepsilon)+C^{-}_{h}(\varepsilon)\right),
\end{align*}
where
\begin{align*}
    C^{+}_{h}(\varepsilon)=\bbe\left[W_{1}^{2}{\bf 1}_{\{W_{1}+\sigma^{-1}h^{-1/2}J_{h}{ >}\sigma^{-1}h^{-1/2}\varepsilon\}}\right],\quad
    C^{-}_{h}(\varepsilon)=\bbe\left[W_{1}^{2}{\bf 1}_{\{W_{1}+\sigma^{-1}h^{-1/2}J_{h}{ <}-\sigma^{-1}h^{-1/2}\varepsilon\}}\right].
\end{align*}
By conditioning on $J$ and using the fact that $\bbe[W_{1}^{2}{\bf
1}_{\{W_{1}>x\}}]=x\phi(x)+\bar{\Phi}(x)$, for all $x\in\bbr$, we
have
\begin{align*}
    C^{\pm}_{h}(\varepsilon)&=\bbe\left[\left(\frac{\varepsilon}{\sigma\sqrt{h}}\mp
    \frac{J_{h}}{\sigma\sqrt{h}}\right)
    \phi\left(\frac{\varepsilon}{\sigma\sqrt{h}}\mp\frac{J_{h}}{\sigma\sqrt{h}}\right)+
    \bar{\Phi}\left(\frac{\varepsilon}{\sigma\sqrt{h}}\mp\frac{J_{h}}{\sigma\sqrt{h}}\right)
    \right].
\end{align*}

\vspace{0.3cm}
\n {The following  Lemmas state the asymptotic behavior of the above quantities under the assumption that $\varepsilon\gg \sqrt{h}$.  Their proofs are in the Appendix.}
\begin{lemma}\label{L1:IJA}
    Suppose that $\{J_{t}\}_{t\geq{}0}$ is a symmetric $Y$-stable process with $Y\in(0,2)$. Then,
{there exist constants ${K_{1}<0}$ and $K_{2}$} such that:
    \begin{align}
    \bbe\left[\phi\left(\frac{{\varepsilon}}{\sigma\sqrt{{h}}}-\frac{J_{{h}}}{\sigma\sqrt{{h}}}\right)\right]
    &=\frac{1}{\sqrt{2\pi}}e^{-\frac{{\varepsilon}^{2}}{2\sigma^{2}{h}}}-K_{1}{\varepsilon}^{-1-Y}{h}^{\frac{3}{2}}+{\rm h.o.t.}\label{NECN}\\
    \bbe\left[J_{{h}}\phi\left(\frac{{\varepsilon}}{\sqrt{{h}}}-\frac{J_{{h}}}{\sqrt{{h}}}\right)\right]&=
    K_{ 2} {h}{\varepsilon}^{1-Y}+{\rm h.o.t.}.\label{NECNc}
\end{align}
\end{lemma}

\begin{lemma}\label{L2:IJA}
    Suppose that $\{J_{t}\}_{t\geq{}0}$ {is} a symmetric strictly stable process with L\'evy measure {$C|x|^{-Y-1}dx$}. Then, the following asymptotics hold:
    \begin{align}
    \bbe\left[\bar{\Phi}\left(\frac{{\varepsilon}}{\sigma\sqrt{{h}}}-\frac{J_{{h}}}{\sigma\sqrt{{h}}}\right)\right]&=
    \frac{C}{Y}{h}{\varepsilon}^{-Y}+O\left({\varepsilon}^{-2Y}{h}^{2}\right)+
    O\left(\bbe\left[{\phi}\left(\frac{{\varepsilon}}{\sigma\sqrt{{h}}}-\frac{J_{{h}}}{\sigma\sqrt{{h}}}\right)\right]\right),\label{NECNL2}\\
        \bbe\left[J^{2}_{h}{\bf 1}_{\{|\sigma W_{h}+J_{h}|\leq{}\varepsilon\}}\right]&= \frac{2C}{2-Y} h\varepsilon^{2-Y}+
    O\left(h^{2}\varepsilon^{2-2Y}\right)+O\left(h^{\frac{4-Y}{2}}\right)+O\left(h^{\frac{2}{Y}}\right).
    \label{NECNcL2}
\end{align}
\end{lemma}

\vspace{0.3cm}\n {As a consequence, the following Theorem states explicitly the asymptotic behavior of $\bbe\left[b_{1}(\varepsilon)\right]$. It's proof is in the Appendix.}
\begin{thm}\label{thrmIJA}
    Let $X_{t}=\sigma W_{t}+J_{t}$, where $W$ is a Wiener process and $J$ is {a} symmetric strictly stable L\'evy process with L\'evy measure $C|x|^{-Y-1}$.
    Then, for any $\ep=\ep(h)$ such that $\ep\to{}0$ and $\ep\gg \sqrt{h}$, as $h\to{}0$, we have
\[
\bbe\left[b_{1}(\varepsilon)\right]=\sigma^{2}h-\frac{2\sigma}{\sqrt{2\pi}}\sqrt{h}\varepsilon
e^{-\frac{\varepsilon^{2}}{2\sigma^{2}h}}+\frac{2C}{2-Y}
h\varepsilon^{2-Y}+{\rm h.o.t.}.
\]
\end{thm}

\subsection{Asymptotic behavior of $\varepsilon^{\star}$}\label{subsec.AsimptEpstarMSE}

We now assume\\

\begin{minipage}[c]{0.9\linewidth} {\bf A3}. %$J\not\equiv 0$ and
The support of any {jump size} $\Delta J_t$ is
$\R$.\\
 \end{minipage}

\n We firstly see that an optimal threshold $\ep^\star=\ep^\star(h)$ has to tend to
0 as $h\to 0$ and in such a way that $\frac{\ep^\star}{\sqrt h} \to +\infty.$ Then we will show the asymptotic behavior of $\ep^\star$ in more detail.

\begin{rem}\label{REMOptThto0}
{Note that under {\bf A3}, if} $\ep^\star(h)$ minimizes MSE, then necessarily
$\ep^\star(h)\to 0$ as $h\to 0$. Indeed, if $\liminf\limits_{h\to 0}
\ep^\star(h)=c>0,$ then on a sequence $\ep^\star(h)$ converging to
$c$ we would have $\hat{IV}_n -IV\to \sum_{s\leq T} \Delta J_s^2
I_{|\Delta J_s|\leq c}$ in probability, rather than $\hat{IV}_n -IV
\to 0$; since $P\{\sum_{s\leq T} \Delta J_s^2 I_{|\Delta J_s|\leq
c}>0\}>0,$ the MSE could not be minimized.
\end{rem}

\begin{lemma}\label{LemEpGrSqrthMSE}
    Suppose $X_{t}=\sigma W_{t}+J_{t}$, where $W$ is a Brownian motion and $J$ is a pure-jump L\'evy process of bounded variation or, more generally,
    such that, for some $Y\in(0,2)$, $h_{n}^{-1/Y}J_{h_{n}} \stackrel{P}{\to} \bar J$, for a real-valued random variable {$\bar J$}. Then, $\varepsilon_{n}^{\star}/\sqrt{h_{n}}\to\infty$, as $n\to\infty$.
\end{lemma}

\n  {\bf Remark}. If $J$ has FA jumps, drift $d$ and $J_t=d t +\sum_{k=1}^{N_t} \gamma_k $, then we have
%any $Y\in(1,2)$ we have
%that $\frac{E[J_h]}{h^{1/Y}}= \frac{Ch}{h^{1/Y}}\to 0,$
%
$h^{-1}J_{h}\toP d$  and, thus, the assumption in Lemma \ref{LemEpGrSqrthMSE} is satisfied with $Y=1$. {If $J$ is a Lévy process with Blumenthal-Getoor index $Y$, then $Y\in (0,2)$ and %$\frac{E[|J_h|]}{h^{1/Y}}= \frac{C\sqrt h}{h^{1/Y}}\to 0$, thus $h_{n}^{-1/Y}J_{h_{n}}\stackrel{P}{\to} 0,$
 for any $\eta\in (Y,2)$ we have $h_{n}^{-1/\eta}J_{h_{n}}\stackrel{a.s.}{\to} 0$,
 and again the assumption is satisfied. }\\

\n We are now ready to show more precisely the asymptotic behavior of $\ep^\star$. {Proposition \ref{ABeps} covers the FA jumps case, while Proposition \ref{ABepsIJA} tackles the case of symmetric strictly stable jumps. Their proofs are deferred to the Appendix.}

\begin{prop}\label{ABeps}
    Let $J$ have FA jumps and satisfy the assumptions of Theorem 3, let $\varepsilon^{\star}=\epst(h)$ be the optimal threshold{.} Then,
    \[
        \varepsilon^{\star}\sim \sqrt{2\sigma^{2}h\ln\frac{1}{h}},  \quad \text{as}\quad h\to{}0.
    \]
\end{prop}

\begin{prop}\label{ABepsIJA}
Under the conditions of Theorem \ref{thrmIJA}, the optimal threshold
$\varepsilon^{\star}=\epst(h)$  is such that
    \[
        \varepsilon^{\star}\sim \sqrt{(2-Y)\sigma^{2}h\ln\frac{1}{h}},  \quad \text{as}\quad h\to{}0.
    \]
\end{prop}
 As explained in the introduction, the proportionality constant $\sqrt{2-Y}$ of the previous result says that the higher the jump activity is, the
lower the optimal threshold has to be if we want  to discard the {higher}
noise represented by the jumps and to catch information about
$IV$.

\section{CONSISTENCY WHEN $\ep_h=\sqrt{2M h\log\frac 1 h}$}

Under the framework described in \cite{Man09}, in the case of
equally spaced observations,
the threshold criterion allows convergence of
$$\hat{IV}\!\!_n := \sum_{i=1}^n \DXq I_{\{ \DXq\leq r(\sigma_{t_{i-1}}, h)\}}$$ to $IV_T=\int_0^T\sigma^2_s ds$
when, for all  $i=1,\dots, n$, we have  $r(\sigma_{t_{i-1}}, h)=r(h)$ and $r(h)$ is a deterministic function of $h$ s.t.
 $r(h)\to 0,$ $\frac{r(h)}{h\log\frac 1 h}\to \infty$, as $h\to 0$.
 Here we show that,  under finite activity jumps,  the same estimator is also consistent in the case where on any $\Ii$ we consider a different truncation level $r_i(\sigma, h)=2M_i h\log\frac 1 h,$ with suitably chosen random variables $M_i$. Concretely, assume the following \\

\begin{minipage}[r]{0.97\linewidth} {\bf A4}.
Let \begin{equation}\label{GnrlFAJModel}
        dX_{t}=a_{t}dt+\sigma_{t}dW_{t}+{dJ}_{t},
\end{equation}
where $J_{t}=\sum_{i=1}^{N_{t}}\gamma_{i}$ for a non-explosive counting process $N$ and real-valued random variables  $\gamma_j$, $a, \sigma$ are càdlàg and
a.s. $\underline{\sigma}^2:=\inf_{s\in[0,T]} \sigma^2_s>0$.\\
 \end{minipage}

\noindent Recall that   a.s. the paths of $a$ and of $\sigma$ are bounded on $[0,T]$. Define $\bar\sigma^2:=\sup_{s\in[0,T]} \sigma^2_s$,
then, the following Proposition and  Corollary hold true. Their proofs are in the Appendix.

\begin{prop}\label{ConvHatIVOptThr}
  Under {\bf A4},  if we choose   $r_i(h)=2M_i h\log\frac 1 h,$ with {any $M_i(\omega)$ such that $M_i(\omega)\in[ \inf_{s\in[t_{i-1},t_i]} \sigma^2_s(\omega),
  \bar\sigma]$,}
  we have:\\ \centerline{a.s. $\forall \eta>0$, for sufficiently small $h$:
 $\forall\,  i=1,\dots,n, \ \ \IDXqleqrieta=\IDNeqZ.$}
  \end{prop}

\begin{cor}\label{CORConvHatIVOptThr}
For all $\eta>0$,\ we have $\sumi {\DXq}\IDXqleqrieta
\stackrel{P}{\to} IV,$ as $h\to 0$.
\end{cor}

\section{CONDITIONAL MEAN SQUARE ERROR: FA jumps case}\label{Sect:CMSE}

\n We now put ourselves under {\bf A1}. The quantity of our interest here, $cMSE(\ep)\doteq E[(\hat{IV} -
IV)^2|\sigma, J]$, is such that $\forall \omega,\  cMSE(0)=IV^2$ and
%as soon as $J\not\equiv 0$ then
 $cMSE(+\infty)>0,$ because
$\hat{IV}\stackrel{\ep\to+\infty}{\to} QV.$ Further, from the proof
of Theorem \ref{expressionMSEingen}, we have
\begin{equation}\label{Ntngi}
cMSE'(\ep)=\ep^2
F(\ep), \mbox{ with } F(\ep)\doteq\sum_{i=1}^n a_i g_i, \quad g_i=
\ep^2 +2\sum_{j\neq i}b_j -2 IV.
\end{equation}
We analyze the sign of
$F(\ep)$: for $n,h$ fixed, $\sigma^2_i$ and $m_i$ also are fixed,
and we have $F(0)= -2 IV \sumi a_i <0,$ since $b_j(0)=0$. Further we have
$F(+\infty)= 0^+$: to see this, first note that, from the expression of $b_i(\ep)$,  $b_i(+\infty) = m_i^2+\sigma_i^2$, then $g_i(\varepsilon)\sim
\ep^2 + 2\sum_{j\neq i} m_j^2 -2\sigma^2_i\sim \ep^2$, as $\varepsilon\to+\infty$.
Moreover, each $a_i\sim
{2}(2\pi)^{-1/2}\sigma_{i}^{-1}\exp\left(-\frac{\ep^2}{2\sigma_i^2}\right)$,
thus, for sufficiently large $\ep$, $F=\sum_{i=1}^n a_i g_i$ is a
finite sum of $n$ positive terms $a_i g_i\leq K
(2\pi)^{-1/2}\sigma_{i}^{-1}\varepsilon^{2}\exp\left(-\frac{\ep^2}{2\sigma_i^2}\right)$
for some constant $K$ and fixed $\sigma_i,$ so
$F(\ep)\to 0^+,$ as $\ep\to+\infty$. % 2/11/16, folder "FA jumps"
Since $F$ is continuous, it follows that an optimal threshold exists and solves $F(\ep)=0.$\\

\n We now assume also {\bf A3}.

\begin{rem} Under {\bf A3}, as in Remark \ref{REMOptThto0}, if $\bar\ep=\bar\ep(h)$ minimizes cMSE, then  it has to be true that $\bar\ep\to 0,$ as $h\to 0$. In Proposition \ref{PropcMSEBarEpBiggerSqrth} below we again also find that under the following {\bf A4'} then  necessarily $\frac{\bar\ep(h)}{\sqrt h}\to+\infty$.\\ %in both the cases of FA or IA jumps.\\
\end{rem}

\begin{minipage}[c]{0.9\linewidth} {\bf A4'}. We assume {\bf A4} %(\ref{GnrlFAJModel})
with $a\equiv 0$, constant $\sigma>0$ and $nh=1.$\\
 \end{minipage}

\n Under FA jumps, when considering $h\to 0,$ we assume to have
a sufficiently small $h$ so that a.s. the number of jumps occurring during $\Ii$ is at most 1; note that for any $t$ we have  $m_i
\to \Delta J_t$, when selecting $i=i(t)$ such that $t_{i-1}<t\leq{}t_{i}$. Thus, when considering a jump time $t$, we assume that $h$ is sufficiently small so that the sign of $m_{i(t)}$ is the same as the one of $\Delta J_t$, in particular if $ \Delta J_t\neq 0$ then the increments $m_i$ approaching it are non-zero.

\subsection{Asymptotic behavior of $b_{i}(\varepsilon), {a_{i}(\varepsilon)}$, and $F$}\label{AsympBiepFA}

\n {The following {result} ensures that, as previously announced, an optimal threshold has to tend to 0, as $h\to 0$,  {but at a slower rate} than $\sqrt{h}.$ Its proof is in {the} Appendix.}

\begin{prop}\label{PropcMSEBarEpBiggerSqrth} Under {\bf A1, A3, A4'}, if $\bar\ep=\bar\ep(h)$ solves $F(\varepsilon)=0$ and $\bar\ep=\bar\ep(h)\to 0$,
%non si dice che \bar\ep is optimal
then $\frac{\bar\ep(h)}{\sqrt h}\to+\infty.$
\end{prop} % pag 39 del fascicolo FA js del gruppo 8/12/16

\n {We now pass to consider the asymptotic behavior of $F(\ep_h)$ for sequences $\ep=\ep(h)=\ep_h$ satisfying the conditions of Proposition \ref{PropcMSEBarEpBiggerSqrth}.} %Concretely,

\begin{prop}\label{PropcMSEFzeroEp}
Under {\bf A4',} if $\ep_h\to 0$ as $h\to 0$ in such a way that $\frac{\ep(h)}{\sqrt h}\to
+\infty$ then $F(\ep_h)= F_0(\ep_h)+{\rm \hot},$ where
$$F_0(\ep_h):=
%n s_h v_h\frac{\sqrt h}{\sigma \sqrt{2\pi}}\Big(v_h  -n  s_h \cdot
%2\sqrt{\frac 2 \pi} \sigma \Big) =
\frac{2\ep_h}{h\sqrt
h}e^{-\frac{\ep_h^2}{2\sigma^2h}} \Big(\ep_h-
\frac{e^{-\frac{\ep_h^2}{2\sigma^2h}}}{\sqrt h}\frac{4\sigma}{\sqrt{2\pi}}\Big)\frac{1}{\sigma\sqrt{2\pi}}. $$
\end{prop}

\vspace{0.5cm}
\n With the notation  $v_{h}:=\frac{\ep_{h}}{\sqrt{h}}$ and $s_h:= \frac{1}{\sqrt{2\pi}}e^{-\frac{v^2_h}{2\sigma^2}}$, we can write
$F_0(\ep_h)= \frac{2}{\sigma} \ep_h \frac{s_h}{h}\Big(v_h- 4\sigma s_h n\Big).$
Note that $v_h \ll n$, but $ s_h \to 0$, so which is the leading term between $v_h$ and $n  s_h$ depends on the choice of $v_h$. {We also remark that a solution $\bar{\bar\ep}$ of $F=0$ not necessarily is such that
$F_0(\bar{\bar\ep})=0$, however if {a sequence $\ep_h$ is such that $F_0(\ep_h)\to 0$} then the whole
$F(\ep_h)\to 0,$ so it has to be true that  $\ep_h$ is close (in a way that will become explicit later) to
one of the solutions $\bar{\bar\ep}$ of $F=0$.}

\begin{rem}\label{RemDriftcMSE}  The {asymptotic behavior of $F(\ep)$ {stated in Proposition \ref{PropcMSEFzeroEp} also holds under the presence of a nonzero drift process $\{a_{t}\}_{t\geq{}0}$} that has almost surely locally bounded paths (recall that any {c\'adl\'ag} process $a$ satisfies such a requirement) and that is independent on $W$. This is shown in {the} Appendix. }\end{rem}

\subsection{Asymptotic behavior of $\bar\ep$}\label{AsymBarEpFA}

{We show here that any cMSE optimal threshold $\bar\ep$ has the same asymptotic behavior as the MSE optimal threshold $\ep^{\star}.$ The proof of the {following result is given in the} Appendix.}

\begin{cor}\label{AsympOptThldFJA} {Under {\bf A1, A3, A4'}} we have that $$\bar\ep\sim \sqrt{2\sigma^{2}h\ln\frac{1}{h}}, \quad \text{as}\quad h\to{}0.$$
\end{cor}

\n {The previous result suggests an approximation for the optimal $\bar\ep:=\bar\ep_{h}$ of the form $\ep_h=\sigma w_{h}\sqrt{2h} $, with $w_{h}=\sqrt{\ln(1/h)}$. It is natural to wonder about other choices for $w_h$. Intuitively,
we should} aim at making {$F(\ep_h)$ to converge to $0$} as quickly as possible: in view of (\ref{condEpStar}) within the proof of Corollary \ref{AsympOptThldFJA},
the only possible way is rendering $v_h$ and $n  s_h$ {within $F_0(\ep)$} of
the same order,
so we choose $w_h$ such that
\beq
\begin{array}{c}
1)\  w_h \to +\infty,\\
2)\  w_h \sqrt h \to 0,\\
3)\  \frac{e^{-w^2_h}}{w_h h} \to \frac{\sqrt \pi}{2},
\end{array}
\eeq
as $h\to 0$.
For example a function of type $w_h= \sqrt{\ln \frac 1 h -\frac 1 2 \ln\ln\frac 1 h - \ln y_h},$
with any continuous function $y_h$ tending to $\frac{\sqrt \pi}{2}$ as $h\to 0$, satisfies the three above  conditions\footnote{We thank Andrey Sarychev
for having provided such nice examples.}.
%Then  $v_h= \sqrt 2 \sigma w_h$ satisfies (\ref{vSimns}).
 However the quickest convergence speed of $F$ to 0 would be reached by choosing a function %$v_h=\sqrt 2 \sigma w_h$ constructed using a
 $w_h$, which satisfies the following three more restrictive conditions, as $h\to 0$,
\beq\label{ephSalvatore}
\begin{array}{c}
1)\  w_h \to +\infty\\
2)\  w_h \sqrt h \to 0\\
\text{3'})\  \frac{e^{-w^2_h}}{w_h h} \equiv \frac{\sqrt \pi}{2},
\end{array}
\eeq where condition 3') means that $F_0(\ep_h)\equiv 0.$ In fact
such a $w_h$ exists, since the following holds true\footnote{We
thank Salvatore Federico for having provided such a nice result. The proof is available upon request.}.
%\end{rem}
\begin{thm}\label{Federico}
There exists a unique deterministic {function %$w_h$ of $h$ such that
$w_h:(0,1]\to (0,+\infty)$ such that the three
conditions 1), 2) and 3') above} are satisfied.
Such a $w_h$ turns out to be differentiable and to satisfy also the ODE $w'_h=\frac{w_h h}{1+2 w^2_h},$ which entails that
$w_h\leq w_1+ \frac{1}{2\sqrt 2}\log\frac 1 h.$\hfill\qed \\
\end{thm}

\n We finally reach the uniqueness of the optimal threshold $\bar\ep$ as a consequence of the following  result, whose proof is in {the} Appendix. We remark that the asymptotic behavior of $\bar\ep$ described in Corollary \ref{AsympOptThldFJA} is obtained after having proved just before (\ref{AsymEqEps}) that it has to satisfy $\bar{\ep}_h\sim4\sigma \frac{s_h}{\sqrt h}$, as $h\to{}0$.\\

\begin{prop}\label{ProfFprimeCMSE}
The first derivative $\frac{d}{d\ep} F(\ep)$ of $F$ is such that,
when evaluated at a function $\ep_h$ of $h$ satisfying $\ep_h\to 0$,
$\frac{\ep_h}{\sqrt h}\to +\infty$, and $\ep_h=4\sigma \frac{s_h}{\sqrt h}+ \hot$,
%\sqrt{2\sigma^{2}h\ln\frac{1}{h}}+ {\rm h.o.t.}$
then, as $h\to 0$,  %$\frac{d}{d\ep} F(\ep)$ satisfies
$$ F'(\ep_h) = F_1(\ep_h) +{\rm h.o.t.}, \mbox{ as } h\to 0, \quad \mbox{ where } \quad
F_1(\ep_h)= \frac{4}{\sigma^2\pi} e^{-\frac{\ep_h^2}{\sigma^2h}}\frac{\ep_h^2}{h^3}.$$
\end{prop}

\begin{rem} Uniqueness of $\bar\ep$. Since $F_1(\ep_h)>0$ for any $\ep_h,$ we reach that for sufficiently small $h$ we have
$\frac{d}{d\ep} F(\ep_h)>0$ on any sequence $\ep_h$ as in the above Proposition.
That entails that for any sufficiently small $h$ the cMSE optimal $\bar\ep$ is unique.
{Indeed,} if there existed two optimal $\bar\ep^{(1)}_h<\bar\ep^{(2)}_h$, we would necessarily have
 that $\bar\ep^{(i)}_h\to 0$,   $\frac{\bar\ep^{(i)}_h}{\sqrt h}\to +\infty$, and $\bar\ep^{(i)}_h=4\sigma \frac{\bar s_h}{\sqrt h}+ \hot$,
 but then, for small $h$,
 on such sequences {$F'(\bar\ep^{(i)}_h)>0$,
%  and then on such sequences $F$ is strictly increasing, and thus
% $F(\bar\ep^{(1)}_h)<F(\bar\ep^{(2)}_h),$
 which} is a contradiction, because in order to be optimal both sequences have to satisfy
 $F(\bar\ep^{(i)}_h)=0.$
\end{rem}

\begin{rem}
{The fact that the asymptotic behavior of the cMSE optimal threshold $\bar\ep=\bar\ep(h)$ is the same as the one of the
MSE optimal threshold $\ep^{\star}$ under FA jumps is due to the fact that   $\bar\ep$ solves $F=0$, $\epst$ solves $G=0$, $F= F_0 + h.o.t.$, $G= G_0 + h.o.t.$, and the leading terms
 in $F$ are the ones with $m_i=0$, which do not depend on $\omega$, thus they are the same as for $G$. It follows that,
in the case of Lévy FA jumps, we have $F= F_0 + h.o.t.=E[F_0] + h.o.t.=G+\hot$.} Also, an alternative heuristic justification is that
we expect that $F(\ep)=\frac{\sumi a_i g_i }{n} \cdot n \sim n E[a_i g_i],$ thus the asymptotic
behavior of the $\epst$ satisfying $G=n E[a_i g_i]=0$ is the same as any $\bar\ep$ satisfying $F(\ep)=0.$
 \end{rem}

\begin{rem}
Comparison with the results in \cite{FigNis13}.
In  \cite{FigNis13}, {a process $X$ with FA jumps} is considered, either of Lévy type, with jumps sizes having  distribution density satisfying given conditions, or of It\^o SM type, with
deterministic absolutely continuous local characteristics (additive process). The estimators  $$\hat{J}_n= \sumi \DX I_{\{ |\DX|> \ep_h\}},\quad \hat{N}_n= \sumi I_{\{ |\DX|> \ep_h\}}$$
are considered, and, as $h\to 0$,  firstly it is shown that the condition $\frac{\ep_h}{\sqrt h}\to +\infty$ is  necessary and sufficient for the convergence to 0 of
both $MSE(\hatIV-IV)$ (stronger condition implying consistency of $\hatIV$) and $MSE(\hat{J}_n-J_T)$.
Secondly, the authors show that
$$MSE(\hat{N}_n-N_T) \to 0 \Leftrightarrow \frac{e^{-\frac{\ep_h^2}{2\sigma^2h}}}{\sqrt h \ep_h}\to 0,$$
meaning that in order to have $L^2(\Omega,P)$ convergence to 0 of the estimation error $\hat{N}_n-N_T$ a stronger condition on $\ep_h$ is needed, implying $\frac{\ep_h}{\sqrt h}\to \infty$. %(3.2) implies (3.6)
Thirdly, existence and uniqueness of an optimal threshold $\check{\ep}(h)$ minimizing
$$E[|\hat{IV}_n-IV|^2 + |\hat{N}_n-N_T|^2] $$ %asymptotically equivalent to min Loss^{(2)}
for fixed $h$ is obtained, and the asymptotic expansion in $h$ of $\check{\ep}(h)$ has leading term
$\sqrt{3\sigma^2 h \log\frac 1 h}$. The factor 3 is higher than the factor 2 of the leading terms of $\bar\ep$ and $\epst$: that is due to the fact that the minimization criterion for  $\check{\ep}(h)$ includes also the error on $N_T,$ which requires that $ \frac{\check{\ep}(h)}{\sqrt h}$ is higher than $\frac{\bar\ep}{\sqrt h}$, and thus $\check{\ep}(h)>\bar\ep(h)$ is necessary.
\end{rem}

\section{A NEW METHOD}\label{FACVP0}
 In this section, we propose a new method {for tuning} the threshold parameter $\ep:=\sqrh$ of the TRV
introduced in (\ref{TRV00}). This is based on the conditional mean square error $cMSE(\ep)= E[(\hat{IV} -
IV)^2|\sigma, J]$ studied in Section \ref{Sect:CMSE}. We illustrate the method for a driftless FA process with constant volatility $\sigma$. As proved
therein, the optimal threshold $\bar\ep$ is such that
\[
    F(\bar\ep)=\sum_{i=1}^n a_i(\bar\ep) g_i(\bar\ep)=0,\quad
     g_i(\bar\ep)= {\bar\ep}^2 +2\sum_{j\neq i}b_j(\bar\ep) -2 nh\sigma^{2},
\]
where $a_{i}(\ep)$ and $b_{i}(\ep)$ are rewritten here for easy
reference:
\begin{align*}
a_i(\ep)&:= a(\ep,m_{i},\sigma):=\frac{e^{-\frac{(\ep-m_i)^2}{2\sigma^2 h}} +
e^{-\frac{(\ep+m_i)^2}{2\sigma^2 h}}}{\sigma \sqrt{2\pi h}},\\
     b_{i}(\varepsilon)&:= b(\ep,m_{i},\sigma):=-\frac{\sigma\sqrt{h}}{\sqrt{2\pi}}\left(e^{-\frac{(\varepsilon-m_{i})^{2}}{2{\sigma^{2}}h}}(\varepsilon+m_{i})+
     e^{-\frac{(\varepsilon+m_{i})^{2}}{2{\sigma^{2}}h}}(\varepsilon-m_{i})\right)+\frac{m_{i}^{2}+{\sigma^{2}}h}{\sqrt{2\pi}}
     \int_{\frac{m_{i}-\varepsilon}{{\sigma}\sqrt{h}}}^{\frac{m_{i}+\varepsilon}{{\sigma}\sqrt{h}}}e^{-x^{2}/2}dx.
\end{align*}
{It is convenient to} set ${\bf m}=(m_{1},\dots,m_{n})$ and
\[
    F(\ep;\sigma,{\bf m}):=\sum_{i=1}^{n}a(\ep,m_i,\sigma)\left(\ep^2 +2\sum_{j\neq i}b(\ep,m_{j},\sigma) -2 nh\sigma^{2}\right){.}
\]
The main issue with the optimal threshold $\bar\ep$ lies on the fact that this depends on $\sigma$ and the
increments ${\bf m}=(m_{1},\dots,m_{n})$ of the
 jump process, which we don't know. Note also that, for $h$ small enough, each $m_{i}$ will be either $0$ or one of the jumps of the process and a good
 proxy of $m_{i}$ is actually $(\Delta_{i}^{n}X) {\bf 1}_{\{|\Delta_{i}^{n}X|>\bar\ep\}}$. The idea is then to iteratively estimating $\bar\ep$,
 $\sigma$, and ${\bf m}$ as follows: 
\begin{enumerate}
    \item Start with some initial `guesses' of $\sigma$ and ${\bf m}$, which we call $\hat{\sigma}_{0}$ and
    $\hat{{\bf m}}_{0}$. There are different possibilities for these initial values, for instance
      $\hat{\sigma}_{RV}$ {(defined in item 1 of Section \ref{SimulFACV}) or $\hat{\sigma}_{BV}$ (defined in item 2 of Section \ref{SimulFACV})  or a truncated $\hat{\sigma}_{TRV}$ (defined in {item 12} of Section \ref{SimulFACV})}  with threshold  $
        \sqrt{2 \hat{\sigma}_{BV}^2 h \log(1/h)}$, for $\sigma$, and
    $\hat{{\bf m}}_{0} =(0,\dots, 0)$ (no jumps) for ${\bf m}$.

   \item Using $\hat{\sigma}_{0}$ and $\hat{{\bf m}}_{0}$, {by solving $F(\ep;\hat{\sigma}_{0},\hat{{\bf m}}_{0})=0$,} we find an initial estimate for the optimum $\bar\ep$ that we denote
    $\ep_{NEW}$. {For instance with the choice of $\hat{{\bf m}}_{0}=(0,\dots,0)${,} $\ep_{NEW}$ solves the equation}:
\begin{equation}\label{BdInGs}
    \ep^2 +2(n-1)\left(-\frac{2{\hat{\sigma}_{0}}\sqrt{h}}{\sqrt{2\pi}}
    {\varepsilon}\,e^{-\frac{\varepsilon^{2}}{2{{\hat{\sigma}_{0}}^{2}}h}}+
    \frac{{{\hat{\sigma}_{0}}^{2}}h}{\sqrt{2\pi}}
     \int_{\frac{-\varepsilon}{{{\hat{\sigma}_{0}}}\sqrt{h}}}^{\frac{\varepsilon}{{{\hat{\sigma}_{0}}}
     \sqrt{h}}}e^{-x^{2}/2}dx\right)
     -2 nh{\hat{\sigma}_{0}}^{2}=0.
\end{equation}
It is easy to see that, in that case, $\ep_{NEW}$ is of the form $v_{n}\hat{\sigma}_{0}\sqrt{h}$, where $v_{n}$ is the unique solution of the equation:
        \begin{equation}\label{SolvnEq}
    v_{n}^{2} +4(n-1)\left(-v_{n}\frac{1}{\sqrt{2\pi}}e^{-\frac{v_{n}^{2}}{2}}+\frac{1}{\sqrt{2\pi}}
     \int_{0}^{v_{n}}e^{-x^{2}/2}dx\right) -2 n=0.
\end{equation}
Figure \ref{Plot1} shows that $v_{n}$ ranges from about $3$ to $4$ when $n$ ranges from 100 to 10000.
\begin{figure}[htp]
    {\par \centering
    \includegraphics[width=8.0cm,height=7.5cm]{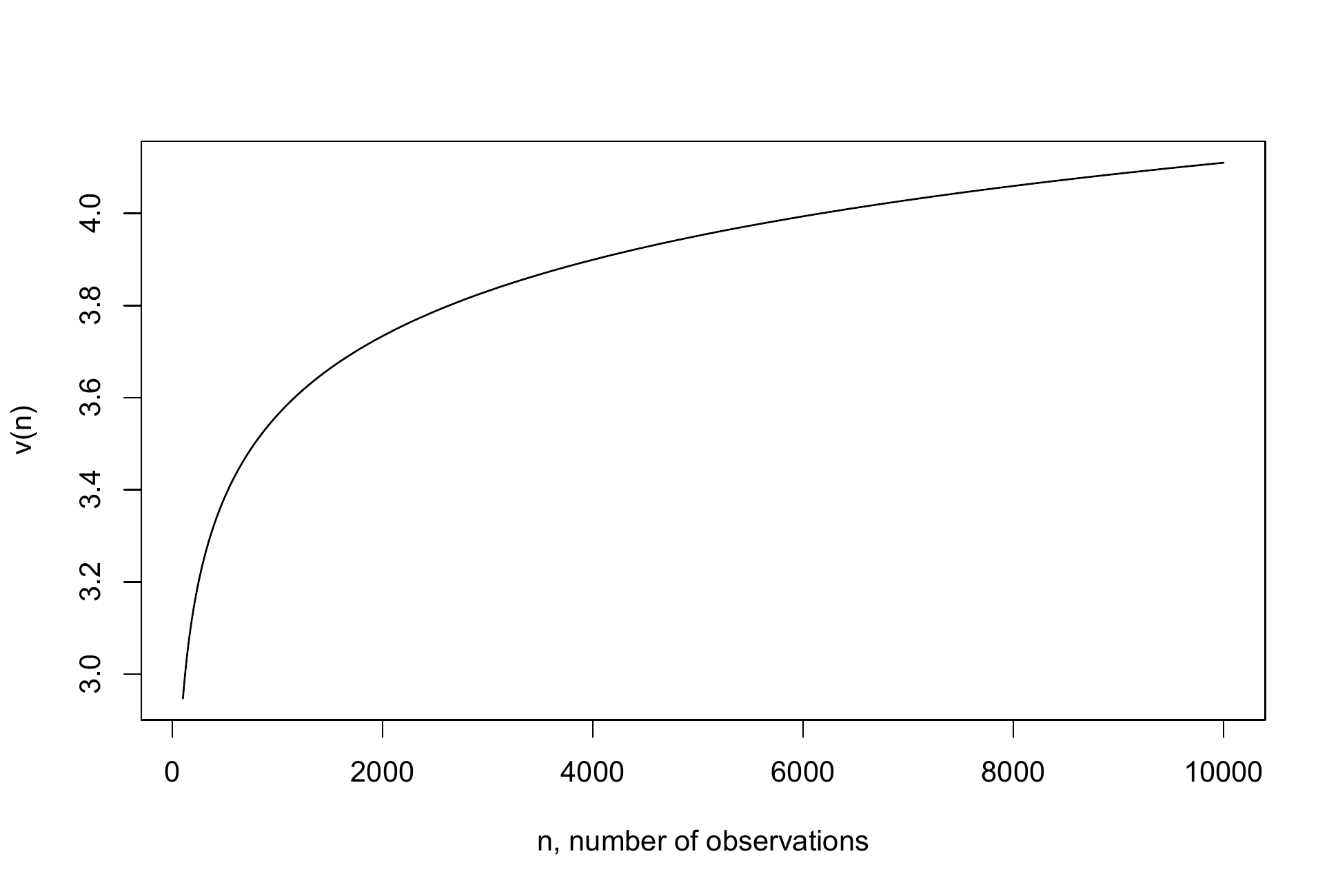}\par}\vspace{-.8 cm}
    \caption{The solution $v_{n}$ of equation (\ref{SolvnEq}) as a function of $n$.}
\label{Plot1}
\end{figure}

\item Once we have an initial estimate of $\bar\ep$, we can update our estimates of $\sigma$ and ${\bf m}$ using:
\begin{equation}\label{DfnOneItr}
    \hat{\sigma}_{NEW}^{2}:=\frac{1}{T} \sum_{i=1}^n \DXq {\bf 1}_{\{|\DX|\leq\ep_{NEW}\}},\quad
    \hat{{\bf m}}_{1}:=((\Delta_{1}^{n}X) {\bf 1}_{\{|\Delta_{1}^{n}X|>\ep_{NEW}\}},\dots,
    (\Delta_{n}^{n}X) {\bf 1}_{\{|\Delta_{n}^{n}X|>\ep_{NEW}\}})
\end{equation}
\item We continue this procedure {iteratively: $\ep_{NEW,0}:= \ep_{NEW}$,
$\hat{\sigma}_{NEW,1}:= \hat{\sigma}_{NEW},$ and for $k\geq 1$
\begin{align}\label{ItrNewMeth}
	&\text{Find}\quad\ep_{NEW,k}\quad\text{s.t.}\quad F(\ep_{NEW,k};\hat{\sigma}_{NEW,k},\hat{{\bf m}}_{k})=0,\\
&\text{set}\quad  \hat{\sigma}_{NEW,k+1}^{2}:= \frac{1}{T}
\sum_{i=1}^n \DXq {\bf 1}_{\{|\DX|\leq\ep_{NEW,k}\}},\\
    &\quad\quad\hat{{\bf m}}_{k+1}:=((\Delta_{1}^{n}X)
    {\bf 1}_{\{|\Delta_{1}^{n}X|>\ep_{NEW,k}\}},\dots,
    (\Delta_{n}^{n}X) {\bf 1}_{\{|\Delta_{n}^{n}X|>\ep_{NEW,k}\}}).\label{ItrNewMethLst}
\end{align}
The algorithm is stopped} when the sequence of estimates $\hat{\sigma}_{NEW,k}$ stabilizes
(e.g., when\\ $|\hat{\sigma}_{NEW,k+1}-\hat{\sigma}_{NEW,k}|/$ $\hat{\sigma}_{NEW,k}\leq{\rm tol}$,
for some desired small tolerance ${\rm tol}$).

\end{enumerate}

The previous procedure resembles the one introduced {in} \cite{FigNis13}, which is based on choosing the threshold $\ep$ so to minimize
%{$E[|\hatIV-IV|^2+ |\hat{N}_n-N_T|^2]$, or equivalently}
the expected number of {jumps} miss-classifications:
\begin{equation}\label{LossFN}
    {\rm {Loss}}(\ep):=E\left[\sum_{i=1}^{n}\left({\bf 1}_{\{|\Delta_{i}^{n}X|>\ep,\Delta_{i}^{n}N=0\}}+
    {\bf 1}_{\{|\Delta_{i}^{n}X|\leq{}\ep,\Delta_{i}^{n}N>0\}}\right)\right].
\end{equation}
It was proved therein that, for {a L\'evy process with FA jumps}, the optimal
{threshold},  hereafter denoted {$\ep_{3mc}$}, is asymptotically
equivalent to $\sqrt{3\sigma^{2}h\ln(1/h)}$, as $h\to{}0$ \footnote{{`mc' in the notation $\ep_{3mc}$ refers to `modulus of continuity' of the Brownian motion.}}. Using
this information, an iterative method was proposed, in which, given
an initial estimate {$\hat{\sigma}_{3mc,0}$ of $\sigma$,  we set, for $ k\geq 1$,
\begin{equation}\label{FNMeth}
    \ep_{3mc,k-1} :=\sqrt{3\hat{\sigma}^{2}_{3mc,k-1}h\ln\frac{1}{h}},\quad
    \hat{\sigma}_{3mc,k}^{2}:={\frac{1}{T}}
    \sum_{i=1}^n \DXq {\bf 1}_{\{|\DX|\leq \ep_{3mc,k-1}\}}.
\end{equation}}

 Since, as proved in Section \ref{Sect:CMSE}, the optimal threshold {$\bar\ep$ that minimizes cMSE for given $h$} has the asymptotic behavior $\sqrt{2\sigma^{2}h\ln(1/h)}$,
as $h\to{}0$, it is natural {to} consider the following  iterative method to estimate $\bar\ep$:
given an initial guess $\hat{\sigma}_{2mc,0}$ for $\sigma$, {we set}
\begin{equation}\label{FMancMeth}
    \ep_{2mc,k-1}:=\sqrt{2\hat{\sigma}^{2}_{2mc,k-1}h\ln\frac{1}{h}},\quad
    \hat{\sigma}_{2mc,k}^{2}:={\frac{1}{T}}
    \sum_{i=1}^n \DXq {\bf 1}_{\{|\DX|\leq\ep_{2mc,k-1}\}},\quad k\geq 1.
\end{equation}
{We can go one step further and consider, as suggested below Corollary \ref{AsympOptThldFJA}, a threshold of the form $\ep_h=\sigma w_h\sqrt{2h} $, with $w_{h}$ given as in (\ref{ephSalvatore})\footnote{In order to obtain $w_h$, we change variable, as $x_h=w_h^2$,  in 3') and then we use a fixed-point algorithm  to find the solution $x_h$,  starting with $x_h(0)=-\log(h)$. The algorithm converges very quickly.}.
This leads us to consider the iterative method:
\begin{equation}\label{FMancMeth2}
    \ep_{mc_{2},k-1}:=w_{h}\sqrt{2\hat{\sigma}^{2}_{mc_{2},k-1}h},\quad
    \hat{\sigma}_{mc_{2},k}^{2}:={\frac{1}{T}}
    \sum_{i=1}^n \DXq {\bf 1}_{\{|\DX|\leq\ep_{mc_{2},k-1}\}},\quad k\geq 1.
\end{equation}
It can be proved that
if we take $\hat{\sigma}_{3mc,0}$, %=\hat{\sigma}_{0}^{2}$
$\hat{\sigma}_{2mc,0}$, and $\hat{\sigma}_{mc_{2},0}$} equal $\hat\sigma_{RV}$ in (\ref{FNMeth}),  (\ref{FMancMeth}), and (\ref{FMancMeth2}), then the {obtained} sequences of estimates
$\{\hat{\sigma}_{2mc,k}\}_{k\geq{}0}$,
$\{\hat{\sigma}_{3mc,k}\}_{k\geq{}0}$, {$\{\hat{\sigma}_{mc_{2},k}\}_{k\geq{}0}$ {are} nonincreasing and, thus,
eventually they reach a constant limiting value}.
So, {for these two estimators}
we can (and will) set the tolerance ${\rm tol}$ to $0$. Even though asymptotically $w_{h}\sim \sqrt{\ln(1/h)}$, there are some differences in finite samples. For instance, for the span of 5 minutes used in our simulations ($h=\frac{1}{252\times 6.5\times 12}$), we have  $w_{h}=2.98$, while  $\sqrt{\ln(1/h)}=3.14$, which means that the $\ep_{mc_{2},k}$ will be smaller than $\ep_{2mc,k}$.

\subsection{Simulation performance: finite actvity jumps and constant volatility}\label{SimulFACV}
We now proceed to assess the methods introduced in this paper and compare them against other popular alternatives. We take a Merton's log-normal model of the form:
\begin{equation}\label{MertonLogNormal}
    X_{t}=\sigma W_{t}+\sum_{j=1}^{N_{t}}\gamma_{j},
\end{equation}
%{\Green $\blacktriangleright$ [$a=0$ in fact]}
where $N$ is a Poisson process with intensity $\lambda$ and $\{\gamma_{i}\}_{i\geq{}1}$ is an independent sequence of independent {normally}
distributed random variables with mean and standard deviation $\mu^{{\rm Jmp}}$ and $\sigma^{{\rm Jmp}}$, respectively.
We consider the following estimators:
\begin{enumerate}
    \item {The Realized quadratic Variation estimator:} ${\hat{\sigma}^2_{RV}}:=
    T^{-1}\sum_{i=1}^{n}(\Delta_{i}^{n}X)^{2}$;

    \item {The realized Bipower Variation (BV) estimator of \cite{BaNShe06}:}
    \[
   	 \hat{\sigma}_{BV}^{2}:=\frac{\pi}{2T}\sum_{i=1}^{n-1}|\Delta_{i}X||\Delta_{i+1}X|;
	\]
	
    \item {The MinRV estimator of \cite{Andersen12}:}
    \[
     \hat{\sigma}_{MinRV}^{2}:=\frac{\pi}{T(\pi-2)}\frac{n}{n-1}\sum_{i=1}^{n-1}\min\{|\Delta_{i}X|,|\Delta_{i+1}X|\}^{2};
	\]
	
	\item { The MedRV estimator of \cite{Andersen12}:}
    \[
    \hat{\sigma}_{MedRV}^{2}:=\frac{\pi}{T(\pi+6-4\sqrt{3})}\frac{n}{n-2}\sum_{i=2}^{n-1}{\rm median}\{|\Delta_{i-1}X|,|\Delta_{i}X|,|\Delta_{i+1}X|\}^{2};
	\]

    \item {{The TRV given in} (\ref{TRV00a})} using a threshold of the form
    $\ep=4 h^{\omega}\hat{\sigma}_{BV}$ with $\omega=0.49$. This was used in the recent
    work of Jacod and Todorov \cite{JcdTodorov} and is denoted $\hat{\sigma}^2_{TRV_{JT}}$;

    \item The estimator $\hat{\sigma}^2_{3mc}$ {as in (\ref{FNMeth}) with $k=1$}, using the {initial} threshold
    ${\ep_{3mc,0}:=\sqrt{3 \hat\sigma_{RV}^2 h \log(1/h)}}$; %\sqrt{T^{-1}\sum_{i=1}^{n}(\Delta_{i}^{n}X)^{2}}$;

    \item The estimator $\hat{\sigma}^2_{3mc,k}$ defined by (\ref{FNMeth}) with {$k\geq 1$
    such that $\hat{\sigma}_{3mc,\ell}=\hat{\sigma}_{3mc,\ell-1}$, for all $\ell\geq k$};

    \item The estimator {$\hat{\sigma}^2_{2mc}$} {as in (\ref{FMancMeth}) with {$k=1$}}, using
    the {initial} threshold {$\ep_{2mc, 0}:=\sqrt{2 \hat\sigma_{RV}^2 h \log(1/h)}$};
    %=\sqrt{T^{-1}\sum_{i=1}^{n}(\Delta_{i}^{n}X)^{2}}$;

    \item  The estimator $\hat{\sigma}^2_{2mc,k}$ defined by {the} iterative formulas
    (\ref{FMancMeth}) and {with $k\geq 1$ such that
    $\hat{\sigma}_{2mc,\ell}=\hat{\sigma}_{2mc,\ell-1}$ for all $\ell\geq k$};

	\item The estimator {$\hat{\sigma}^2_{mc_{2}}$} {as in (\ref{FMancMeth2}) with {$k=1$}}, using
    the {initial} threshold {$\ep_{mc_{2}, 0}:= w_{h}\sqrt{2 \hat\sigma_{RV}^2 h} $};
	
	\item {The estimator $\hat{\sigma}^2_{mc_{2},k}$ defined by {the} iterative formulas
    (\ref{FMancMeth2}) and {with $k\geq 1$ such that
    $\hat{\sigma}_{mc_{2},\ell}=\hat{\sigma}_{mc_{2},\ell-1}$ for all $\ell\geq k$};}
	
    \item The estimator $\hat{\sigma}^2_{NEW}$ as defined in (\ref{DfnOneItr}) where $\ep_{NEW}$
    is such that $F(\ep_{NEW};\hat{\sigma}_{0},\hat{{\bf m}}_{0})=0$, with initial
    guesses $\hat{{\bf m}}_{0}=(0,\dots,0)$ and
    $\hat{\sigma}_{TRV}^{2}:=T^{-1}\sum_{i=1}^{n}(\Delta_{i}^{n}X)^{2}
    {\bf 1}_{\{|\Delta_{1}^{n}X|\leq \ep_{as}\}}$, with
    $\ep_{as}:=\ep_{2mc,0}=\sqrt{2 \hat{\sigma}_{BV}^2 h \log(1/h)}$;%$\bar{\ep}_{0}^{\star}$

    \item $\hat{\sigma}_{NEW,k}$ found with the new method described by the iterative formulas
    (\ref{ItrNewMeth})-(\ref{ItrNewMethLst}), with initial guesses %$\hat{\sigma}_{0}$ and
    %$\hat{{\bf m}}_{0}$
    given as in the previous item and $k$ determined by the stopping rule
    $|\hat{\sigma}_{NEW,k}-\hat{\sigma}_{NEW,k-1}|/\hat{\sigma}_{NEW,k-1}\leq {\rm tol}=10^{-5}$;

    \item {An Oracle type estimator of the form
        \[
    		\hat{\sigma}_{{\rm Orc}}^{2}:=
       \sum_{i=1}^n \DXq {\bf 1}_{\{|\DX|\leq\ep_{{\rm Orc}}\}},
	   \]
    where} $\ep_{{\rm Orc}}$ is such that $F(\ep_{{\rm Orc}};{\sigma},{{\bf m}})=0$, using the true
    values of the volatility $\sigma$ and of the  jump vector
    ${\bf m}=(m_{1},\dots,m_{n})=(\Delta_{1}J,\dots,\Delta_{n}J)$;

    \item  The following estimator based on the Threshold Bipower Variation (TBV):
    \[
    \hat{\sigma}^2_{TBV}:=\frac{\pi}{2T}
    \sum_{i=1}^{n-1}|\Delta_{i}X||\Delta_{i+1}X|
    {\bf 1}_{\{|\Delta_{i}X|\leq\ep_{TBV}\}}{\bf 1}_{\{|\Delta_{i+1}X|\leq\ep_{TBV}\}},
	\]
	using a threshold of the form $\ep_{TBV}:=4 h^{\omega}\hat{\sigma}_{BV}$ with $\omega=0.49$:

	 \item  The iterated TBV estimator given by: $\hat{\sigma}_{TBV,1}:= \hat{\sigma}_{TBV}$,
    \[
    		\ep_{TBV,k}:=4 h^{\omega}\hat{\sigma}_{TBV,k},\quad
     \hat{\sigma}^2_{TBV, k+1}:=\frac{\pi}{2T}\sum_{i=1}^{n-1}|\Delta_{i}X||\Delta_{i+1}X|
     {\bf 1}_{\{|\Delta_{i}X|\leq\ep_{TBV,k}\}}{\bf 1}_{\{|\Delta_{i+1}X|\leq\ep_{TBV,k}\}},
     \quad k\geq 1,
	\]
	using $\omega=0.49$ and $\hat{\sigma}_{TBV}^{2}$ as defined in the previous
    item.\footnote{The estimators in {items 15 and 16} were suggested by {an anonymous}  referee.}
    We stop when $|\hat{\sigma}_{TBV,k}-\hat{\sigma}_{TBV,k-1}|/
    \hat{\sigma}_{TBV,k-1}\leq{\rm tol}=10^{-5}$.
\end{enumerate}

\begin{rem}
	Different variations of the above estimators, that are not shown here for sake of brevity,
were also analyzed in our simulations.  For instance, the 3 alternative  thresholds $\ep=h^{\omega}$ and
$\ep=2h^{\omega}$, with $\omega=0.495$, were implemented; each one of the estimators
in  items 15 and 16 was also implemented with thresholds $3 h^{\omega}\hat{\sigma}_{BV}$ and
$ 5 h^{\omega}\hat{\sigma}_{BV}$. The results of these variations were suboptimal to those shown
here. We also implemented the estimators in the items 6 to 11 starting with an initial threshold of the form $\sqrt{2 \hat\sigma_{BV}^2 h \log(1/h)}$ (i.e., using $\hat{\sigma}_{BV}$  rather than $\hat{\sigma}_{RV}$ as an initial guess for  $\sigma$), and  the same stopping condition as therein: in these cases we  obtained the same performances  for the liming estimators.
\end{rem}

The adopted time unit of measure is 1 year {(252 days)} and we consider 5 minute observations over a 1 month time horizon with a 6.5 {hours per day}
open market. For our
first simulation experiment, we use the following parameters:
\begin{equation}\label{Prmtr1}
    \sigma=0.4,\quad \sigma^{Jmp}=3\sqrt{h}, \quad \mu^{Jmp}=0, \quad
    \lambda=100, \quad h=\frac{1}{252\times 6.5\times 12}.
\end{equation}
{The dependence {of $\sigma^{Jmp}$ on} $\sqrt{h}$ was done for {an} easier comparison with {the} standard deviation of the increments of the continuous component, which is $0.4\sqrt{h}$. So, the standard deviation of the jumps is about 7.5 times the standard deviation of the continuous component increment. The parameter values in (\ref{Prmtr1})} yield an expected annualized volatility of 0.45, which is reasonable.
{Table \ref{Table1} below shows the sample biases,
standard deviations, and MSE's based on 5000 simulations. We also show the sample {version} of Loss, i.e., the {expected} number of jump misclassifications as defined by (\ref{LossFN}), with its standard deviation; the sample average of $N$, i.e. the number of iterations needed to find the estimator's value, with its standard deviation; and, for the methods using truncation, the average threshold of the last step of the iteration used to obtain the estimate of $\sigma.$}

As expected, the unfeasible oracle estimator, which is shown as a benchmark for the other estimators,
performs the best, followed by the estimators $\hat{\sigma}_{NEW}$ and $\hat{\sigma}_{NEW,k}$ based on
finding the root of $F(\ep;\sigma,{\bf m})$. The iterative {estimators $\hat{\sigma}_{2mc,k}$ and $\hat{\sigma}_{mc_{2},k}$, based on the
thresholds $\sqrt{2 \sigma^{2} h\ln(1/h)}$ and $\sqrt{2 \sigma^{2} h}w_{h}$, also have a good performance and significantly improve} on the
estimator $\hat{\sigma}_{3mc,k}$ (number 7 above) proposed in \cite{FigNis13} and based on
$\sqrt{3 \sigma^2 h\ln(1/h)}$. The estimator $\hat{\sigma}_{TRV_{JT}}$ proposed by Jacod and Todorov
\cite{JcdTodorov} also performs quite well {in terms
     of MSE, but the estimation relative error {is comparatively large}.}
     The estimators based on TBV (namely, the estimators $\hat{\sigma}_{TBV}$ and $\hat{\sigma}_{TBV,k}$
     of items {15 and 16 above}) as well as the MinRV and MedRV are suboptimal for the considered parameters
     choice.

\begin{table}[h]
    {\par \centering
\begin{tabular}{ccrcrccccll}
\toprule
        &       &  ${\rm mean}$  & ${\rm std}$  & ${\rm MSE}(\hat{\sigma})$ &
${\rm mean}$   &  ${\rm std}$  & %Loss
${\rm mean}$   &  ${\rm std}(\ep)$  & %\ep
${\rm mean}$   &  ${\rm std}$  \\ % N
& \rm{Estimator} & $\frac{\hat\sigma^2- \sigma^2}{\sigma^2}$ &
$\frac{\hat\sigma^2- \sigma^2}{\sigma^2}$  & $\times 10^{5}$ &
$\text{Loss} $& $\text{Loss}$ &
$\ep$ & {$\times 10^{3}$}  &
$N$ &  $N$ \\
\hline
1& $\hat{\sigma}_{RV}$  & 0.28625 & 0.17562         & 288.7300\\
2& $\hat{\sigma}_{BV}$  & 0.06664 & 0.05517        & 19.1650 \\
3& $\hat{\sigma}_{{\rm MinRV}}$& 0.01563 & 0.05117  & 7.3287 \\
4& $\hat{\sigma}_{{\rm MedRV}}$  & 0.01799& 0.04593 & 6.2292\\
5& $\hat{\sigma}_{TRV_{JT}}$  & 0.00992 & 0.03712   & 3.7799 &  3.825 &  1.948 & 0.0130 &  0.34 & 1 & 0\\
6 & $\hat{\sigma}_{3mc}$  & 0.02971 & 0.04262 & 6.9121 & 4.905 & 2.257 & 0.0176 &  1.20 &  1 &  0\\
7& $\hat{\sigma}_{3mc,k}$  & 0.02033 & 0.03978 & 5.1097 & 4.488 & 2.154 & 0.0157 &  0.30 & 2.30 & 0.52\\
8& $\hat{\sigma}_{2mc}$  & 0.01500 & 0.03822 & 4.3174  & 4.161 & 2.060 & 0.0144 & 0.98 & 1 & 0\\
9& $\hat{\sigma}_{2mc,k}$  & 0.00908 & 0.03698 & {3.7127} & 3.776 & 1.929 & 0.0127 & 0.23 & 2.30 & 0.52 \\
10& $\hat{\sigma}_{mc_{2}}$  & 0.01190 & 0.03712 & 3.8920 & 3.981 & 1.974 &  0.0136 & 0.93 & 1 & 0\\
11& $\hat{\sigma}_{mc_{2},k}$  & 0.00654 & 0.03646 & 3.5133 & 3.622 & 1.886 & 0.0120 & 0.21 & 2.31 & 0.49\\
12& $\hat{\sigma}_{NEW}$ & -0.00046 &  {0.03623} &  \textbf{3.3605} & 3.532 & 1.812 & 0.0105 & 0.46 &  1 &  0 \\
13& $\hat{\sigma}_{NEW,k}$ & -0.00048 &  {0.03622} &  \textbf{3.3593} &  3.552 & 1.819 & 0.0106 & 0.54 & 1.70 & 0.56\\
14& $\hat{\sigma}_{Orc}$ &  -0.00373 &  0.03463 &  \textbf{3.1072} &  3.647 &  2.012 &  0.0102 &  0.58 &  1 &  0\\
15& $\hat{\sigma}_{TBV}$ & 0.00185 & 0.04130 & 4.3759 & 3.825 & 1.948 & 0.0130 & 0.34 & 1 & 0\\
16& $\hat{\sigma}_{TBV,k}$ & 0.00110 & 0.04124 & 4.3586 & 3.722 & 1.908 & 0.1260 & 0.25 & 2.08 & 0.32\\
\bottomrule
\end{tabular}
\caption{
{Estimation of the volatility $\sigma=0.4$ for a log-normal Merton model, based on simulated
5-minutes observations of 5000 paths} over a 1 month
time horizon. The jump parameters are $\lambda=100$, {$\sigma^{Jmp}={3}\sqrt{h}$} and $\mu^{Jmp}=0$.
}
\label{Table1}
}
\end{table}

\begin{table}[h]
    {\par \centering
\begin{tabular}{ccrcrrcccll}
\toprule
        &       &  ${\rm mean}$  & ${\rm std}$  & ${\rm MSE}(\hat{\sigma})$ &
${\rm mean}$   &  ${\rm std}$  & %Loss
${\rm mean}$   &  ${\rm std}(\ep)$  & %\ep
${\rm mean}$   &  ${\rm std}$  \\ % N
& \rm{Estimator} & $\frac{\hat\sigma^2- \sigma^2}{\sigma^2}$ &
$\frac{\hat\sigma^2- \sigma^2}{\sigma^2}$  & $\times 10^{5}$ &
$\text{Loss} $& $\text{Loss}$ &
$\ep$ & {$\times 10^{3}$}  &
$N$ &  $N$ \\
\hline
1& $\hat{\sigma}_{RV}$  & 0.57126  & 0.24671 & 991.2600\\
2& $\hat{\sigma}_{BV}$  & 0.13690 & 0.06899 & 60.1700\\
3& $\hat{\sigma}_{{\rm MinRV}}$  &  0.03533 & 0.05833 & 11.9000\\
4& $\hat{\sigma}_{{\rm MedRV}}$  &  0.04192 & 0.05592 & 12.5000\\
5& $\hat{\sigma}_{TRV_{JT}}$ & 0.02341 & 0.03962 & 5.4235 & 7.78 & 2.80 & 0.0134 & 0.40 & 1 & 0\\
6& $\hat{\sigma}_{3mc}$  &0.08219 & 0.05847 & 26.0480 & 10.58 & 3.39 & 0.0194  & 1.50 & 1 & 0\\
7& $\hat{\sigma}_{3mc,k}$  & 0.04364 & 0.04500 & 10.0620 & 9.01 & 3.05 & 0.0158 & 0.34 & 2.85 &  0.62\\
8& $\hat{\sigma}_{2mc}$  & 0.04353 & 0.04507 & 10.0500 & 8.97 & 3.08 & 0.0158 & 1.20 & 1 & 0\\
9& $\hat{\sigma}_{2mc,k}$  & 0.01941 & 0.03893 & 4.8453 & 7.49 & 2.75 & 0.0128 & 0.24 & 2.80 & 0.57 \\
10& $\hat{\sigma}_{mc_{2}}$  & 0.03561 & 0.04278 & 7.9338 & 8.53 & 3.05 & 0.0150 & 1.15 & 1 & 0\\
11& $\hat{\sigma}_{mc_{2},k}$  & 0.01471 & 0.03806 & 4.2643 & 7.14 & 2.74 & {0.0121} & 0.22 & 2.78 & 0.55 \\
12 & $\hat{\sigma}_{NEW}$ & 0.00389 & 0.03762 & \textbf{3.6632} & 6.70 & 2.59 & 0.0106 & 0.50 &  1 & 0 \\
13& $\hat{\sigma}_{NEW,k}$ & 0.00500 & 0.03766 & \textbf{3.6955} & 6.73 & 2.59 & 0.0109 & 0.76 & 1.97 & 0.53\\
14& $\hat{\sigma}_{{\rm Orc}}$ & -0.00347 &  0.03509 &  \textbf{3.1832} &  6.80 &  2.77 &  0.0100 & 0.60 &  1 & 0\\
15& $\hat{\sigma}_{TBV}$ & 0.00647 & 0.04181 & 4.5837 & 7.78 & 2.80 & 0.0134 & 0.40 & 1 & 0\\
16& $\hat{\sigma}_{TBV,k}$ & 0.00348 & 0.04165 & 4.4726 & 7.37 & 2.72 & 0.0126 & 0.26 & 2.32 & 0.48\\
\bottomrule
\end{tabular}
\caption{
{Estimation of the volatility $\sigma=0.4$ for a log-normal Merton model based on
simulations of 5-minutes observations of 5000 paths over a 1 month}
time horizon. The jump parameters are $\lambda=200$, {$\sigma^{Jmp}={3}\sqrt{h}$} and $\mu^{Jmp}=0$.
}
\label{Table2}
}
\end{table}

We now double the intensity of jumps and consider the following parameter setting:
\begin{equation}\label{Setting2}
    \sigma=0.4,\quad {\sigma^{Jmp}={3}\sqrt{h}}, \quad \mu^{Jmp}=0, \quad
    \lambda=200, \quad h=\frac{1}{{252\times 6.5\times 12}},
\end{equation}
which yields an expected annualized volatility of 0.5. The results are shown in Table \ref{Table2}. We
again notice that the Oracle estimator performs the best followed by the {new}  estimators based on
finding the root of $F(\ep;\sigma,{\bf m})$. As before, the estimators based on  the MinRV, the MedRV, and $\hat{\sigma}_{TBV}$  underperform compared to $\hat{\sigma}_{NEW}$ and $\hat{\sigma}_{NEW,k}$; $\hat{\sigma}_{TBV, k}$ has a small relative estimation error, but a comparatively high MSE.

\section{Extensions}

In this section we assess our results on models with stochastic volatility and leverage and on models with infinite activity jumps. We now mention the main ideas that we are pursuing in the theoretical ongoing analysis in the presence of stochastic volatility and then we show  on simulated data that the performance of our new methods is promising also in such extended contexts.

In the presence of stochastic volatility without leverage we can deal with cMSE as described in the subsequent paragraph. If also leverage is present, then  we can use a similar approach under MSE or, alternatively, we can work at minimizing cMSE by assuming that $d\sigma_t=\gamma_t dB_t$, with $B$ a Brownian motion correlated with $W$, and by splitting $W$ into a term completely dependent on $\sigma$ and an  independent one\footnote{We thank Alexei Kolokolov for having suggested to consider such an approach}. Conditioning then on $\sigma$ and $J$, the term independent of $\sigma$ can be dealt exactly as in this paper.

A popular approach to deal with the case of stochastic volatility is ``localization".  Assuming continuity of the paths of $\sigma$, the idea is that the volatility is approximately constant in a small time interval. So we can divide the time horizon into k intervals $\Ii$ and apply our methods (that assume constant volatility) to each interval. More specifically, we want to consider an estimator of the form
\begin{equation}\label{TRV00c}
\hat{IV}\!\!_n(\bm{\varepsilon}) := \sum_{i=1}^{k}\sum_{\ell=1}^{n_{i}}(\Delta_{i,\ell}X)^{2} I_{\{|\Delta_{i,\ell}X|\leq{}\varepsilon_{i}\}},
\end{equation}
where $\bm{\ep}:=[\ep_{1},\dots,\ep_{k}]$ and, for $i=1,\dots,k$, each $\Ii$ is divided into $n_i$ subintervals {$]t_{i,\ell-1},t_{i,\ell}],\  \ell=1,.., n_i,$} with $t_{i-1}=t_{i,0}<t_{i,1}<\dots<t_{i,n_{i}}=t_{i}$,
$\Delta_{i,\ell}X:=X_{t_{i,\ell}}-X_{t_{i,\ell-1}}$,  and the threshold $\ep_{i}$ is uniform on $\Ii$. In the case that $n_{i}=1$ for all $i$, we have the extreme case of one different threshold for each subinterval. When  $\sigma$ is independent on $W,$ we can consider the cMSE of $\hat{IV}\!\!_n(\bm{\varepsilon})$, denoted by $cMSE(\bm{\ep})$, and use this to determine the optimal thresholding levels $\ep_{i}$ for the different intervals. {We define
\[
	cMSE_{i}(\ep_{i}):=E[(\hat{IV}_{i}- IV_{i})^2|J, \sigma],
\]
where $IV_{i}=\bar{\bar{\sigma}}_{i}h,$   with $\bar{\bar{\sigma}}_{i}$  a random  number depending on the path of $\sigma$ over the interval $\Ii$ (e.g., $h^{-1}\int_{t_{i-1}}^{t_{i}}\sigma_{s}^{2}ds$),
and $\hat{IV}_{i}:=\sum_{\ell=1}^{n_{i}}(\Delta_{i,\ell}X)^{2} I_{\{|\Delta_{i,\ell}X|\leq{}\varepsilon_{i}\}}$. Then it turns out that minimizing $cMSE(\bm{\ep})$, as $\bm{\ep}$ varies while $k$ and $n_1, .., n_k$ are fixed, is asymptotically equivalent to solve the $k$ problems %{we have that,
%\[
%	\min_{\bm{\ep}}MSE(\bm{\ep})\approx
%	\min_{\ep_{1}}MSE_{1}(\ep_{1})+\dots+\min_{\ep_{k}}MSE_{k}(\ep_{k}),
%\]
$$\min\limits_{\ep_{i}}cMSE_{i}(\ep_{i}), \quad \mbox{i=1..k},$$
which can be treated at once and justify why we tackled the minimization of cMSE by assuming constant volatility.

Although the theoretical analysis of  cMSE under stochastic volatility and leverage and  under infinite activity jumps are still ongoing, in the rest of this section, we illustrate on simulated data the behavior of our newly proposed methods. We find in fact that
again they outperform the methods currently  used in the literature, at least in the realistic scenarios that we considered here.

\subsection{Simulation performance: stochastic volatility models with leverage}
{Even though the new method presented in Section \ref{FACVP0} was originally designed for a model
with constant volatility (and thus no leverage), it can still be applied for the more general stochastic
volatility model (\ref{Mod}). In this part, we examine by simulations the performance of the
{same} estimators introduced in Section \ref{SimulFACV} in the presence of stochastic volatility and leverage.
For the continuous part of the process, we take {the popular Heston model}  \cite{Heston_Model}
{and consider}:
\begin{equation}\label{HestonTransformed}
\begin{split}
dX_t = & \mu_t dt + \sqrt{V_t} dB_t+dJ_{t}, \quad {X_{0}=1},\\
dV_t = & \kappa(\theta-V_t)dt + \xi \sqrt{V_t} dW_t ,\quad {V_{0}=\theta},
\end{split}
\end{equation}
where $B$ and $W$ are correlated Wiener processes such that $\mathbb{E}(dB_{t}\cdot dW_{t})=\rho dt$ while, in accordance with our Assumption  \textbf{A1}, we take the jump component $J$ independent of $(W,B)$. For  $J$ we adopt the Merton's log-normal model studied in Section \ref{SimulFACV}.
We consider the following settings,
where  $h$ will be set to 5 minutes (i.e., {$h=1/(252\times 6.5\times 12)$}):
\begin{center}
\begin{tabular}{|cccccc|ccc|}
\hline
 \multicolumn{6}{|c|}{Continuous Component Parameters} &  \multicolumn{3}{|c|}{Jump Component }\\
\hline
$\mu_{t}$ & $\kappa$ & $\xi$ & $\theta$ & $\rho$ &  $V_{0}$&
$\sigma^{Jmp}$ & $\mu^{Jmp}$ & $\lambda$ \\
\hline
$0$ & $5$ & $0.5$ & $0.16$ & {$0$ or $-0.5$} & $0.16$ & ${3}\sqrt{h}$ & 0 &  200 \\
\hline
\end{tabular}
\end{center}
{The values of $\kappa$ and $\xi$, which are standard in the literature, are the same as those used in \cite{Two_Time_Scale}, where they also propose $\rho=-0.5$ and $\theta=0.04$. We adopt here the value of $\theta=0.16$ for easier comparison with the constant volatility case of Section \ref{SimulFACV}, where $\sigma$ is taken to be $0.4$. {We remark however that
 we checked the performance of all the estimators in the case $\theta=0.04$, and there are no significant changes, except that it is easier to identify jumps because the variance of the jump part is bigger compared to that of the continuous component, so the MSEs are smaller. }
 %I am sending you the word file where I have the details for the FA Levy case.

%\n {\Green $\blacktriangleright$ [motivation for the choice of the parameters?]}

\n Since the volatility changes from simulation to simulation, to assess the accuracy of the different methods, we compute the relative error,
\[
	\mathcal{E}:=\frac{\hat{IV}\!\!_n-IV}{IV},
\]
for each {simulated path}, where $IV$ is given as below Eq.~(\ref{TRV00}) and $\hat{IV}\!\!_n$ is an estimator of the integrated variance. The sample mean and standard deviation of the error over 5000 simulations for each of the estimators considered in Section \ref{SimulFACV}  are reported in Table {\ref{Table3}}. We also show the sample mean of $(\hat{IV}\!\!_n-IV)^{2}$. As in Section \ref{SimulFACV}, the ``Oracle" is obtained by  the formula
\[
    		\hat{\sigma}_{{\rm Orc}}^{2}=\sum_{i=1}^n \DXq {\bf 1}_{\{|\DX|\leq{\ep}_{{\rm Orc}}\}},
	\]
where {${\ep}_{{\rm Orc}}$} is such that
$F({\ep}_{{\rm Orc}};{\sigma}_{Avg},{{\bf m}})=0$, using the true increments of the jump component,
${\bf m}=(m_{1},\dots,m_{n})=(\Delta_{1}J,\dots,\Delta_{n}J)$, and the {true} average volatility value ${\sigma}_{Avg}^{2}:=T^{-1}\int_{0}^{T}\sigma_{s}^{2}ds$.

The results are consistent with those obtained in Section \ref{SimulFACV}. The new estimators $\hat{\sigma}_{NEW}$ and $\hat{\sigma}_{NEW,k}$ based on finding the root of $F(\ep;\sigma,{\bf m})$ perform the best. Also, the iterative {estimators $ \hat{\sigma}_{2mc,k}$ and  $ \hat{\sigma}_{mc_{2},k}$ based on the threshold $\sqrt{2 \sigma^{2} h\ln(1/h)}$ and  $\sqrt{2 \sigma^{2} h}w_{h}$ perform quite well and significantly improve} on the estimator $\hat{\sigma}_{3mc,k}$ proposed in \cite{FigNis13} and based on $\sqrt{3 \sigma^{2} h\ln(1/h)}$. {In particular}, the leverage factor seems to have  a minor effect {on} the performance of all the estimators, {while stochastic volatility seems not to have any adverse effects,}
%on the performance of the estimators
compared to the constant volatility case. Thus, for instance, for $\lambda=200$ and a long-run average volatility level of $\sqrt{\theta}=\sqrt{0.16}=0.4$, the sample mean and standard deviation of $\mathcal{E}(\hat{\sigma}_{NEW})$ {are   $0.00186$ and $0.03720$}, respectively, which are smaller that those attained by $\mathcal{E}(\hat{\sigma}_{NEW})$ in the constant volatility case of $\sigma=0.4$ (namely,
$0.00389$ and $0.03762$ as seeing in Table \ref{Table2}).}

%\n {\Green Note that $\hat{\sigma}_{TBV,k}$ is worse than $\hat{\sigma}_{NEW,k}$ in terms of MSE, but better in terms of mean relative error.}

\begin{table}[h]
    {\par \centering
\begin{tabular}{cc|rcr|rcr}
\toprule
 & &\multicolumn{3}{c|}{$\rho=0$} & \multicolumn{3}{c}{$\rho=-0.5$} \\
 &               &  ${\rm mean}$  & ${\rm std}$  & ${\rm MSE}(\hat{IV})$ &
                 ${\rm mean}$  & ${\rm std}$  & ${\rm MSE}(\hat{IV})$ \\ % N
& \rm{Estimator} & $\frac{\hat{IV}-IV}{IV}$ &
 $\frac{\hat{IV}-IV}{IV}$  & $\times 10^{7}$ &$\frac{\hat{IV}-IV}{IV}$ &
 $\frac{\hat{IV}-IV}{IV}$  & $\times 10^{7}$\\
\hline
1& $\hat{\sigma}_{RV}$&  0.58420 & 0.28119 & 680.590 & 0.59211 & 0.28610 & 693.970\\
2& $\hat{\sigma}_{BV}$ & 0.13736 & 0.07135 & 41.478 & 0.13910 & 0.07373 & 42.394\\
3& $\hat{\sigma}_{MinRV}$ & 0.03471 & 0.05910 & 8.562 &  0.03478 & 0.05981 & 8.527\\
4& $\hat{\sigma}_{MedRV}$ & 0.04091 & 0.05637 & 8.734 & 0.04114 & 0.05745 & 8.823\\
5& $\hat{\sigma}_{TRV_{JT}}$ & 0.02198 & 0.03951 & 3.850 & 0.02275 & 0.04032 &  4.000\\
6& $\hat{\sigma}_{3mc}$ & 0.08083 & 0.05834 & 17.715 &  0.08215 & 0.06073 & 18.556\\
7& $\hat{\sigma}_{3mc,k}$ & 0.04174 & 0.04555 & 7.301 & 0.04266 & 0.04528 & 7.364\\
8& $\hat{\sigma}_{2mc}$ & 0.04239 & 0.04586 & 7.061 & 0.04388 & 0.04637 & 7.277\\
9& $\hat{\sigma}_{2mc,k}$ & 0.01815 & 0.03888 & 3.477 & 0.01850 & 0.03984 & 3.624\\
10& $\hat{\sigma}_{mc_{2}}$ & 0.03616 & 0.04280 & 5.644 & 0.036717 & 0.04360 & 5.800 \\
11& $\hat{\sigma}_{mc_{2},k}$ & 0.01376 & 0.03784 & 2.997 & 0.01423 & 0.03849 & 3.108\\
12& $\hat{\sigma}_{NEW}$ & 0.00186 & 0.03720 & \textbf{2.527} &  0.00228 & 0.03771 & \textbf{2.629}\\
13& $\hat{\sigma}_{NEW,k}$ & 0.00295 &  0.03719 & \textbf{2.536} & 0.00342 & 0.03765 & \textbf{2.633}\\
14& $\hat{\sigma}_{{\rm Orc}}$ & -0.00584 &  0.03459 &  \textbf{2.293}
& -0.00582 &  0.03506 &  \textbf{2.324}\\
15& $\hat{\sigma}_{TBV}$ & 0.00525 & 0.04218 & 3.373 & 0.00574 & 0.04296 & 3.466\\
16& $\hat{\sigma}_{TBV,k}$ &  0.00225 & 0.04199 & 3.297 & 0.00254 & 0.04274 & 3.381\\
\bottomrule
\end{tabular}
\caption{For each estimator we report: sample mean and standard deviation of the estimation percentage error
$\mathcal{E}:=(\hat{IV}\!\!_n-IV)/IV$, and MSE, i.e. the sample mean of $(\hat{IV}\!\!_n-IV)^{2}$.
The means and standard deviations are based on simulations of 5-minutes observations of 5000 paths from a jump-diffusion model for $X$ obtained by adding log-normal jumps to the
Heston model. The  time horizon of the paths is 1 month.
}
\label{Table3}
}
\end{table}

\subsection{Simulation performance: infinite jump activity}
It is natural to wonder about the robustness of the estimators introduced in this article against jumps of infinite activity (IA). To this end, in this section, we consider one of the most popular models of this kind: the Variance Gamma model (VG) of \cite{Madan1}. 
Concretely, we assume the model
\[
	X_{t}=at+\sigma W_{t}+J_{t}:=at+\sigma W_{t}+\sigma^{Jmp}B_{S_t}+\theta S_t,
\]
where $W$ and $B$ are independent Wiener processes and {$\{S_t\}_{t\geq{}0}$ is an independent L\'evy subordinator
such that $S_t$ is Gamma distributed
with scale  parameter $\beta:=\kappa$ and shape parameter $\alpha:=t/\kappa$. Note that, in that case, $\mathbb{E}[S_t]=t$ and ${\rm Var}(S_t)=\kappa t$.
For the parameter values, we take the following (the time unit is one day):
\[
	\sigma=\frac{0.2}{\sqrt{252}}=0.0126, \quad \sigma^{Jmp}=0.01,
	\quad \kappa=0.7; \quad {a}=\theta=0.
\]
The values of $\sigma^{Jmp}$ and $\kappa$ are consistent with the empirical results of \cite{FigLee17}.

The results are shown in Table \ref{Table4}. Basically, the estimators that use {truncation}
{($\hat{\sigma}_{3mc, k}$, $\hat{\sigma}_{2mc, k}$,  $\hat{\sigma}_{mc_{2}}$,
$ \hat{\sigma}_{NEW, k}$, $\hat{\sigma}_{TBV, k}$, {but not $\hat{\sigma}_{TRV_{JT}}$})} perform better
than those without it ($\hat{\sigma}_{RV}$, $\hat{\sigma}_{BV}$, $\hat{\sigma}_{MedRV}$, and $\hat{\sigma}_{MinRV}$).
As expected, the Oracle estimator performs the best, followed by the estimators $\hat{\sigma}_{2mc,k}$ and $\hat{\sigma}_{mc_{2},k}$, which are based on the
respective asymptotic {thresholds} {$\sqrt{2\sigma^{2}h\ln(1/h)}$ and $\sqrt{2\sigma^{2}h}w_{h}$}. Indeed, their MSEs are less or equal to a quarter of any
other feasible threshold estimator{s}. The iterative estimators $\hat{\sigma}_{NEW,k}$ (based on finding the root  of
$F({\epsilon};\sigma,{\bf m})$) and $\hat{\sigma}_{TBV,k}$ (based on truncated bipower variation) have similar
performances. \\

\begin{table}[h]
    {\par \centering
\begin{tabular}{cc|rcccccr}
\toprule
&                &  ${\rm mean}$  & ${\rm std}$  & ${\rm MSE}(\hat{\sigma})$
               &  ${\rm mean}$  &  ${\rm std}(\varepsilon)$ & mean & std\\
& \rm{Estimator} & $\frac{\hat\sigma^2- \sigma^2}{\sigma^2}$ &
 $\frac{\hat\sigma^2- \sigma^2}{\sigma^2}$  & $\times 10^{10}$ & $\varepsilon$ & $\times 10^{4}$ & $N$ & $N$\\
\hline
1&$\hat{\sigma}_{RV}$  & 0.6318 & 0.21300 & 112.000 & \\
2&$\hat{\sigma}_{BV}$  & 0.2523 & 0.07308 & 17.390 & \\
3&$\hat{\sigma}_{{\rm MinRV}}$  &  0.1258 & 0.06276 & 4.978 & \\
4&$\hat{\sigma}_{{\rm MedRV}}$  &  0.1404 & 0.05940 & 5.859 & \\
5&$\hat{\sigma}_{TRV_{JT}}$ & 0.1582 & 0.05347 & 7.032 & 0.00661 & 1.911 & 1 & 0\\
6&$\hat{\sigma}_{3mc}$  & 0.1547 & 0.05444 &  6.777 &  0.00657 & 4.244 & 1 & 0 \\
7& $\hat{\sigma}_{3mc,k}$  & 0.0987 & 0.04764 & 3.030 & 0.00540 & 1.171 & 3.50 & 0.69\\
8&$\hat{\sigma}_{2mc}$  &  0.0955 & 0.04748 & 2.870 &  0.00536 & 3.465 &  1 & 0\\
9&$\hat{\sigma}_{2mc,k}$  & 0.0171 & 0.04518 & \textbf{0.588}  & 0.00424 & 0.942 & 4.26 & 0.88\\
10&$\hat{\sigma}_{mc_{2}}$  & 0.0760 & 0.04652 & 2.004 &  0.00502 & 3.244 &  1 &0\\
11&$\hat{\sigma}_{mc_{2},k}$  &  -0.0260 &  0.04614 &  \textbf{0.707} &  0.00388 &  0.921 &  4.87 & 1.04\\
12&$\hat{\sigma}_{NEW}$ & 0.0957 & 0.04860 & 2.903 & 0.00537 & 2.599 &  1 & 0\\
13&$\hat{\sigma}_{NEW,k}$ & 0.0953 & 0.04870 & 2.887 &  0.00537 & 2.773 &  2.01 & 2.20\\
14&$\hat{\sigma}_{{\rm Orc}}$ & -0.00015 &  0.04461 &  \textbf{0.501} &   0.00412 &  2.394 &  1 &0\\
15&$\hat{\sigma}_{TBV}$ & 0.1038 & 0.05151 & 3.387 &  0.00666 & 1.911 & 1 &0\\
16&$\hat{\sigma}_{TBV,k}$ & 0.0927 & 0.05172 & 2.839 &  0.00622 & 1.473 & 2.79 &  0.61\\
\bottomrule
\end{tabular}
\caption{
 Estimation of the volatility $\sigma=0.2/\sqrt{252}=0.0126$ for a Gauss-VG model, based on simulations of 5-minutes observations of 5000 paths for each model, over a 1 month time horizon.
The jump parameters are  {$\sigma^{Jmp}=0.01$} and ${\theta=0}$.
}
\label{Table4}
}
\end{table}

\section{Conclusions}
{We consider the problem of estimating the integrated variance $IV$
of a semimartingale model $X$ with jumps for the log price of a
financial asset. In view of adopting the truncated realized variance
of $X$, we look for {a} theoretical and practical way to select an
optimal threshold in finite samples. We consider the following
two optimality criteria: minimization of MSE, the expected quadratic
error in the estimation of IV; and minimization of cMSE, the
expected quadratic error conditional to the realized paths of the
jump process $J$ and of the volatility process
$(\sigma_s)_{s\geq{}0}$. Under given assumptions, we find that for
each criterion an optimal TH exists, is unique and is a solution of
an explicitly given equation, the equation being different under the
two criteria. Also, under each criterion, an asymptotic expansion
with respect to the step $h$ between the observations is possible
for the optimal TH. The leading terms
of {both} the two expansions turn out to be
proportional to the modulus of continuity of the Brownian motion
paths and to the spot volatility of X, with  proportionality
constant  $\sqrt{2-Y}$,  $Y$ being  the jump activity index of $X$.
{Further, we show that} the threshold estimator of  $IV$ constructed with
the {leading term of the} optimal TH is consistent, at least in the finite activity jumps
case, even if it does not satisfy the classical assumptions.\\
The results obtained for the cMSE criterion allow for a novel numerical
way to tuneup the threshold parameter in finite samples.
{Based on simulated data,} we illustrate the superiority of the new method {on other broadly used estimators in the literature}.
%in the finite jump activity case with constant volatility.
 Minimization of {cMSE in the presence of infinite activity jumps in $X$ and in the presence of stochastic volatility and leverage are object of  ongoing research, but the newly proposed estimators are implemented on simulated data under such frameworks, and again are superior.}

%\newpage
%\appendix

\section{Appendix: proofs}\label{ApndA}

\vp \n {\bf Proof of Theorem \ref{expressionMSEingen}}.  % Proof Thm 1
Under {\bf A1} we have that conditionally to $(\sigma, J)$ the
increment $\DX= \intIi\sigma_s dW_s + \DiJ$ is a Gaussian r.v. with
law $\mathcal{N}(m_i, \sigma^2_i)$, which allows to compute the
conditional expectation $E[\hatIV| \sigma, J].$ We have
%which will be indicated for simplicity by $E[\hatIV].$

$$E[\hatIV|\sigma, J]=\sum_{i=1}^n b_i(\ep)=
\sum_{i=1}^n -\Big(e^{-\frac{(\ep-m_i)^2}{2\sigma^2_i}}(\ep + m_i)
+e^{-\frac{(\ep+m_i)^2}{2\sigma^2_i}}(\ep -
m_i)\Big)\frac{\sigma_i}{\sqrt{2\pi}}$$
$$+ \frac{m_i^2+\sigma_i^2}{\sqrt{\pi}}
    \Big(\int_0^{\frac{\ep-m_i}{\sqrt 2\sigma_i}} e^{-t^2} dt +
    \int_0^{\frac{\ep+m_i}{\sqrt 2\sigma_i}} e^{-t^2} dt\Big),
$$
and
$$E[(\hatIV(\ep))^2| \sigma, J]=
\sum_i E[(\DX_\star)^4| \sigma, J] +
2\sum_i\sum_{j>i}E[(\DX_\star)^2(\Delta_j X_\star)^2| \sigma, J]  $$
$$ =\sum_i \Big[ -  e^{-\frac{(\ep-m_i)^2}{2\sigma^2_i}} \sigma_i\Big( \ep^3+ m_i  \ep^2 + m_i^2 \ep
+ m_i^3 + 5 m_i \sigma_i^2 + 3 \sigma_i^2\ep\Big)
$$
$$-  e^{-\frac{(\ep+m_i)^2}{2\sigma^2_i}} \sigma_i\Big( \ep^3- m_i  \ep^2 + m_i^2 \ep
- m_i^3 - 5 m_i \sigma_i^2 + 3 \sigma_i^2\ep\Big)$$
 \beq\label{Sumbibj}+\Big(\int_0^{\frac{\ep-m_i}{\sqrt 2\sigma_i}} e^{-t^2} dt + \int_0^{\frac{\ep+m_i}{\sqrt 2\sigma_i}} e^{-t^2} dt\Big)
 \sqrt{2} \Big( m_i^4+ 6m_i^2\sigma_i^2 + 3 \sigma_i^4\Big)\Big]\frac{1}{\sqrt{2\pi}}
% + 2\sum_i\sum_{j>i}E[(\DX_\star)^2| \sigma, J]E[(\DjX_\star)^2|\sigma, J],
 +{2\sum_i\sum_{j>i}b_ib_j},
\eeq having used that conditionally to $\sigma$ and  $J$,
$\DX_\star$ and $\DjX_\star$ are independent. It  follows that
$$MSE(\ep)=E\left[\sum_i \Big[ -
e^{-\frac{(\ep-m_i)^2}{2\sigma^2_i}} \sigma_i\Big( \ep^3+ m_i \ep^2
+ m_i^2 \ep + m_i^3 + 5 m_i \sigma_i^2 + 3 \sigma_i^2\ep\Big)\right.
$$
$$-  e^{-\frac{(\ep+m_i)^2}{2\sigma^2_i}} \sigma_i\Big( \ep^3- m_i  \ep^2 + m_i^2 \ep
- m_i^3 - 5 m_i \sigma_i^2 + 3 \sigma_i^2\ep\Big)$$
$$+\Big(\int_0^{\frac{\ep-m_i}{\sqrt 2\sigma_i}} e^{-t^2} dt + \int_0^{\frac{\ep+m_i}{\sqrt 2\sigma_i}} e^{-t^2} dt\Big)
 \sqrt{2} \Big( m_i^4+ 6m_i^2\sigma_i^2 + 3 \sigma_i^4\Big)\Big]\frac{1}{\sqrt{2\pi}}$$
 $$ + 2\sum_i\sum_{j>i}b_i(\ep)b_j(\ep) -2 IV \sum_{i=1}^n
\Big[-\Big(e^{-\frac{(\ep-m_i)^2}{2\sigma^2_i}}(\ep + m_i)
+e^{-\frac{(\ep+m_i)^2}{2\sigma^2_i}}(\ep -
m_i)\Big)\frac{\sigma_i}{\sqrt{2\pi}}$$
$$\left.+ \frac{m_i^2+\sigma_i^2}{\sqrt{\pi}}
    \Big(\int_0^{\frac{\ep-m_i}{\sqrt 2\sigma_i}} e^{-t^2} dt +
    \int_0^{\frac{\ep+m_i}{\sqrt 2\sigma_i}} e^{-t^2} dt\Big)\Big] + IV^2\right].$$
% nota che fin qui non serve J FA
%The two  expressions for $E[\hatIV|\sigma, J]$ and $E[(\hatIV(\ep))^2| \sigma, J]$
{$MSE(\ep)$ is} a differentiable function of $\ep$, therefore to find the minimum on $[0, + \infty[$ of
$MSE(\ep)$ we can study the sign of its first derivative
$MSE'(\ep).$ Since $MSE'(\ep)= \frac{d}{d\ep}E[(\hatIV(\ep))^2]- 2
IV\frac{d}{d\ep}E[\hatIV(\ep)]$, we begin to compute
$\frac{d}{d\ep}E[\hatIV(\ep)| \sigma, J]$.
%\frac{d}{d\ep}E[(\hatIV(\ep))^2| \sigma, J]- 2 IV E[\hatIV\!'(\ep)| \sigma, J]$
% NOta: \hatIV\!'(\ep) non si puo' fare prima di integrare, perche' in \ep hatIV non derivb
Note that
$$\frac{d}{d\ep}b_i(\ep)=
\Big(e^{-\frac{(\ep-m_i)^2}{2\sigma^2_i}} +
 e^{-\frac{(\ep+m_i)^2}{2\sigma^2_i}}\Big)\frac{(\ep + m_i)(\ep - m_i)}{\sigma_i\sqrt{2\pi}}
$$
$$ -  \Big(e^{-\frac{(\ep-m_i)^2}{2\sigma^2_i}}  +  e^{-\frac{(\ep+m_i)^2}{2\sigma^2_i}}\Big)
\frac{\sigma_i}{\sqrt{2\pi}} +
\frac{m_i^2+\sigma_i^2}{\sigma_i\sqrt{2\pi}} \Big(
e^{-\frac{(\ep-m_i)^2}{2\sigma^2_i}} +
e^{-\frac{(\ep+m_i)^2}{2\sigma^2_i}}\Big) =\ep^2
\frac{e^{-\frac{(\ep-m_i)^2}{2\sigma^2_i}} +
  e^{-\frac{(\ep+m_i)^2}{2\sigma^2_i}}}{\sigma_i\sqrt{2\pi}}= \ep^2a_i(\ep),$$
  so that
\beq\label{derivEDXqIDXqleqr} \frac{d}{d\ep}E[\hatIV(\ep)| \sigma,
J]=\ep^2 \sum_{i=1}^n a_i(\ep) \eeq is strictly greater than zero
for all values of $\ep>0$.
{As for $\frac{d}{d\ep}E[(\hatIV(\ep))^2| \sigma, J]$, note that the term
$2\sum_i \sum_{j>i} b_i b_j$ in (\ref{Sumbibj}) can be written as $\sum_i \sum_{j\neq
i} b_i b_j$, so its derivative coincides with
$\sum_i\sum_{j\neq i} (\ep^2 a_i b_j + b_i \ep^2 a_j),$ however
$$ \sum_i b_i\sum_{j\neq i} a_j =
\Big(\sum_i b_i \sum_j a_j - \sum_i b_i a_i\Big)$$
%= (\sum_i b_i) (\sum_j a_j) - \sum_i b_i a_i
$$=\Big( \sum_i a_i \sum_j b_j - \sum_i a_i b_i\Big)=  \sum_i a_i \sum_{j\neq i} b_j$$
so that $\sum_i\sum_{j\neq i} (\ep^2 a_i b_j + b_i \ep^2 a_j)= 2 \sumi \sum_{j\neq i} \ep^2 a_i b_j,$
\beq\frac{d}{d\ep}E[(\hatIV(\ep))^2| \sigma, J]=
\ep^4\sum_ia_i(\ep) + 2 \Big(\sumi \sum_{j>i} b_i(\ep) b_j(\ep)\Big)'\eeq
$$
=\sum_i \Big[\ep^4 a_i + 2 \ep^2 a_i \sum_{j\neq i} b_j \Big]$$ and
$$\frac{d}{d\ep}MSE(\ep)=\ep^2\sum_i E\Big[\ep^2 a_i + 2  a_i \sum_{j\neq i}b_j - 2 IV  a_i \Big].$$  \beq\label{derivhatIVqsemplif}
\begin{array}{cl}
 & = \ \ep^2\sum_i E\Big[a_i \Big(\ep^2 + 2  \sumjni b_j  - 2 IV \Big)\Big]\\
 \\
 & = \ \ep^2G(\ep).\hfill
\end{array}\eeq}

\vspace{-0.9cm}\qed\\

\n
{\bf Proof of Corollary \ref{OptThrExists}.} Note that $a_i(\ep)$ and $b_i(\ep)$ are continuously
differentiable functions of $\ep$, and, with fixed $h=\frac T n$,
$$ a_i(0)=\frac{2 e^{-\frac{m_i^2}{2\sigma^2_i}}}{\sigma_i\sqrt{2\pi}}, \quad b_i(0)=0,$$
$$a_i(+\infty)=0,\quad  b_i(+\infty)=E[(\DX_\star)^2|\sigma,J]= m_i^2 + \sigma_i^2, $$
$$ a_i'(\ep)= -\frac{1}{\sigma_i^3\sqrt{2\pi}}\Big[e^{-\frac{(\ep-m_i)^2}{2\sigma^2_i}}(\ep- m_i)+
e^{-\frac{(\ep+m_i)^2}{2\sigma^2_i}}(\ep+ m_i)\Big], \quad
b_i'(\ep)=\ep^2 {a_i(\varepsilon)},$$
so we find that $G(0)=- \sum_i
E\Big[ IV \cdot\frac{4}{\sigma_i\sqrt{2\pi}}e^{-\frac{m_i^2}{2\sigma_i^2}}\Big]<0$, and
$\lim_{\ep\ri +\infty} G(\ep)=0^+,$ so there exists $\ep_+>0:$
$MSE'(\ep)>0$ on  $[\ep_+, +\infty)$. On the compact set $[0,
\ep_+]$ the continuous function  $MSE$ has necessarily  absolute
minimum value $\underline{MSE}$, and since on $[\ep_+, +\infty)$
$MSE$ is increasing we have that on $[0, +\infty)$ the absolute
minimum is $\underline{MSE}.$\\
$MSE'(\ep)$ is continuous and assumes both negative and positive
values, thus equation $ G(\ep)=0$ has {a} solution. Any minimum point of
$MSE$ on $[0, +\infty)$ has to be a stationary point, so it has to
solve the equation.\qed\\

\n
{\bf Proof of Theorem \ref{TeoEsUNicOptThLevycase}.} For $\ep>0$ we have $MSE'(\ep)>0$ if and only if $G(\ep)>0$,
which in turn is true if and only if
$$g(\ep):=\ep^2+ 2(n-1)E[b_1]- 2 IV>0$$
where, setting $m:=m_1= \DuJ$,  we recall that we have
\begin{align*}
E[b_1]&= E\Big[-\Big(e^{-\frac{(\ep-m)^2}{2\sigma^2 h}}(\ep + m)
+e^{-\frac{(\ep+m)^2}{2\sigma^2 h}}(\ep - m)\Big)\frac{\sigma\sqrt
h}{\sqrt{2\pi}}\\
&\quad\qquad+ \frac{m^2+\sigma^2 h}{\sqrt\pi}
    \Big(\int_0^{\frac{\ep-m}{\sqrt 2\sigma \sqrt{h}}} e^{-t^2} dt +
    \int_0^{\frac{\ep+m}{\sqrt 2\sigma\sqrt{h}}} e^{-t^2} dt\Big)
 \Big].
 \end{align*}
The sign of $g(\ep)$ is studied as follows:
 $$g(0)= -2\sigma^2T<0,$$
 $$\lim_{\ep\ri+\infty} g(\ep)= +\infty,$$
 $$g'(\ep)= 2\ep(1+(n-1)\ep E[a_1])$$
 so that $g'(\ep)>0$ for all $\ep >0$, $n>1$.
 That implies that $g(\ep)$ starts at $\ep=0$ from a negative value and strictly increases towards $+\infty$, as
 $\ep$ increases,  so that
 there exists a unique $\ep^\star$ such that $g(\ep)<0$ for $\ep\in [0, \ep^\star[$, $g(\ep^\star)=0$ and
$g(\ep)>0$ for $\ep\in ] \ep^\star, + \infty[$. That implies in turn
that $MSE(\ep)$ has a unique minimum point in $\ep^\star$, which is
then the optimal threshold we were looking for: $\ep^\star$ is the
unique solution of equation (\ref{eqOptThreshXLevy}), corresponding
to $g(\ep) =G(\ep)=0$.\qed\\

\n
{\bf Proof of Theorem \ref{thrmKA}.}
By definition,
\begin{equation}\label{DcmEb1}
    \bbe\left[b_{1}(\varepsilon)\right]=\bbe\left[(\Delta_{1}^{n}X)^{2}{\bf 1}_{\{|\Delta_{1}^{n}X|<\varepsilon,\Delta_{1}^{n}N=0\}}\right]+
    \bbe\left[(\Delta_{1}^{n}X)^{2}{\bf 1}_{\{|\Delta_{1}^{n}X|<\varepsilon,\Delta_{1}^{n}N\neq{}0\}}\right]=: \mathcal{G}+\mathcal{L}.
\end{equation}
By Lemma S.2  and Lemma S.5 {with $k=2$ in \cite{FigNis16}}, provided that $\varepsilon\to{}0$, we have
\begin{equation}\label{FINSM}
    \mathcal{L}:=\bbe\left[(\Delta_{1}^{n}X)^{2}{\bf 1}_{\{|\Delta_{1}^{n}X|<\varepsilon,\Delta_{1}^{n}N\neq{}0\}}\right]\sim
\lambda h \frac{\varepsilon^{3}}{3}C(f), \quad (h\to{}0),
\end{equation}
\[
    \mathcal{G}:=\sigma^{2}h-\frac{2}{\sqrt{2\pi}}\sigma\varepsilon\sqrt{h} e^{-\frac{\varepsilon^{2}}{2\sigma^{2}h}}+
O\left(h^{2}\right)+o\left(\varepsilon\sqrt{h}
e^{-\frac{\varepsilon^{2}}{2\sigma^{2}h}}\right),
\]
which shows the result.
\qed\\

\n
 {\bf Proof of Lemma \ref{L1:IJA}.}
 Throughout, $p_{t}$ denotes the density of $J_{t}$ and recall that the characteristic function of $J_{t}$ is of the form $\bbe\left[e^{iuJ_{t}}\right]=e^{-c t |u|^{Y}}$. Let us also recall that the Fourier transform and its inverse are defined by $\mathcal{F}g(x)=\frac{1}{\sqrt{2\pi}}\int_{\bbr}g(z)e^{-izx}dz$ and $\mathcal{F}^{-1}G(x)=\frac{1}{\sqrt{2\pi}}\int_{\bbr}G(z)e^{izx}dz$. In what follows, we set
 \begin{align*}
     &h(u):=\left(\mathcal{F}^{-1}\phi\left(\frac{\cdot}{\sigma\sqrt{{h}}}-\frac{{\varepsilon}}{\sigma\sqrt{{h}}}\right)\right)(u)
    =\frac{1}{\sqrt{2\pi}}\int \phi\left(\frac{x}{\sigma\sqrt{{h}}}-\frac{{\varepsilon}}{\sigma\sqrt{{h}}}\right)e^{iux}dx.
\end{align*}
Let us start by noting that
\begin{align*}
     \bbe\left[\phi\left(\frac{{\varepsilon}}{\sigma\sqrt{{h}}}-\frac{J_{{h}}}{\sigma\sqrt{{h}}}\right)\right]&=\int \phi\left(\frac{x}{\sigma\sqrt{{h}}}-\frac{{\varepsilon}}{\sigma\sqrt{{h}}}\right)p_{{h}}(x)dx=\int (\mathcal{F}h)\left(x\right)p_{{h}}(x)dx=\int h\left(u\right)(\mathcal{F}p_{{h}})(u)du,
\end{align*}
where, since $J$ is a symmetric stable process, $(\mathcal{F}p_{{h}})(u)=(2\pi)^{-1/2}e^{-c{h}|u|^{Y}}$. Therefore, we obtain the representation
    \begin{align}\label{RFT0}
    \bbe\left[\phi\left(\frac{{\varepsilon}}{{\sigma}\sqrt{{h}}}\pm \frac{J_{{h}}}{{\sigma}\sqrt{{h}}}\right)\right]&=
    \frac{\sigma {h}^{1/2}}{{2\pi}}\int e^{-c{h}|u|^{Y}-\frac{\sigma^{2}{h}u^{2}}{2}+i{\varepsilon}u}du.
\end{align}
In order to prove (\ref{NECN}), let us make the change of variables $w=\sigma {h}^{1/2}u$ and, then, expand in a Taylor's expansion
$\exp(-c{\sigma^{-Y}}{h}^{1-Y/2}|w|^{Y})$ as follows:
\begin{align*}
    \frac{1}{2\pi}\int e^{-c\sigma^{-Y}{h}^{1-Y/2}|w|^{Y}-\frac{w^{2}}{2}+i\frac{{\varepsilon}}{\sigma {h}^{1/2}}w}dw&=
    \frac{1}{2\pi}\int e^{-\frac{w^{2}}{2}+i\frac{{\varepsilon}}{\sigma {h}^{1/2}}w}dw +\sum_{k=1}^{\infty} I_{k,n},
%   &\quad-c\sigma^{-Y}{h}^{1-Y/2}
%   \frac{1}{\sqrt{2\pi}}\int |w|^{Y}e^{-\frac{w^{2}}{2}+i\frac{{\varepsilon}}{\sigma {h}^{1/2}}w}dw+\rm{h.o.t.}.
\end{align*}
where
\begin{align*}
    I_{k,n}&:=\frac{1}{k!}(-c)^{k}\sigma^{-kY}{h}^{k(1-Y/2)}
    \frac{1}{\sqrt{2\pi}}\int |w|^{kY}e^{-\frac{w^{2}}{2}+i\frac{{\varepsilon}}{\sigma {h}^{1/2}}w}dw\\
    &=\frac{1}{k!}(-c)^{k}\sigma^{-kY}{h}^{k(1-Y/2)}
    \frac{2}{\sqrt{2\pi}}\int_{0}^{\infty} w^{kY}e^{-\frac{w^{2}}{2}}\cos\left(\frac{{\varepsilon}}{\sigma {h}^{1/2}}w\right)dw.
\end{align*}
The first term of (\ref{NECN}) is then clear. For the subsequent terms, let us apply the formula for the cosine integral transformation of $w^{kY}e^{-w^2/2}$ as well as the asymptotics for the generalized hypergeometric series or Kummer's function $M(a,b,z)$:
\begin{align*}
    I_{k,n}
    &=\frac{1}{k!}(-c)^{k}\sigma^{-kY}{h}^{k(1-Y/2)}
     \frac{2}{\sqrt{2\pi}}\left\{\frac{1}{2}2^{\frac{1}{2}(1+kY)}\Gamma\left(\frac{1}{2}+\frac{kY}{2}\right)
     M\left(\frac{1}{2}+\frac{kY}{2};\frac{1}{2};-\frac{{\varepsilon}^{2}}{2\sigma^{2}{h}}\right)\right\}\\
    &=\frac{1}{k!}(-c)^{k}\sigma^{-kY}{h}^{k(1-Y/2)}
     \frac{2}{\sqrt{2\pi}}\left(\frac{1}{2}2^{\frac{1}{2}(1+kY)}\Gamma\left(\frac{1}{2}+\frac{kY}{2}\right)\right)\\
    &\quad\qquad\quad \times \left(\frac{\Gamma\left(\frac{1}{2}\right)}{\Gamma\left(-\frac{kY}{2}\right)}
    \left(\frac{{\varepsilon}^{2}}{2\sigma^{2}{h}}\right)^{-\frac{1}{2}-\frac{kY}{2}}+
    \frac{\Gamma\left(\frac{1}{2}\right)}{\Gamma\left(\frac{1}{2}+\frac{kY}{2}\right)}
    e^{-\frac{{\varepsilon}^{2}}{2\sigma^{2}{h}}}\left({\frac{{\varepsilon}^{2}}{2\sigma^{2}{h}}}\right)^{\frac{kY}{2}}\right)+{\rm h.o.t.}.
\end{align*}
%{I think the sign in $\left(-\frac{{\varepsilon}^{2}}{2\sigma^{2}{h}}\right)^{\frac{kY}{2}}$ would have to be +} \\
In the asymptotic formula for the Kummer's function above, the first term (respectively, second term)
vanishes if $\Gamma(-kY/2)$ (respectively, $\Gamma(1/2 +kY/2)$) are infinity. This happens when $-kY/2$ or 1/2 +kY/2 are nonpositive integers.
It is now evident that there exist nonzero constants $a_{k}$ and $b_{k}$ such that
\[
    I_{k,n}= \frac{a_{k}}{\Gamma\left(-\frac{kY}{2}\right)} {\varepsilon}^{-1-kY}{h}^{k+\frac{1}{2}}
     +\frac{b_{k}}{\Gamma\left(\frac{1}{2}+\frac{kY}{2}\right)}e^{-\frac{{\varepsilon}^{2}}{2\sigma^{2}{h}}}
    \varepsilon^{kY}h^{k(1-Y)}+{\rm h.o.t.}.
\]
Note that
\begin{align*}
    &{\varepsilon}^{-1-kY}{h}^{k+\frac{1}{2}}\gg {\varepsilon}^{-1-(k+1)Y}{h}^{k+1+\frac{1}{2}}\;\Longleftrightarrow\;{\varepsilon}\gg {h}^{1/Y}\;\Longleftarrow\;{\varepsilon}\gg {h}^{1/2},\\
    &{\varepsilon}^{-1-Y}{h}^{1+\frac{1}{2}}\gg {\varepsilon}^{-1-kY}{h}^{k+\frac{1}{2}}\gg
    e^{-\frac{{\varepsilon}^{2}}{2\sigma^{2}{h}}}
    \varepsilon^{kY}h^{k(1-Y)}.
\end{align*}
Therefore, ${\varepsilon}^{-1-Y}{h}^{1+\frac{1}{2}}\gg I_{k,n}$, for all ${k>1}$.
%Note that, in general, the relation $I_{1,n}\gg e^{-{\varepsilon}^{2}/2\sigma^{2}{h}}$ does not hold since we don't know whether
%\[
%    \left(\frac{{\varepsilon}^{2}}{2\sigma^{2}{h}}\right)^{-\frac{1}{2}-\frac{Y}{2}}{h}^{1-\frac{Y}{2}}\gg e^{-{\varepsilon}^{2}/2\sigma^{2}{h}}
%\]
%holds or not.

We now show (\ref{NECNc}). {Note that
\begin{align*}
    \bbe\left[J_{{h}}\phi\left(\frac{{\varepsilon}}{\sigma \sqrt{{h}}}-\frac{J_{{h}}}{\sigma \sqrt{{h}}}\right)\right]&=\int \phi\left(\frac{x}{\sigma \sqrt{{h}}}-\frac{{\varepsilon}}{\sigma \sqrt{{h}}}\right)xp_{{h}}(x)dx
%   =\int (\mathcal{F}h)\left(u\right)u p_{{h}}(u)dx
    =\int h\left(u\right)\mathcal{F}(xp_{{h}}(x))(u)du,
\end{align*}
where
\begin{align*}
    &\mathcal{F}(xp_{{h}}(x))(u)=i\frac{d}{du}(\mathcal{F}p_{{h}})(u)=
    \frac{i}{\sqrt{2\pi}}\frac{d}{du}e^{-c{h}|u|^{Y}}=\frac{-i}{\sqrt{2\pi}}e^{-c{h}|u|^{Y}} Y {\rm sign}(u) c{h}|u|^{Y-1}.
\end{align*}
Therefore, we have the following representation:
\begin{align*}
    \bbe\left[J_{{h}}\phi\left(\frac{{\varepsilon}}{\sigma \sqrt{{h}}}-\frac{J_{{h}}}{\sigma \sqrt{{h}}}\right)\right]=\sigma \frac{-iYc}{\sqrt{2\pi}}{h}^{3/2}\int \text{sign}(u)|u|^{Y-1}e^{-c{h}|u|^{Y}-\frac{\sigma^{2}{h}u^{2}}{2}+i{\varepsilon}u}du.
\end{align*}
Furthermore,
\begin{align*}
     \bbe\left[J_{{h}}\phi\left(\frac{{\varepsilon}}{\sqrt{{h}}}-\frac{J_{{h}}}{\sqrt{{h}}}\right)\right]&=2\sigma \frac{Yc}{\sqrt{2\pi}}{h}^{3/2}\int_{0}^{\infty} u^{Y-1}e^{-c{h}u^{Y}-\frac{\sigma^{2}{h}u^{2}}{2}}\sin\left({\varepsilon}u\right)du\\
    &=2\sigma^{-(Y-1)}\frac{Yc}{\sqrt{2\pi}}{h}^{\frac{3-Y}{2}}\int_{0}^{\infty} w^{Y-1}e^{-c\sigma^{-Y}h^{1-Y/2}w^{Y}-\frac{w^{2}}{2}}\sin\left(\sigma^{-1}{\varepsilon}{h}^{-1/2}w\right)dw.
\end{align*}
Next, we expand in a Taylor's expansion $\exp(-c\sigma^{-Y}{h}^{1-Y/2}w^{Y})$ as follows:
\[
    \frac{1}{\sqrt{2\pi}}\int_{0}^{\infty} w^{Y-1}e^{-c\sigma^{-Y}{h}^{1-Y/2}w^{Y}-\frac{w^{2}}{2}}\sin\left(\sigma^{-1}{\varepsilon}{h}^{-1/2}w\right)dw=
%   \frac{1}{\sqrt{2\pi}}\int_{0}^{\infty} w^{Y-1}e^{-\frac{w^{2}}{2}}\sin\left({\varepsilon}{h}^{-1/2}w\right)dw+
    \sum_{k=0}^{\infty}I_{k,n},
\]
where
\[
    I_{k,n}:=\frac{1}{k!}(-c)^{k}\sigma^{-Yk}{h}^{k(1-Y/2)}
    \frac{1}{\sqrt{2\pi}}\int_{0}^{\infty} w^{(k+1)Y-1}e^{-\frac{w^{2}}{2}}
    \sin\left({\varepsilon}{h}^{-1/2}w\right)dw.
\]
Then,  we again apply the following formula for the sine integral transformation of $w^{(k+1)Y-1}e^{-w^2/2}$:
\begin{align*}
    I_{k,n}
    &=\frac{1}{k!}(-c)^{k}\sigma^{-Yk}{h}^{k(1-Y/2)}
     \frac{1}{\sqrt{2\pi}}\left\{\frac{1}{2}2^{\frac{1}{2}(1+(k+1)Y)}\Gamma\left(\frac{1}{2}+\frac{(k+1)Y}{2}\right)
     \left(\frac{{\varepsilon}^{2}}{{h}}\right)M\left(\frac{1}{2}+\frac{(k+1)Y}{2};\frac{3}{2};-\frac{{\varepsilon}^{2}}{2{h}}\right)\right\}.
%   \\
%   &={h}^{k(1-Y/2)}
%    \frac{1}{\sqrt{2\pi}}\left\{\frac{1}{2}2^{\frac{1}{2}(1+(k+1)Y)}\Gamma\left(\frac{1}{2}+\frac{(k+1)Y}{2}\right)
%    \left(\frac{{\varepsilon}^{2}}{{h}}\right)\right\}\left(\frac{\Gamma(\frac{3}{2})}{\Gamma\left(1-\frac{(k+1)Y}{2}\right)}\left(\frac{{\varepsilon}^{2}}{2{h}}\right)^{-\frac{1}{2}-\frac{(k+1)Y}{2}}\right)+{\rm h.o.t.}.
\end{align*}
Finally, we use the relationship
\begin{align*}
    M\left(\frac{1}{2}+\frac{(k+1)Y}{2};\frac{3}{2};-\frac{{\varepsilon}^{2}}{2{h}}\right)&=
\frac{\Gamma(\frac{3}{2})}{\Gamma\left(1-\frac{(k+1)Y}{2}\right)}
\left(\frac{{\varepsilon}^{2}}{2{h}}\right)^{-\frac{1}{2}-\frac{(k+1)Y}{2}}\\
&\quad+\frac{\Gamma\left(\frac{3}{2}\right)}{
\Gamma\left(\frac{1}{2}+\frac{(k+1)Y}{2}\right)}
e^{-\frac{{\varepsilon}^{2}}{2\sigma^{2}{h}}}\left({\frac{{\varepsilon}^{2}}{2\sigma^{2}{h}}
}\right)^{-1+\frac{(k+1)Y}{2}}+{\rm h.o.t.},
\end{align*}
which, in turn shows that,
\[
    I_{k,n}\ll I_{1,n}\ll {h}{\varepsilon}^{1-Y}.
\]
We then conclude the result of the Lemma.}
\qed\\

\n
{\bf Proof of Lemma \ref{L2:IJA}.}
{Let
\[
     I^{\pm}_{n}:=\bbe\left[\bar{\Phi}\left(\frac{{\varepsilon}}{\sigma\sqrt{{h}}}-\frac{J_{{h}}}{\sigma\sqrt{{h}}}\right){\bf 1}_{\left\{\pm\left(\frac{{\varepsilon}}{\sigma\sqrt{{h}}}-\frac{J_{{h}}}{\sigma\sqrt{{h}}}\right)\geq{}0\right\}}\right]
\]
For $I^{+}_{n}$, let us note that for a constant $K$, $\bar{\Phi}(z)\leq{}K\phi(z)$ for all $z\geq{}0$ and, thus,
\begin{align*}
    I_{n}^{+}\leq K\bbe\left[{\phi}\left(\frac{{\varepsilon}}{\sigma\sqrt{{h}}}-\frac{J_{{h}}}{\sigma\sqrt{{h}}}\right){\bf 1}_{\left\{\frac{{\varepsilon}}{\sigma\sqrt{{h}}}-\frac{J_{{h}}}{\sigma\sqrt{{h}}}\geq{}0\right\}}\right]=O\left(\bbe\left[{\phi}\left(\frac{{\varepsilon}}{\sigma\sqrt{{h}}}-\frac{J_{{h}}}{\sigma\sqrt{{h}}}\right)\right]\right).
\end{align*}
For the other term, we decompose it as follows:
\begin{align*}
     I_{n}^{-}&=\int_{\bbr}\phi(u)\bbp\left[0\geq\frac{{\varepsilon}}{\sigma\sqrt{{h}}}-\frac{J_{{h}}}{\sigma\sqrt{{h}}},u\geq{}\frac{{\varepsilon}}{\sigma\sqrt{{h}}}-\frac{J_{{h}}}{\sigma\sqrt{{h}}}\right]du\\
     &=\int_{0}^{\infty}\phi(u)\bbp\left[0\geq\frac{{\varepsilon}}{\sigma\sqrt{{h}}}-\frac{J_{{h}}}{\sigma\sqrt{{h}}}\right]du+\int_{-\infty}^{0}\phi(u)\bbp\left[u\geq{}\frac{{\varepsilon}}{\sigma\sqrt{{h}}}-\frac{J_{{h}}}{\sigma\sqrt{{h}}}\right]du\\
    &=\frac{1}{2}\bbp\left[J_{1}\geq {h}^{-\frac{1}{Y}}{\varepsilon}\right]+\int_{-\infty}^{0}\phi(u)\bbp\left[J_{1}\geq{}{h}^{-\frac{1}{Y}}{\varepsilon}-\sigma u{h}^{\frac{1}{2}-\frac{1}{Y}}\right]du.
\end{align*}
The first term above is well-known to be $\bbp\left[J_{1}\geq {h}^{-1/Y}{\varepsilon}\right]=Y^{-1}{C}\left({h}^{-1/Y}{\varepsilon}\right)^{-Y}+O\left({\varepsilon}^{-2Y}{h}^{2}\right)$. For the second term, let us first recall that there exists a constant $K$ such that for all $x>0$,
\begin{equation}\label{EWKI}
    |\mathcal{E}(x)|:=\left|\bbp\left[J_{1}\geq{}x\right]-\frac{{C}}{Y}x^{-Y}\right|\leq{}K x^{-2Y}.
\end{equation}
Therefore,
\begin{align*}
    \int_{-\infty}^{0}\phi(u)\bbp\left[J_{1}\geq{}{h}^{-\frac{1}{Y}}{\varepsilon}-\sigma u{h}^{\frac{1}{2}-\frac{1}{Y}}\right]du&=\frac{{C}}{Y}
    \int_{-\infty}^{0}\phi(u)\left({h}^{-\frac{1}{Y}}{\varepsilon}-\sigma u{h}^{\frac{1}{2}-\frac{1}{Y}}\right)^{-Y}du\\
    &\quad+\int_{-\infty}^{0}\phi(u)\mathcal{E}\left({h}^{-\frac{1}{Y}}{\varepsilon}-\sigma u{h}^{\frac{1}{2}-\frac{1}{Y}}\right)du.
\end{align*}
For the first term above, note that
\begin{align*}
     \frac{1}{{h}{\varepsilon}^{-Y}}\int_{-\infty}^{0}\phi(u)\left({h}^{-\frac{1}{Y}}{\varepsilon}-\sigma u{h}^{\frac{1}{2}-\frac{1}{Y}}\right)^{-Y}du
    &=\int_{-\infty}^{0}\phi(u)\left(1-\sigma u{\varepsilon}^{-1}{h}^{1/2}\right)^{-Y}du,
\end{align*}
which, by the dominated convergence theorem, converges to $1/2$, because ${\varepsilon}^{-1}{h}^{1/2}\to{}0$, as $n\to\infty$. Similarly, using (\ref{EWKI}), we have
\[
\left|\int_{-\infty}^{0}\phi(u)\mathcal{E}\left({h}^{-\frac{1}{Y}}{\varepsilon}-\sigma u{h}^{\frac{1}{2}-\frac{1}{Y}}\right)du\right|\leq{}K\int_{-\infty}^{0}\phi(u)\left({h}^{-\frac{1}{Y}}{\varepsilon}-\sigma u{h}^{\frac{1}{2}-\frac{1}{Y}}\right)^{-2Y}du=O\left({\varepsilon}^{-2Y}{h}^{2}\right).
\]
Therefore, we finally conclude that $I_{n}^{-}=Y^{-1}C{h}{\varepsilon}^{-Y}+O\left({\varepsilon}^{-2Y}{h}^{2}\right)$, which implies (\ref{NECNL2}).

We now show (\ref{NECNcL2}). To this end, let us first consider
\begin{align*}
    E_{1,h}(\varepsilon)&:=\bbe\left[J^{2}_{h}{\bf 1}_{\{0\leq\sigma W_{h}+J_{h}\leq{}\varepsilon,J_{h}\geq{}0,W_{h}\geq{}0\}}\right]\\
    &=h^{2/Y}\int_{0}^{\varepsilon \sigma^{-1}h^{-\frac{1}{2}}}\phi(x)\int_{0}^{h^{-\frac{1}{Y}}\varepsilon-\sigma h^{\frac{1}{2}-\frac{1}{Y}} x}u^{2}p_{1}(u)dudx\\
    &=h^{\frac{2}{Y}}\left(\frac{\varepsilon}{\sigma h^{\frac{1}{2}}}\right)\int_{0}^{1}\phi\left(\frac{\varepsilon}{\sigma h^{\frac{1}{2}}}w\right)\int_{0}^{h^{-\frac{1}{Y}}\varepsilon(1-w)}u^{2}p_{1}(u)dudx.
\end{align*}
Let $\mathcal{E}(u):=p_{1}(u)-C u^{-Y-1}$ and let us recall that, for a constant $K$, $|\mathcal{E}(u)|\leq{}K\left(u^{-Y-1}\wedge u^{-2Y-1}\right)\leq{}K u^{-2Y-1}$, for all $u>0$. Next,
\begin{align*}
    E_{1,h}(\varepsilon)&=Ch^{\frac{2}{Y}}\left(\frac{\varepsilon}{\sigma h^{\frac{1}{2}}}\right)\int_{0}^{1}\phi\left(\frac{\varepsilon}{\sigma h^{\frac{1}{2}}}w\right)\int_{0}^{h^{-\frac{1}{Y}}\varepsilon(1-w)}u^{1-Y}dudx\\
    &\quad +h^{\frac{2}{Y}}\left(\frac{\varepsilon}{\sigma h^{\frac{1}{2}}}\right)\int_{0}^{1}\phi\left(\frac{\varepsilon}{\sigma h^{\frac{1}{2}}}w\right)\int_{0}^{h^{-\frac{1}{Y}}\varepsilon(1-w)}u^{2}\mathcal{E}(u)dudw
\end{align*}
For the first term above, note that
\begin{align*}
    \frac{1}{2-Y}\int_{0}^{1}\phi\left(\frac{\varepsilon}{\sigma h^{\frac{1}{2}}}w\right)\left(h^{-\frac{1}{Y}}\varepsilon(1-w)\right)^{2-Y}dw&=
     \frac{h^{-\frac{2-Y}{Y}}\varepsilon^{2-Y}}{2-Y}\int_{0}^{1}\phi\left(\frac{\varepsilon}{\sigma h^{\frac{1}{2}}}w\right)\left(1-w\right)^{2-Y}dw\\
    &\sim 2^{-1}\frac{h^{-\frac{2-Y}{Y}}\varepsilon^{2-Y}}{2-Y}\left(\frac{\sigma h^{\frac{1}{2}}}{\varepsilon}\right).
\end{align*}
We divide the second term in two cases. If $Y\leq{}1$, then
\begin{align*}
\left|\int_{0}^{1}\phi\left(\frac{\varepsilon}{\sigma h^{\frac{1}{2}}}w\right)\int_{0}^{h^{-\frac{1}{Y}}\varepsilon(1-w)}u^{2}\mathcal{E}(u)dudx\right|&\leq{}
K\frac{1}{2-2Y}\int_{0}^{1}\phi\left(\frac{\varepsilon}{\sigma h^{\frac{1}{2}}}w\right)\left(h^{-\frac{1}{Y}}\varepsilon(1-w)\right)^{2-2Y}dw\\
&\leq{}K\frac{h^{-\frac{2-2Y}{Y}}\varepsilon^{2-2Y}}{2-2Y}\int_{0}^{1}\phi\left(\frac{\varepsilon}{\sigma h^{\frac{1}{2}}}w\right)\left(1-w\right)^{2-2Y}dw\\
&\sim 2K\frac{h^{-\frac{2-2Y}{Y}}\varepsilon^{2-2Y}}{2-2Y}\left(\frac{\sigma h^{\frac{1}{2}}}{\varepsilon}\right).
\end{align*}
Note that the last limit is valid provided that $\int_{0}^{1}\left(1-w\right)^{2-2Y}dw<\infty$, which holds true when $Y\leq{}1$. For $Y>1$, let us first observe that
\begin{equation}\label{USI0}
    \int_{0}^{z}u^{2}\left(u^{-Y-1}\wedge{}u^{-2Y-1}\right)du\leq \frac{1}{2-Y}+{\bf 1}_{\{z>1\}}\frac{1-z^{2(1-Y)}}{2(Y-1)}\leq{}\frac{1}{2-Y}+\frac{1}{2(Y-1)}.
\end{equation}
Therefore, for a constant $K$,
\begin{align*}
\left|\int_{0}^{1}\phi\left(\frac{\varepsilon}{\sigma h^{\frac{1}{2}}}w\right)\int_{0}^{h^{-\frac{1}{Y}}\varepsilon(1-w)}u^{2}\mathcal{E}(u)dudx\right|&\leq{}
K\int_{0}^{1}\phi\left(\frac{\varepsilon}{\sigma h^{\frac{1}{2}}}w\right)dw
\sim K\left(\frac{\sigma h^{\frac{1}{2}}}{\varepsilon}\right).
\end{align*}
We conclude that
%\[
%   h^{-\frac{2-Y}{Y}}\varepsilon^{2-Y}\gg h^{-\frac{2-2Y}{Y}}\varepsilon^{2-2Y}\text{ since }\varepsilon\gg h^{1/Y}.
%\]
%Putting together the previous observations,
\begin{align*}
    E_{1,h}(\varepsilon)=\frac{2^{-1}C}{2-Y}h \varepsilon^{2-Y}+O\left(h^{2}\varepsilon^{2-2Y}\right)+O\left(h^{\frac{2}{Y}}\right).
\end{align*}
Next, we consider
\begin{align*}
    E_{2,h}(\varepsilon)&:=\bbe\left[J^{2}_{h}{\bf 1}_{\{0\leq{}\sigma W_{h}+J_{h}\leq{}\varepsilon,J_{h}\geq{}0,W_{h}\leq{}0\}}\right]\\
    &=h^{2/Y}\int_{-\infty}^{0}\phi(x)\int_{-\sigma h^{\frac{1}{2}-\frac{1}{Y}} x}^{h^{-\frac{1}{Y}}\varepsilon-\sigma h^{\frac{1}{2}-\frac{1}{Y}} x}u^{2}p_{1}(u)dudx\\
    &=C h^{2/Y}\int_{-\infty}^{0}\phi(x)\int_{-\sigma h^{\frac{1}{2}-\frac{1}{Y}} x}^{h^{-\frac{1}{Y}}\varepsilon-\sigma h^{\frac{1}{2}-\frac{1}{Y}} x}u^{1-Y}dudx\\
    &\quad+h^{2/Y}\int_{-\infty}^{0}\phi(x)\int_{-\sigma h^{\frac{1}{2}-\frac{1}{Y}} x}^{h^{-\frac{1}{Y}}\varepsilon-\sigma h^{\frac{1}{2}-\frac{1}{Y}} x}u^{2}\mathcal{E}(u)dudx.
%   \\
%   &=h^{\frac{2}{Y}}\left(\frac{\varepsilon}{\sigma h^{\frac{1}{2}}}\right)\int_{-\infty}^{0}\phi\left(\frac{\varepsilon}{\sigma h^{\frac{1}{2}}}w\right)\int_{0}^{h^{-\frac{1}{Y}}\varepsilon(1-w)}u^{2}p_{1}(u)dudx.
\end{align*}
The first term on the right-hand side above can be written as
\begin{align*}
%   &\frac{C}{2-Y} h^{2/Y}\int_{-\infty}^{0}\phi(x)\left\{\left(h^{-\frac{1}{Y}}\varepsilon-\sigma h^{\frac{1}{2}-\frac{1}{Y}} x\right)^{2-Y}-\left(-\sigma h^{\frac{1}{2}-\frac{1}{Y}} x\right)^{2-Y}\right\}dx\\
%   &=
    \frac{C}{2-Y} h^{2/Y}\left(h^{-\frac{1}{Y}}\varepsilon\right)^{2-Y}\int_{-\infty}^{0}\phi(x)\left\{\left(1-\frac{\sigma h^{\frac{1}{2}}}{\varepsilon} x\right)^{2-Y}-\left(-\frac{\sigma h^{\frac{1}{2}}}{\varepsilon} x\right)^{2-Y}\right\}dx
    \sim 2^{-1}\frac{C}{2-Y} h \varepsilon^{2-Y},
\end{align*}
where the last asymptotic relationship follows from dominated convergence theorem and the facts that $h^{1/2}/\varepsilon\to{}0$ and $\int_{-\infty}^{0}(1-x)^{2-Y}\phi(x)dx<\infty$. For the second term of $E_{2,h}(\varepsilon)$, we have two cases. For $Y\leq{}1$, we have
\begin{align*}
    &h^{\frac{2}{Y}}\left|\int_{-\infty}^{0}\phi(x)\int_{-\sigma h^{\frac{1}{2}-\frac{1}{Y}} x}^{h^{-\frac{1}{Y}}\varepsilon-\sigma h^{\frac{1}{2}-\frac{1}{Y}} x}u^{2}\mathcal{E}(u)dudx\right|
    \leq K h^{\frac{2}{Y}}\int_{-\infty}^{0}\phi(x)\int_{-\sigma h^{\frac{1}{2}-\frac{1}{Y}} x}^{h^{-\frac{1}{Y}}\varepsilon-\sigma h^{\frac{1}{2}-\frac{1}{Y}} x}u^{1-2Y}dudx\\
    &\quad=\frac{K}{2(1-Y)} h^{\frac{2}{Y}}\left(h^{-\frac{1}{Y}}\varepsilon\right)^{2-2Y}\int_{-\infty}^{0}\phi(x)\left\{\left(1-\frac{\sigma h^{\frac{1}{2}}}{\varepsilon} x\right)^{2(1-Y)}-\left(-\frac{\sigma h^{\frac{1}{2}}}{\varepsilon} x\right)^{2(1-Y)}\right\}dx
    \sim K h^{2} \varepsilon^{2-2Y},
\end{align*}
where again we used dominated convergence and use the fact that $\int_{-\infty}^{0}\phi(x)(1-x)^{2(1-Y)}dx<\infty$. For $Y>1$, we just use (\ref{USI0}) to deduce that
\begin{align*}
    &h^{\frac{2}{Y}}\int_{-\infty}^{0}\phi(x)\int_{-\sigma h^{\frac{1}{2}-\frac{1}{Y}} x}^{h^{-\frac{1}{Y}}\varepsilon-\sigma h^{\frac{1}{2}-\frac{1}{Y}} x}u^{2}|\mathcal{E}(u)|dudx\leq{}K' h^{\frac{2}{Y}}\int_{-\infty}^{0}\phi(x)dx,
\end{align*}
for a constant $K'$. Finally, we conclude that
\[
    E_{2,h}=2^{-1}\frac{C}{2-Y} h \varepsilon^{2-Y}+O\left( h^{2} \varepsilon^{2-2Y}\right)+O\left( h^{\frac{2}{Y}}\right).
\]
Finally, let us consider
\begin{align*}
    E_{3,h}(\varepsilon)&:=\bbe\left[J^{2}_{h}{\bf 1}_{\{0\leq{}\sigma W_{h}+J_{h}\leq{}\varepsilon,J_{h}\leq{}0,W_{h}\geq{}0\}}\right]\\
%   &=h^{2/Y}\int_{0}^{\infty}\phi(x)\int_{-\sigma h^{\frac{1}{2}-\frac{1}{Y}} x}^{h^{-\frac{1}{Y}}\varepsilon-\sigma h^{\frac{1}{2}-\frac{1}{Y}} x}u^{2}p_{1}(u){\bf 1}_{u\leq{}0}dudx\\
    &=h^{2/Y}\int_{0}^{\sigma^{-1}h^{-\frac{1}{2}}\varepsilon}\phi(x)\int_{-\sigma h^{\frac{1}{2}-\frac{1}{Y}} x}^{0}u^{2}p_{1}(u)dudx\\
    &\quad + h^{2/Y}\int_{\sigma^{-1}h^{-\frac{1}{2}}\varepsilon}^{\infty}\phi(x)\int_{-\sigma h^{\frac{1}{2}-\frac{1}{Y}} x}^{h^{-\frac{1}{Y}}\varepsilon-\sigma h^{\frac{1}{2}-\frac{1}{Y}} x}u^{2}p_{1}(u)dudx.
\end{align*}
Using the fact that $p_{1}(u)\leq{}K u^{-Y-1}$ for a constant $K$ and all $u>0$, the first term above is such that
\begin{align*}
    h^{2/Y}\int_{0}^{\sigma^{-1}h^{-\frac{1}{2}}\varepsilon}\phi(x)\int_{0}^{\sigma h^{\frac{1}{2}-\frac{1}{Y}} x}u^{2}p_{1}(u)dudx&\leq{}K h^{2/Y}\int_{0}^{\sigma^{-1}h^{-\frac{1}{2}}\varepsilon}\phi(x)\int_{0}^{\sigma h^{\frac{1}{2}-\frac{1}{Y}} x}u^{1-Y}dudx\\
    &=\frac{K}{2-Y}\left({\sigma h^{\frac{1}{2}-\frac{1}{Y}}}\right)^{2-Y}\int_{0}^{\sigma^{-1}h^{-\frac{1}{2}}\varepsilon}\phi(x)x^{2-Y}dx\\
    &\sim \frac{K}{2-Y}h^{\frac{4-Y}{2}}\int_{0}^{\infty}\phi(x)x^{2-Y}dx=o\left(h \varepsilon^{2-Y}\right).
\end{align*}
Similarly, the second term can be written as
\begin{align*}
    h^{2/Y}\int_{\sigma^{-1}h^{-\frac{1}{2}}\varepsilon}^{\infty}\phi(x)\int^{\sigma h^{\frac{1}{2}-\frac{1}{Y}} x}_{\sigma h^{\frac{1}{2}-\frac{1}{Y}} x-h^{-\frac{1}{Y}}\varepsilon}u^{2}p_{1}(u)dudx&\leq{}\frac{K}{2-Y}\left({\sigma h^{\frac{1}{2}-\frac{1}{Y}}}\right)^{2-Y}\int_{\sigma^{-1}h^{-\frac{1}{2}}\varepsilon}^{\infty}\phi(x)x^{2-Y}dx\\
    &=o\left(h^{\frac{4-Y}{2}}\right)=o\left(h \varepsilon^{2-Y}\right).
\end{align*}
Putting together the previous results, we obtain that
\begin{align*}
    E_{h}(\varepsilon)&=2\bbe\left[J^{2}_{h}{\bf 1}_{\{0\leq{}\sigma W_{h}+J_{h}\leq{}\varepsilon\}}\right]=2E_{1,h}(\varepsilon)+2E_{2,h}(\varepsilon)+2E_{3,h}(\varepsilon)\\
    &=\frac{2C}{2-Y}h \varepsilon^{2-Y}+O\left(h^{2}\varepsilon^{2-2Y}\right)+O\left(h^{\frac{4-Y}{2}}\right)+O\left(h^{\frac{2}{Y}}\right).
\end{align*}
}

 \vspace{-1.3cm}\qed\\

{\bf Proof of Theorem \ref{thrmIJA}.}
From Lemmas \ref{L1:IJA} {and \ref{L2:IJA}},
\begin{align*}
    C_{h}^{+}(\varepsilon)&=\bbe\left[\left(\frac{\varepsilon}{\sigma\sqrt{h}}-\frac{J_{h}}{\sigma\sqrt{h}}\right)
    \phi\left(\frac{\varepsilon}{\sigma\sqrt{h}}-\frac{J_{h}}{\sigma\sqrt{h}}\right)+\bar{\Phi}\left(\frac{\varepsilon}{\sigma\sqrt{h}}-\frac{J_{h}}{\sigma\sqrt{h}}\right)\right]\\
    &=\frac{\varepsilon}{\sigma\sqrt{h}}\left(\frac{1}{\sqrt{2\pi}}e^{-\frac{\varepsilon^{2}}{2\sigma^{2}h}}-K_{1}\varepsilon^{-1-Y}h^{\frac{3}{2}}\right)-\frac{1}{\sigma\sqrt{h}}\left(K_{2} h\varepsilon^{1-Y}\right)+\frac{C}{Y}h\varepsilon^{-Y}+{\rm h.o.t.}\\
    &=\frac{\varepsilon}{\sigma\sqrt{h}\sqrt{2\pi}}e^{-\frac{\varepsilon^{2}}{2\sigma^{2}h}}-\frac{K_{2}}{\sigma}h^{1/2}\varepsilon^{1-Y}+{\rm h.o.t.},
\end{align*}
where above we used that $\varepsilon^{-Y}h\ll  h^{1/2}
\varepsilon^{1-Y}$. Therefore, using that $D_{h}=0$ and Lemma
 \ref{L2:IJA}, with $K_3=\frac{2C}{2-Y}$,
        \begin{align*}
        \bbe[b_{1}\left(\varepsilon\right)]&=\bbe\left[\left(\sigma W_{h}+J_{h}\right)^{2}{\bf 1}_{\{|\sigma W_{h}+J_{h}|\leq{}\varepsilon\}}\right]=C_{h}(\varepsilon)+D_{h}(\varepsilon)+E_{h}(\varepsilon)\\
        &=\sigma^{2}h-2\sigma^{2}h\left(\frac{\varepsilon}{\sigma\sqrt{h}\sqrt{2\pi}}e^{-\frac{\varepsilon^{2}}{2\sigma^{2}h}}-\frac{K_{2}}{\sigma}h^{1/2}\varepsilon^{1-Y}\right)+K_{3} h\varepsilon^{2-Y}+{\rm h.o.t.}\\
        &=\sigma^{2}h-\frac{2\sigma}{\sqrt{2\pi}}\sqrt{h}\varepsilon e^{-\frac{\varepsilon^{2}}{2\sigma^{2}h}}+K_{3} h\varepsilon^{2-Y}+{\rm h.o.t.},
    \end{align*}
where above we used that $h\varepsilon^{2-Y}\gg
h^{3/2}\varepsilon^{1-Y}$.
\qed\\

\n
{\bf Proof of Lemma \ref{LemEpGrSqrthMSE}.}
%The key property of a bounded variation process that we want to take advantage of is that $h^{-1}J_{h}\stackrel{\bbp}{\to}\gamma$, when $h\to{}0$, where $\gamma$ is the drift of $J$.
 We show the result by contradiction. Suppose that $\liminf_{n\to{}\infty}\frac{\varepsilon_{n}^{\star}}{\sqrt{h_{n}}}<\infty$. For simplicity and without loss of generality, we further assume that $\lim_{n\to{}\infty}\frac{\varepsilon_{n}^{\star}}{\sqrt{h_{n}}}=:L<\infty$ as all the statements below are valid on a subsequence $\{n_{k}\}_{k\geq{}0}$.
Let $M\in(0,\infty)$ be such that
$\sup_{n}\frac{\varepsilon_{n}^{\star}}{\sqrt{h_{n}}}\leq{}M$. {Also, for simplicity, let us write $\varepsilon_{n}$ for
$\varepsilon_{n}^{\star}$ and assume that $T=1$ so that $h_{n}=1/n$.} Consider
the decomposition
    \begin{align*}
        \bbe[b_{1}\left(\varepsilon\right)]&=\bbe\left[\left(\sigma W_{h}+J_{h}\right)^{2}{\bf 1}_{\{|\sigma W_{h}+J_{h}|\leq{}\varepsilon\}}\right]\\
        &=\sigma^{2}\bbe\left[W_{h}^{2}{\bf 1}_{\{|\sigma W_{h}+J_{h}|\leq{}\varepsilon\}}\right]+
        2\sigma\bbe\left[W_{h}J_{h}{\bf 1}_{\{|\sigma W_{h}+J_{h}|\leq{}\varepsilon\}}\right]+
        \bbe\left[J_{h}^{2}{\bf 1}_{\{|\sigma W_{h}+J_{h}|\leq{}\varepsilon\}}\right]\\
        &=:c_{h}(\varepsilon)+d_{h}(\varepsilon)+e_{h}(\varepsilon).
    \end{align*}
Note that dominated convergence implies that
\[
        \frac{1}{h_{n}}c_{h_{n}}(\varepsilon_{n})=\sigma^{2}\bbe\left[W_{1}^{2}{\bf 1}_{\{|\sigma W_{1}+h^{-1/2}_{n}J_{h_{n}}|\leq{}h_{n}^{-1/2}\varepsilon_{n}\}}\right]
        \stackrel{n\to{}\infty}{\longrightarrow}\sigma^{2}\bbe\left[W_{1}^{2}{\bf 1}_{\{|W_{1}|\leq{}L/\sigma\}}\right]<\sigma^{2},
\]
since
$h_{n}^{-1/2}J_{h_{n}}=h_{n}^{\frac{1}{Y}-\frac{1}{2}}(h_{n}^{-1/Y}J_{h_{n}})\to{}0$,
in probability. For $d_{h}$ note that
\[
    {\sigma}|W_{1}h_{n}^{-1/2}J_{h_{n}}|{\bf 1}_{\{|\sigma W_{1}+h_{n}^{-1/2}J_{h_{n}}|\leq{}h_{n}^{-1/2}\varepsilon_{n}\}}\leq
    {\sigma^{2}|W_{1}|^{2}+\sigma|W_{1}|h_{n}^{-1/2}\varepsilon_{n}}\leq{}{\sigma^{2}|W_{1}|^{2}+\sigma|W_{1}|M},
\]
therefore, again by dominated convergence
\[
    h_{n}^{-1}d_{h_{n}}(\varepsilon_{n})={2\sigma}\bbe\left[W_{1}h_{n}^{-1/2}J_{h_{n}}{\bf 1}_{\{|\sigma W_{1}+h_{n}^{-1/2}J_{h_{n}}|\leq{}h_{n}^{-1/2}\varepsilon_{n}\}}\right]\stackrel{n\to{}\infty}{\longrightarrow}{}0.
\]
Similarly, since $(h_{n}^{-1/2}J_{h_{n}})^{2}{\bf 1}_{\{|\sigma
W_{1}+h_{n}^{-1/2}J_{h_{n}}|\leq{}h_{n}^{-1/2}\varepsilon_{n}\}}\leq
    {2\sigma^{2}W_{1}^{2}+2h_{n}^{-1}\varepsilon_{n}^{2}}\leq{}{2W_{1}^{2}+2M^{2}}$,
\[
    h_{n}^{-1}e_{h_{n}}(\varepsilon_{n})=\bbe\left[\left(h_{n}^{-1/2}J_{h_{n}}\right)^{2}{\bf 1}_{\{|\sigma W_{1}+h_{n}^{-1/2}J_{h_{n}}|\leq{}h_{n}^{-1/2}\varepsilon_{n}\}}\right]\stackrel{n\to{}\infty}{\longrightarrow}{}0.
\]
Finally, let us write the equation
$\varepsilon_{n}^{2}+2(n-1)\bbe[b_{1}(\varepsilon_{n})]-2nh_{n}\sigma^{2}=0$
as
%\[
%    \varepsilon_{n}^{2}+2\frac{n-1}{n}\left(\frac{c_{h_{n}}(\varepsilon_{n})}{h_{n}}+\frac{d_{h_{n}}(\varepsilon_{n})}{h_{n}}+\frac{e_{h_{n}}(\varepsilon_{n})}{h_{n}}\right)-2\sigma^{2}=0.
%\]
\begin{equation}\label{NEH0}
    \varepsilon_{n}^{2}+2\frac{n-1}{n}\left(\frac{d_{h_{n}}(\varepsilon_{n})}{h_{n}}+\frac{e_{h_{n}}(\varepsilon_{n})}{h_{n}}\right)=2\sigma^{2}-2\frac{n-1}{n}\frac{c_{h_{n}}(\varepsilon_{n})}{h_{n}}.
\end{equation}
{The} right-hand side of the equation converges to
$2\sigma^{2}\left(1-\bbe\left[W_{1}^{2}{\bf
1}_{\{|W_{1}|\leq{}L/\sigma\}}\right]\right)>0$, while the left hand
side converges to $0$ and this leads to a contradiction and
therefore
$\lim_{n\to{}\infty}\frac{\varepsilon_{n}^{\star}}{\sqrt{h_{n}}}=\infty$.
\qed\\

\n
{\bf Proof of Proposition \ref{ABeps}.}
For simplicity, in what follows we take $T=1$ so that $h=1/n$.
Again, recall that $\varepsilon^{\star}$ is the solution of
\[
    (\varepsilon^{\star})^{2}+2(n-1)\bbe[b_{1}(\varepsilon^{\star})]-2nh\sigma^{2}=0.
\]
Throughout, we shall use that $\varepsilon^{\star}\gg \sqrt{h}$, as
proved in the above lemma.  For simplicity, we write $\varepsilon$
instead of $\varepsilon^{\star}$. By the asymptotic behavior of
$\bbe\left[b_{1}(\varepsilon)\right]$ described above,
\begin{align*}
     \varepsilon^{2}+2(n-1)\left(\sigma^{2}h-\frac{2}{\sqrt{2\pi}}\sigma\varepsilon\sqrt{h} e^{-\frac{\varepsilon^{2}}{2\sigma^{2}h}}+
\lambda h \frac{\varepsilon^{3}}{3}C(f)+
O\left(h^{2}\right)+o\left(\varepsilon\sqrt{h}
e^{-\frac{\varepsilon^{2}}{2\sigma^{2}h}}\right)+o\left(h
\varepsilon^{3}\right)\right)-2nh\sigma^{2}=0,
\end{align*}
and, thus, using that $h=1/n$,
\begin{align}\label{EqN1H}
    \varepsilon^{2}-{2}\sigma^{2}h-\frac{4}{\sqrt{2\pi}}\sigma\frac{\varepsilon}{\sqrt{h}} e^{-\frac{\varepsilon^{2}}{2\sigma^{2}h}}+
2\lambda  \frac{\varepsilon^{3}}{3}C(f)+
O\left(h\right)+o\left(\frac{\varepsilon}{\sqrt{h}}
e^{-\frac{\varepsilon^{2}}{2\sigma^{2}h}}\right)+o\left(\varepsilon^{3}\right)=0.
\end{align}
Now, since $h=o(\varepsilon^{2})$ (as assumed at the beginning), we
can write the previous equation as
\begin{align*}
    \varepsilon^{2}-\frac{4}{\sqrt{2\pi}}\sigma\frac{\varepsilon}{\sqrt{h}} e^{-\frac{\varepsilon^{2}}{2\sigma^{2}h}}+o\left(\frac{\varepsilon}{\sqrt{h}} e^{-\frac{\varepsilon^{2}}{2\sigma^{2}h}}\right)+o\left(\varepsilon^{2}\right)=0.
\end{align*}
Dividing by $\varepsilon$ and rearranging the terms,
\begin{align}\label{AsymEqEps}
    \varepsilon\left(1+o(1)\right)=\frac{4}{\sqrt{2\pi}}\sigma\frac{1}{\sqrt{h}} e^{-\frac{\varepsilon^{2}}{2\sigma^{2}h}}\left(1+o(1)\right).
\end{align}
Then, taking logarithms of both sides and since
$\ln(1+o(1))=o(1)$,
\begin{equation}
    \ln \varepsilon +o(1)=-\frac{\varepsilon^{2}}{2\sigma^{2}h}-\frac{1}{2}\ln h +\ln \left(\frac{4\sigma}{\sqrt{2\pi}}\right)+o(1).
\end{equation}
which can be written as
\[
     \ln\left(\frac{\varepsilon^{2}}{\sigma^{2}h}\right)+o(1)=-\frac{\varepsilon^{2}}{\sigma^{2}h}-2\ln h +\ln \left(\frac{8}{\pi}\right)+o(1)
\]
Defining $\varpi=\varepsilon^{2}/(\sigma^{2}h)$, we can write
\[
    -\frac{\ln \varpi}{\varpi}+\frac{2\ln \frac{1}{h}}{\varpi}-
    \frac{\ln\frac{\pi}{8}}{\varpi}-\frac{o(1)}{\varpi}=1+\frac{o(1)}{\varpi}.
\]
Therefore, making $h\to{}0$ and using that $\varpi\to\infty$ (since
$\varepsilon\gg\sqrt{h}$),
\[
    \frac{2\ln \frac{1}{h}}{\varpi}\;\stackrel{h\to{}0}{\longrightarrow}\; 1.
\]
Recalling that $\varpi=\varepsilon^{2}/(\sigma^{2}h)$, we conclude the result.
\qed\\

\n
{\bf Proof of Proposition \ref{ABepsIJA}.}
{For simplicity, we again take $T=1$ so that $h=1/n$ and write $\varepsilon$
instead of $\varepsilon^{\star}$.
By the asymptotic behavior of
$\bbe\left[b_{1}(\varepsilon)\right]$ described in Theorem
\ref{thrmIJA}, we can write $(\varepsilon^{\star})^{2}+2(n-1)\bbe[b_{1}(\varepsilon^{\star})]-2nh\sigma^{2}=0$ as}
\begin{align*}
     \varepsilon^{2}+2(n-1)\left(\sigma^{2}h-\frac{2\sigma}{\sqrt{2\pi}}\varepsilon\sqrt{h} e^{-\frac{\varepsilon^{2}}{2\sigma^{2}h}}+\frac{2C}{2-Y} h\varepsilon^{2-Y}+{\rm h.o.t.}\right)-2nh\sigma^{2}=0,
\end{align*}
and, thus, using that {$h=o(\varepsilon^{2})$} and
$\varepsilon^{2}=o\left(\varepsilon^{2-Y}\right)$, {we have}
\begin{align}\label{EqN1H}
    \frac{4 C}{2-Y}\varepsilon^{2-Y}-\frac{4}{\sqrt{2\pi}}\sigma\frac{\varepsilon}{\sqrt{h}} e^{-\frac{\varepsilon^{2}}{2\sigma^{2}h}}+o\left(\frac{\varepsilon}{\sqrt{h}} e^{-\frac{\varepsilon^{2}}{2\sigma^{2}h}}\right)+o\left(\varepsilon^{2-Y}\right)=0.
\end{align}
Dividing by $\varepsilon$ and rearranging the terms,
\begin{align*}
    \varepsilon^{1-Y}\left(1+o(1)\right)=\frac{2-Y}{C\sqrt{2\pi}}\sigma\frac{1}{\sqrt{h}} e^{-\frac{\varepsilon^{2}}{2\sigma^{2}h}}\left(1+o(1)\right).
\end{align*}
Then, taking logarithms {of} both sides and since
$\ln(1+o(1))=o(1)$,
\[
    (1-Y)\ln \varepsilon +o(1)=-\frac{\varepsilon^{2}}{2\sigma^{2}h}-\frac{1}{2}\ln h +\ln \left(\frac{(2-Y)\sigma}{C\sqrt{2\pi}}\right)+o(1),
\]
which can be written as
%\[
%   \frac{1}{2}\ln \varepsilon^{2} +o(1)=-\frac{\varepsilon^{2}}{2\sigma^{2}h}-\frac{1}{2}\ln h +\ln \left(\frac{4\sigma}{\sqrt{2\pi}}\right)+o(1)
%\]
%\[
%   \ln \varepsilon^{2} +o(1)=-\frac{\varepsilon^{2}}{\sigma^{2}h}-\ln h +2\ln \left(\frac{4\sigma}{\sqrt{2\pi}}\right)+o(1)
%\]
%\[
%    \ln\left(\frac{\varepsilon^{2}}{\sigma^{2}h}\right)+\ln\left(\sigma^{2}\right)+\ln h +o(1)=-\frac{\varepsilon^{2}}{\sigma^{2}h}-\ln h +\ln \left(\frac{8\sigma^{2}}{\pi}\right)+o(1)
%\]
\[
     \frac{1-Y}{2}\ln\left(\frac{\varepsilon^{2}}{\sigma^{2}h}\right)+\frac{1-Y}{2}\ln\left( \sigma^{2}\right)+\frac{1-Y}{2}\ln\left(h\right)+o(1)=-\frac{\varepsilon^{2}}{2\sigma^{2}h}-\frac{1}{2}\ln h +\ln \left(\frac{(2-Y)\sigma}{C\sqrt{2\pi}}\right)+o(1).
\]
Equivalently,  writing $\varpi=\varepsilon^{2}/(\sigma^{2}h)$
%,
%%\[
%%     (1-Y)\ln\left(\varpi\right)+(1-Y)\ln\left(2\sigma^{2}\right)+(1-Y)\ln\left(h\right)+o(1)=-\varpi-\ln h +2\ln \left(\frac{(2-Y)\sigma}{C\sqrt{2\pi}}\right)+o(1),
%%\]
%%or, for a constant $K$,
%\[
%     (1-Y)\ln\left(\varpi\right)=-\varpi+(Y-2)\ln h +K+o(1),
%\]
%for a constant $K$.}
and dividing by $-\varpi$,
\[
    -(1-Y)\frac{\ln \varpi}{\varpi}+\frac{(2-Y)\ln \frac{1}{h}}{\varpi}-\frac{K}{-\varpi}=1+\frac{o(1)}{\varpi}.
\]
 and using that $\varpi\to\infty$ (since $\varepsilon\gg\sqrt{h}$), we get
\[
    \frac{(2-Y)\ln \frac{1}{h}}{\varpi}\;\stackrel{h\to{}0}{\longrightarrow}\; 1.
\]
Recalling that $\varpi=\varepsilon^{2}/(\sigma^{2}h)$, we conclude the result.
%that
%\[
%    \varepsilon\sim \sqrt{(2-Y)\sigma^{2}h\ln\frac{1}{h}}, \quad \text{as}\quad h\to{}0.
%\]
\qed\\

\n
{\bf Proof of Proposition \ref{ConvHatIVOptThr}.} In order to prove the  proposition, we follow and modify the proof of
Theorem 1 in \cite{Man09}, in that we show that a.s.,
for all $\eta>0 $, for sufficiently small $h${,}  we have\\
1) $\forall i=1,\dots,n,  \IDNeqZ\leq \IDXqleqrieta$\\
2) $\forall i=1,\dots,n, \IDNeqZ\geq \IDXqleqrieta.$\\
Then %1) implies that for all $\eta>0 $, for sufficiently small $h$ we have $\forall i=1..n, \IDNeqZ\leq \IDXqleqrieta, $ and
%$\forall i=1..n, \IDNeqZ=\IDXqleqrieta $
the thesis follows.

Call $\Delta_i X_0=\intIi a_s ds + \intIi \sigma_s dW_s,$
$\bar a= \sup_{s\in[0,T]} |a_s|$,
$\bar \sigma= \sup_{s\in[0,T]} \sigma_s$ and
$\underline{\gamma}(\omega)= \min_{\ell: \Delta N_\ell\neq 0} |\gamma_\ell (\omega)|, $ and note that under our
assumptions $P(\underline{\gamma}\neq 0)=1$.
To show 1) a 2) we use the following key fact:
$$
\sup_{i\in \{ 1,...,n \} } \frac{|\DXz|}{\sqrt{2M_i h
\log\frac{1}{h} }}\leq \sup_i\frac{\bar a \sqrt h}{\sqrt{2 M_i \log
\frac 1 h}} +$$
$$\sup_i \frac{|B_{IV_{t_{i}}}-B_{IV_{t_{i-1}}}|}{\sqrt{2\Delta_{i} IV\log\frac{1}{\Delta_{i} IV }} }\
\sup_{i}\frac{\sqrt{2\Delta_{i}IV \log\frac{1}{\Delta_{i} IV
}}}{\sqrt{2M_ih \log\frac{1}{M_ih}} } \sup_{i\in \{ 1,...,n \}
}\frac{\sqrt{2h \log\frac{1}{M_ih}} }{\sqrt{2h \log\frac{1}{h}} },
$$
where $B$ is a standard Brownian motion and we used the fact that
$\sigma\!\cdot W$ is  a time changed Brownian motion (\cite{RevYor01}, theorems 1.9 and 1.10), meaning that we can represent $
\Delta_{i}\left(\sigma\!\cdot
W\right)=B_{IV_{t_{i}}}-B_{IV_{t_{i-1}}}.$ By the Paul Lévy law on
the modulus of continuity of the BM paths ({\cite{KarShr99}},
theorem 9.25) and the monotonicity of the function $x\ln (1/x)$ on
$(0, 1/e),$ it follows that for sufficiently small $h$ the first two
factors of the last line of last display are bounded above by 1, so that
$$
\sup_{i} \frac{|\DXz|}{\sqrt{2M_i h \log\frac{1}{h} }}\leq
\sup_i\frac{\bar a \sqrt h}{\sqrt{2 M_i \log \frac 1 h}} +
\sup_i\sqrt{\frac{ \log \frac{1}{M_i}}{\log\frac 1 h} + 1}
$$
%since the function $x\log\frac 1 x$ ha on $(0, +\infty)$ an absolute maximum value of $1/e,$ we have
$$\leq
M_h:=\frac{\bar a \sqrt h}{\sqrt{2 \underline\sigma^2 \log \frac 1
h}} + \sqrt{\frac{ \log \frac{1}{\underline \sigma^2}}{\log\frac 1
h} + 1}$$ which tends to 1, as $h\to 0$.

Now, in order to show 1), we  define {$\{J\}=\{i\in \{1, 2, ..., n\}:\
\DN\neq 0\},$ and it is sufficient to prove that for $h$ small enough
$\sup_{i\not\in \{J\}} \frac{|\DX|}{\sqri}\leq 1+\eta.$ % pag 4
Indeed, $\sup_{i\not\in \{J\}} \frac{|\DX|}{\sqri} = \sup_{i\not\in \{J\}}
\frac{|\DXz|}{\sqri}\leq \sup_{i\in \{1,.., n\}}
\frac{|\DXz|}{\sqri}\leq M_h \to 1,$ thus for all $\eta>0$ for
sufficiently small $h$, it is ensured that $\sup_{i\not\in \{J\}}
\frac{|\DX|}{\sqri}< 1+\eta$, that is: for all $i$,
 if $\DN=0$ then necessarily we have $|\DX| < (1+\eta)\sqri$, and 1) follows.

In order to show 2) we prove that, for sufficiently small $h$, $\inf_{i\in \{J\}}
\frac{|\DX|}{\sqri}>1+\eta.$ In fact firstly  note that for sufficiently
small $h$ all the increments of $N$ are either 0 or 1. It follows
that if $\DN\neq 0,$ then $\DN=1,$ and $\Delta_i J$ coincides with
the size, say $\gamma_{\ell_i},$ of a single jump
$\Delta_iJ=\gamma_{\ell_i}.$ Then $ \frac{|\DX|}{\sqri}\geq
\frac{|\gamma_{\ell_i}|}{\sqri}-  \frac{|\DXz|}{\sqri}$ and
$$ \inf_{i\in \{J\}}\frac{|\DX|}{\sqri}
\geq \frac{\underline{\gamma}}{\bar\sigma\sqrt{2h\log\frac 1 h}}-\sup_{i\in\{J\}}  \frac{|\DXz|}{\sqrt{2M_ih\log\frac 1 h}} \geq
\frac{\underline{\gamma}}{\bar\sigma\sqrt{2h\log\frac 1 h}}
-(1+\eta) $$ and this tends to $+\infty$ when $h\to 0,$ thus $
\inf_{i\in \{J\}}\frac{|\DX|}{\sqri}>1+\eta$, meaning that if $\DN\neq 0$
then necessarily $|\DX|>\sqri(1+\eta)$,
as we needed.}\qed\\

\n
{\bf Proof of  Corollary \ref{CORConvHatIVOptThr}}.
The proof of the Corollary is straightforward, in that a.s. we fix
any $\eta>0$, and for sufficiently small $h$ we have
$$\sum_{i=1}^n \DXq \IDXqleqrieta=\sum_{i=1}^n \DXq \IDNeqZ =\sum_{i=1}^n (\DXz)^2 - \sum_{i=1}^n (\DXz)^2\IDNneqZ \toP IV_T,$$
since the last term tends to 0 in probability, as $E[\sum_{i=1}^n (\DXz)^2\IDNneqZ]\leq N_T O(h)\to 0.$\qed\\

\n
{\bf Proof of Proposition \ref{PropcMSEBarEpBiggerSqrth}.} {Recall that} $\bar\ep$ is such that $\sum_{i=1}^n a_i g_i=0,$ i.e. $\sum_{i=1}^n a_i(\bar\ep^2 + 2\sum_{j\neq i} b_j -2 IV)=0.$ For simplicity let us rename
$\bar\ep$ by $\ep.$  If
$\liminf_{h\to 0} \frac{\ep(h)}{\sqrt h}= L\in [0, +\infty)$ we can find a subsequence such that
$\lim \frac{\ep(h)}{\sqrt h}= L$. Note that
$$ 0= \sumi a_i \left(\frac{\ep^2}{h} + \frac{2 \sum_{j\neq i} b_j}{h}-2\sigma^2n\right)=
\frac{\ep^2}{h}\sumi a_i + \frac 2 h \sumi a_i \sum_{j\neq i} b_j - 2 \sigma^2 n \sumi a_i,
$$
i.e.
\beq\label{epLargerThansqrthFA} \frac{\ep^2}{h} = 2 \sigma^2 n-\frac 2 h \frac{ \sumi a_i \sum_{j\neq i} b_j }{\sumi a_i}=
 2n \Big[\sigma^2 - \frac{ \sumi a_i \sum_{j\neq i} b_j }{\sumi a_i}\Big].\eeq
Now we show that %  $\sum_{j\neq i} b_j <\sigma^2$ for all $i=1..n$, which implies that
$\sigma^2-\frac{ \sumi a_i \sum_{j\neq i} b_j }{\sumi a_i}$ tends to a strictly positive constant, which in turn means that
equality (\ref{epLargerThansqrthFA}) is impossible, since on any sequence $\ep(h)$ such that $\frac{\ep(h)}{\sqrt h}\to L$
the left term tends to $L^2,$ while the right one tends to $+\infty$.

Let us then check that $\sigma^2-\frac{ \sumi a_i \sum_{j\neq i} b_j }{\sumi a_i}$ tends to a strictly positive constant.
Since $J$ has FA, a.s. we only have finitely many
$\Delta J_t \neq 0,$ and, for small $h$,
$N_T$ coincides with $\sumi I_{m_i\neq 0}$. Recalling the explicit expression of $b_j$ (also reported below), we have
 $$\sum_{j\neq i} b_j=\sum_{j\neq i, m_j=0} b_j +\sum_{j\neq i, m_j\neq 0} b_j \leq
-( n-N_T)\frac{\sigma\sqrt h}{\sqrt{2\pi}}2\ep e^{-\frac{\ep^2}{2\sigma^2h}}
+ (n-N_T)\frac{\sigma^2 h}{\sqrt{2\pi}}\int_{-\frac{\ep}{\sigma\sqrt h}}^{\frac{\ep}{\sigma\sqrt h}}e^{-\frac{x^2}{2}}dx$$
$$
- \!\!\!\!\sum_{j\neq i, m_j\neq 0}\frac{\sigma\sqrt h}{\sqrt{2\pi}}\left(
\ep\Big(e^{-\frac{(\ep-|m_j|)^2}{2\sigma^2h}}+
e^{-\frac{(\ep+|m_j|)^2}{2\sigma^2h}}\Big) + |m_j|\Big(
e^{-\frac{(\ep-|m_j|)^2}{2\sigma^2h}}-e^{-\frac{(\ep+|m_j|)^2}{2\sigma^2h}}\Big)\right)
+\!\!\!\!\sum_{j\neq i, m_j\neq 0}\frac{m_j^2 +\sigma^2
h}{\sqrt{2\pi}}\int_{\frac{m_j -\ep}{\sigma\sqrt h}}^{\frac{m_j
+\ep}{\sigma\sqrt h}} e^{-\frac{x^2}{2}}dx.$$ Now, the factors $\ep
e^{-\frac{\ep^2}{2\sigma^2h}}$ and
$\ep\Big(e^{-\frac{(\ep-|m_j|)^2}{2\sigma^2h}}+
e^{-\frac{(\ep+|m_j|)^2}{2\sigma^2h}}\Big) + |m_j|\Big(
e^{-\frac{(\ep-|m_j|)^2}{2\sigma^2h}}-e^{-\frac{(\ep+|m_j|)^2}{2\sigma^2h}}\Big)$
of $\frac{\sigma\sqrt h}{\sqrt{2\pi}}$ are strictly positive, so
$$\sum_{j\neq i} b_j\leq
(n-N_T)\frac{\sigma^2 h}{\sqrt{2\pi}}\int_{-\frac{\ep}{\sigma\sqrt
h}}^{\frac{\ep}{\sigma\sqrt h}}e^{-\frac{x^2}{2}}dx +\sum_{j\neq i,
m_j\neq 0}\frac{m_j^2 +\sigma^2 h}{\sqrt{2\pi}}\int_{\frac{m_j
-\ep}{\sigma\sqrt h}}^{\frac{m_j +\ep}{\sigma\sqrt h}}
e^{-\frac{x^2}{2}}dx,$$ where if $\frac{\ep(h)}{\sqrt h}\to L$
as $h\to 0$ then the first term of the rhs above tends to
$d:=\frac{\sigma^2
}{\sqrt{2\pi}}\int_{-\frac{L}{\sigma}}^{\frac{L}{\sigma}}e^{-\frac{x^2}{2}}dx<\sigma^2,
$ while each term of the latter finite sum tends to 0, since
$\frac{|m_j|}{\sqrt h} \to \infty,$
so the finite sum tends to 0. It follows that, for all $i$,
$\sum_{j\neq i} b_j\leq d+o(1),$ where $d<\sigma^2,$ so $\frac{
\sumi a_i \sum_{j\neq i} b_j }{\sumi a_i}\leq d+o(1),$ and
$\sigma^2-\frac{ \sumi a_i \sum_{j\neq i} b_j }{\sumi a_i}\geq
\sigma^2-d+o(1)\to\sigma^2-d>0,$
as we wanted.\qed\\

\n
 {\bf Proof of Proposition \ref{PropcMSEFzeroEp}.}
{Let us start by checking the asymptotic behavior of  $b_i(\ep)$ and $a_{i}(\ep)$ when
$\ep=\ep(h)$ tends to 0 as $h\to 0$ in such a way that
$\frac{\ep}{\sqrt h}\to +\infty$.  To this end, for fixed
$\sigma$, we define
\begin{align}\label{Dfnb}
 b(\varepsilon,m,h)&:=-\frac{{\sigma}\sqrt{h}}{\sqrt{2\pi}}\left(e^{-\frac{(\varepsilon-m)^{2}}{2{{\sigma}^{2}}h}}(\varepsilon+m)+e^{-\frac{(\varepsilon+m)^{2}}{2{{\sigma}^{2}}h}}(\varepsilon-m)\right)+\frac{m^{2}+{{\sigma}^{2}}h}{\sqrt{2\pi}}
     \int_{\frac{m-\varepsilon}{{{\sigma}}\sqrt{h}}}^{\frac{m+\varepsilon}{{{\sigma}}\sqrt{h}}}e^{-x^{2}/2}dx\\
     a(\varepsilon,m,h)&:=\frac{e^{-\frac{(\ep-m)^2}{2\sigma^2 h}} +
e^{-\frac{(\ep+m)^2}{2\sigma^2
h}}}{\sigma\sqrt{h}\sqrt{2\pi}},\label{Dfna}
\end{align}
and note that $b_j(\ep)=b(\varepsilon,m_j,h)$ and $a_j(\ep)=a(\varepsilon,m_j,h)$. Let us also remark that
\begin{align}\label{Rel1}
    b(\varepsilon,m,h)&=\left\{\begin{array}{ll}
   \sigma^{2} h -\frac{2\sigma}{\sqrt{2\pi}}\varepsilon\sqrt{h} e^{-\frac{\varepsilon^{2}}{2\sigma^{2}h}}+{\rm h.o.t.},&\quad \text{if}\quad m=0,\\
    \\
     \frac{\sigma}{|m|\sqrt{2\pi}}\varepsilon^{2}\sqrt{h}e^{-\frac{(|m|-\varepsilon)^{2}}{2\sigma^{2}h}}+{\rm h.o.t.},&\quad \text{if}\quad m\neq{}0.\end{array}\right.\\
     a(\varepsilon,m,h)&=\left\{\begin{array}{ll}
     \frac{2}{\sigma\sqrt{2\pi}} \frac{1}{\sqrt h}e^{-\frac{\varepsilon^{2}}{2\sigma^{2}h}},&\text{if}\quad m=0,\\
    \\
    \frac{1}{\sigma\sqrt{2\pi}} \frac{1}{\sqrt h}e^{-\frac{(|m|-\varepsilon)^{2}}{2\sigma^{2}h}}+{\rm h.o.t.},&\text{if}\quad m\neq{}0.\end{array}\right.
    \label{Rel2}
\end{align}
 The asymptotic behavior for $a(\ep,m,h)$ is direct, while that for $b(\ep,m,h)$ is shown below.
For simplicity, in what follows, we omit the dependence on $h$ in the functions $a(\varepsilon,m,h)$ and $b(\varepsilon,m,h)$ defined above.
 Let us recall that, under Assumption {\bf A4'}, $N_{t}$ is the number of jumps by time $t,$ $\{\gamma_{\ell}\}_{\ell\geq{}1}$ are the consecutive jumps
 of $J$ and  $\{J\}=\{J\}_{(n)}:=\{i: \Delta^{n}_{i}N\neq{}0\}$. It follows that, for $h$ is small enough, $F(\ep_h) =\sum_{i=1}^n a(\ep,m_{i})g_{i}(\ep)$ can be written as}
\begin{align*}
F(\ep_h)
&=\sum_{i\notin \{J\}} a(\ep,m_{i})\left(\varepsilon^{2}+2\sum_{j\neq{}i:j\in\{J\}}b(\varepsilon,m_{j})+
{2}\sum_{j\neq{}i:j\notin\{J\}}b(\varepsilon,m_{j})-2IV\right) \\
&\quad +\sum_{i\in \{J\}} a(\ep,m_{i})\left(\varepsilon^{2}+2\sum_{j\neq{}i:j\in\{J\}}b(\varepsilon,m_{j})+
{2}\sum_{j\neq{}i:j\notin\{J\}}b(\varepsilon,m_{j})-2IV\right) \\
&=(n-N_{T})a(\ep,0)\left[\ep^{2} - 2 h\sigma^2 (N_{T} + 1)+
2 \left( \sum_{k=1}^{N_{T}}b(\ep,\gamma_{k})
+ (n-N_{T} -1)(b(\ep,0)-\sigma^{2}h)\right) \right]+\\
&\quad + \sum_{\ell=1}^{N_{T}}a(\ep,\gamma_{\ell})
\left[\eps^{2} - 2 h\sigma^2N_{T} + 2
\left(\sum_{k\neq \ell} b(\ep,\gamma_{k}) +(n-N_{T})(b(\ep,0)-\sigma^{2}h)\right)\right]\\
&=(n-N_{T})\frac{2}{\sigma\sqrt{h}\sqrt{2\pi}}e^{-\frac{\ep^{2}}{2\sigma^{2}h}}\Bigg[\ep^{2} - 2 h\sigma^2 (N_{T} + 1)- 4 (n-N_{T} -1)\frac{{\sigma}\varepsilon\sqrt{h}}{\sqrt{2\pi}}e^{-\frac{\ep^{2}}{2\sigma^{2}h}}\\
&\qquad \qquad \qquad \qquad \qquad \qquad \quad +
2 \sum_{k=1}^{N_{T}} \frac{\sigma}{|\gamma_{k}|}\frac{\varepsilon^{2}\sqrt{h}}{\sqrt{2\pi}}e^{-\frac{(|\gamma_{k}|-\varepsilon)^{2}}{2\sigma^{2}h}}
\Bigg]+\\
&\quad + \sum_{\ell=1}^{N_{T}}  \frac{1}{\sigma\sqrt{h}\sqrt{2\pi}}e^{-\frac{(|\gamma_{k}|-\varepsilon)^{2}}{2\sigma^{2}h}}
\Bigg[\eps^{2} - 2 h\sigma^2N_{T}- 4 (n-N_{T} )\frac{{\sigma}\varepsilon\sqrt{h}}{\sqrt{2\pi}}e^{-\frac{\ep^{2}}{2\sigma^{2}h}}\\
&\qquad \qquad \qquad \qquad \qquad \qquad \qquad +
2 \sum_{k\neq{}\ell} \frac{\sigma}{|\gamma_{k}|}\frac{\varepsilon^{2}\sqrt{h}}{\sqrt{2\pi}}e^{-\frac{(|\gamma_{k}|-\varepsilon)^{2}}{2\sigma^{2}h}}
\Bigg]+{\rm h.o.t.}.
\end{align*}
In what follows we use the following notation:
\[
    v_{h}=\frac{\ep_{h}}{\sqrt{h}},\quad u_{\ell h}= \frac{1}{\sqrt{2\pi}}e^{-\frac{\left(v_h - \frac{|\gamma_{\ell}|}{\sqrt h}\right)^2}{2\sigma^2}},\quad s_h= \frac{1}{\sqrt{2\pi}}e^{-\frac{v^2_h}{2\sigma^2}},\quad
    p_{\ell h}=
e^{-\frac{|\gamma_{\ell}|}{\sigma^2h}\left( \frac{|\gamma_{\ell}|}{2}- \sqrt h
v_h\right)}.
\]
Now, since $u_{\ell h}= s_h p_{\ell h}$ and $p_{\ell h}\to{}0$, as $h\to{}0$,

\begin{align}\nonumber
F(\ep_h)
&=(n-N_{T})\frac{2}{\sigma}\sqrt{h} s_{h}\Bigg[  v_{h}^{2} - 2 \sigma^2 (N_{T} + 1)+
2\sigma v_{h}s_{h} \left(\varepsilon \sum_{k=1}^{N_{T}} \frac{1}{|\gamma_{k}|}p_{kh}
- 2 (n-N_{T} -1)\right)
\Bigg]+\\
\nonumber
\nonumber
&\quad +\frac{1}{\sigma}\sqrt{h}s_{h} \sum_{\ell=1}^{N_{T}}  p_{\ell h}
\Bigg[v_{h}^{2} - 2 \sigma^2N_{T}+
2 {\sigma}v_{h}s_{h} \left(\varepsilon \sum_{k\neq{}\ell} \frac{1}{|\gamma_{k}|} p_{kh}
- 2 (n-N_{T} )\right)
\Bigg]+{\rm h.o.t.}\\
\nonumber
\nonumber
&=(n-N_{T})\frac{2}{\sigma}\sqrt{h} s_{h}\Bigg[  v_{h}^{2} -
4\sigma v_{h}s_{h}n
\Bigg]+\frac{1}{\sigma}\sqrt{h}s_{h} \sum_{\ell=1}^{N_{T}}  p_{\ell h}
\Bigg[v_{h}^{2} -
4 {\sigma}v_{h}s_{h} n
\Bigg]+{\rm h.o.t.}\\
\nonumber
\nonumber
&=\left(n-N_{T}+\frac{1}{2} \sum_{\ell=1}^{N_{T}}  p_{\ell h}\right)\frac{2}{\sigma}\sqrt{h} s_{h}\Bigg[  v_{h}^{2} -
4\sigma v_{h}s_{h}n
\Bigg]+{\rm h.o.t.}\\
\nonumber
%\end{align}
%\begin{align}
&=\frac{2n}{\sigma}\sqrt{h} s_{h}v_{h}\Bigg[  v_{h}-
4\sigma s_{h}n
\Bigg]+{\rm h.o.t.}\\
&=\frac{2\ep_h}{h\sqrt
h}e^{-\frac{\ep_h^2}{2\sigma^2h}} \Big(\ep_h-
\frac{e^{-\frac{\ep_h^2}{2\sigma^2h}}}{\sqrt h}\frac{4\sigma}{\sqrt{2\pi}} \Big)\frac{1}{\sigma\sqrt{2\pi}}+\hot. \qed \label{AsympEq1}
\end{align}

\n
{\bf Proof of (\ref{Rel1})}.
Let
\[
    \bar{N}(x)=\int_{x}^{\infty}\phi(z)dz,\quad
    R(x)=\int_{x}^{\infty}\phi(z)dz-\frac{\phi(x)}{x},
\]
and recall that, for $x>0$,
\[
    \bar{N}(x)\leq{}\frac{1}{x}\phi(x),\quad |R(x)|\leq{}\frac{\phi(x)}{x^{3}}.
\]
Then,  for fixed $m>0$ and $h$ small enough such that $\varepsilon_{h}<m$, we have
\begin{align*}
     b(\varepsilon,m,h)&=
     \sigma\sqrt{h}\phi\left(\frac{m-\varepsilon}{{{\sigma}}\sqrt{h}}\right)\left(\frac{m^{2}}{m-\varepsilon}-m\right)
     -\sigma\sqrt{h}\phi\left(\frac{m+\varepsilon}{{{\sigma}}\sqrt{h}}\right)\left(\frac{m^{2}}{m+\varepsilon}-m\right)\\
     &\quad-\sigma\sqrt{h}\phi\left(\frac{m-\varepsilon}{{{\sigma}}\sqrt{h}}\right)\varepsilon
     -\sigma\sqrt{h}\phi\left(\frac{m+\varepsilon}{{{\sigma}}\sqrt{h}}\right)\varepsilon\\
     &\quad +
     \sigma^{3}h^{3/2}\phi\left(\frac{m-\varepsilon}{{{\sigma}}\sqrt{h}}\right)\left(\frac{1}{m-\varepsilon}\right)
     -\sigma^{3}h^{3/2}\phi\left(\frac{m+\varepsilon}{{{\sigma}}\sqrt{h}}\right)\left(\frac{1}{m+\varepsilon}\right)
     \pm {(m^2+\sigma^2 h)} R\left(\frac{m\mp \varepsilon_{h}}{{{\sigma}}\sqrt{h}}\right)\\i
                   &=\frac{\sigma}{m}\sqrt{h}\phi\left(\frac{m-\varepsilon}{{{\sigma}}\sqrt{h}}\right)\varepsilon^{2}
     -\frac{\sigma}{m}\sqrt{h}\phi\left(\frac{m+\varepsilon}{{{\sigma}}\sqrt{h}}\right)\varepsilon^{2}\\
     &\quad +\frac{\sigma}{m(m-\varepsilon)}\sqrt{h}\phi\left(\frac{m-\varepsilon}{{{\sigma}}\sqrt{h}}\right)\varepsilon^{3}
     -\frac{\sigma}{m(m+\varepsilon)}\sqrt{h}\phi\left(\frac{m+\varepsilon}{{{\sigma}}\sqrt{h}}\right)\varepsilon^{3}\\
     &\quad +
     \frac{\sigma^{3}}{m-\varepsilon}h^{3/2}\phi\left(\frac{m-\varepsilon}{{{\sigma}}\sqrt{h}}\right)
     -\frac{\sigma^{3}}{m+\varepsilon}h^{3/2}\phi\left(\frac{m+\varepsilon}{{{\sigma}}\sqrt{h}}\right)\pm
      {(m^2+\sigma^2 h)} R\left(\frac{m\mp \varepsilon_{h}}{{{\sigma}}\sqrt{h}}\right)
\end{align*}
It is now clear that (\ref{Rel1}) holds true. We can similarly deal with the case $m<0$.\qed \\

\n
{\bf Proof of the statement in Remark \ref{RemDriftcMSE}.} For nonzero drift, by conditioning also on the drift process $a$, we have that
\begin{align*}
F(\ep_h)
&=\sum_{i\notin\{J\}}a(\ep,h\bar{a}_{i})\left[\ep^{2} - 2 h\sigma^2 (N_{T} + 1)+
2 \left( \sum_{k=1}^{N_{T}}b(\ep,\gamma_{k}+h\bar{a}_{i_{k}})
+ \sum_{j\neq{}i:j\notin\{J\}}(b(\ep,h\bar{a}_{j})-\sigma^{2}h)\right) \right]+\\
&\quad + \sum_{\ell=1}^{N_{T}}a(\ep,\gamma_{\ell}+h\bar{a}_{i_{\ell}})
\left[\eps^{2} - 2 h\sigma^2N_{T} + 2
\left(\sum_{k\neq \ell} b(\ep,\gamma_{k}+h\bar{a}_{i_{k}}) +\sum_{j\neq{}i_{\ell}:j\notin\{J\}}(b(\ep,h\bar{a}_{j})-\sigma^{2}h)\right)\right],
\end{align*}
 where {$\bar{a}_{i}=\int_{t_{i-1}}^{t_{i}}a_{s}ds /h$ and the indices $i_{1}<i_{2}<\dots< i_{N_T}$ are defined such that $\Delta_{i_{k}}J\neq 0$, while $\Delta_{i}J=0$ for any other $i\notin\{i_{1},i_{2},\dots, i_{N_T}\}.$} Next, we follow the same arguments {as those used in the proof of Proposition \ref{PropcMSEFzeroEp} but, instead of (\ref{Rel1})-(\ref{Rel2}), we exploit the following asymptotics:}
\begin{align*}
     &{ a(\varepsilon,h\bar{a}_{i})=
     \frac{2}{\sigma} h^{-1/2}\phi\left(\frac{\varepsilon}{{{\sigma}}\sqrt{h}}\right)+{\rm h.o.t.},\quad
    a(\varepsilon,\gamma_{k}+h\bar{a}_{i_{k}})=\frac{1}{\sigma} h^{-1/2}\phi\left(\frac{|\gamma_{k}|-\varepsilon}{{{\sigma}}\sqrt{h}}\right)e^{-\frac{\gamma_{k}\bar{a}_{i_{k}}}{\sigma^{2}}}+{\rm h.o.t.}.}\\
    & b(\varepsilon,h\bar{a}_{i})=
    \sigma^{2}h -2{\sigma}\varepsilon\sqrt{h}\phi\left(\frac{\varepsilon}{\sigma\sqrt{h}}\right)+{\rm h.o.t.},\quad
    b(\varepsilon,\gamma_{k}+h\bar{a}_{i_{k}})=
     \frac{\sigma}{|\gamma_{k}|}\varepsilon^{2}\sqrt{h}\phi\left(\frac{|\gamma_{k}|-\varepsilon}{{{\sigma}}\sqrt{h}}\right)e^{-\frac{\gamma_{k}\bar{a}_{i_{k}}}{\sigma^{2}}}+{\rm h.o.t.}.\qed
\end{align*}

\vspace{0.2cm}
\n
{\bf Proof of Corollary \ref{AsympOptThldFJA}.} In fact, from
 Proposition 4 and (\ref{AsympEq1}), we have that
\[
    F(\bar{\ep}_h)=\frac 2 \sigma n \bar{s}_h \bar{v}_h\sqrt h\Big(\bar{v}_h  -n  \bar{s}_h \cdot 4 \sigma \Big)+{\rm h.o.t.}=0,
\]
where $\bar{v}_{h}:=\bar\ep_{h}/\sqrt{h}$ and $\bar{s}_h= \frac{e^{-\frac{\bar{\ep}^2_h}{2h\sigma^2}}}{\sqrt{2\pi}}$. Thus,
\beq\label{condEpStar}
\bar{v}_h  -n  \bar{s}_h \cdot 4 \sigma+{\rm h.o.t.}=0,
\eeq
or, equivalently,
\[
    \bar{\ep}_h  -\frac{ e^{-\frac{\bar{\ep}^2_h}{2h\sigma^2}}}{\sqrt h} \frac{4 \sigma}{\sqrt{2\pi}} + {\rm h.o.t.}=0,
\]
which is exactly the condition in (\ref{AsymEqEps}), entailing that as $h\to 0$ \\

\hfill $\bar{\ep}_h\sim \sqrt{2\sigma^{2}h\ln\frac{1}{h}}.$\hfill \qed\\

\n
{\bf Proof of Proposition \ref{ProfFprimeCMSE}} %(p.26, 31 ss del 8/12/16)
We use the same notation as in (\ref{Ntngi}). Let us fix $h$, and $nh=1$, then $\frac{d}{d\ep} F(\ep)= \sumi [a_i' g_i + a_ig_i']$
$$ = - \sumi \frac{1}{\sigma^3 h^{\frac 3 2} \sqrt{2\pi}}\Big[e^{-\frac{(\ep-|m_i|)^2}{2\sigma^2h}}(\ep-|m_i|) +
e^{-\frac{(\ep+|m_i|)^2}{2\sigma^2_i}} (\ep + |m_i|)\Big]g_i
+ \sumi \frac{e^{-\frac{(\ep-|m_i|)^2}{2\sigma^2h}} +
e^{-\frac{(\ep+|m_i|)^2}{2\sigma^2h}}}{\sigma\sqrt h \sqrt{2\pi}}\Big[2\ep+2\sumjni \ep^2 a_j\Big].$$
We now evaluate $F'(\ep)$ at $\ep_h$ such that $\ep_h\to 0$ with $\ep_h\gg \sqrt h,$ as $h\to 0.$
Since again when $m_i\neq 0$ we have  $e^{-\frac{(\ep-|m_i|)^2}{2\sigma^2 h}}\gg e^{-\frac{(\ep+|m_i|)^2}{2\sigma^2 h}}$ and $\ep \ll m_i$, then
$$F'(\ep)\sqrt{2\pi}= \sum_{i\in \{J\}} \frac{1}{\sigma \sqrt h} e^{-\frac{(\ep-|m_i|)^2}{2\sigma^2h}}\Big[\frac{|m_i|}{\sigma^2 h}g_i  +
2\ep(1+\ep\sumjni  a_j)\Big]
+
\sum_{i\not\in\{J\}}  \frac{2\ep}{\sigma \sqrt h} e^{-\frac{\ep^2}{2\sigma^2h}}\Big[-\frac{g_i}{\sigma^2 h} +
2(1+\ep\sumjni  a_j)\Big]
+\hot
$$
Note that within $g_i=
\ep^2 +2\sum_{j\neq i}b_j -2 IV$, we have that the finite sum
    $\frac{1}{\sqrt{2\pi}}\sum_{j\neq i: j\in\{J\}} \frac{\sigma}{|m_j|}\ep e^{-\frac{(|m_{j}|-\varepsilon)^{2}}{2\sigma^{2}h}}=
    \sum_{j\neq{}i: j\in\{J\}}  \frac{\sigma}{|m_j|}\ep u_{jh}$\\
    $= s_h\ep\sum_{j\neq i:  j\in\{J\}}  \frac{\sigma}{|m_j|} p_{jh}$
is negligible wrt
    $s_h \ll\frac{1}{\sqrt{2\pi}}\sum_{j\neq{}i: j\not\in\{J\}} e^{-\frac{{\varepsilon}^{2}}{2\sigma^{2}h}}=  [(n-N_T)I_{\{ i\in\{J\}\}} +
    (n-N_T-1)I_{\{i\not\in\{J\}\}} ] s_h,$
since
     $\ep\sum_{j\neq{}i: j\in\{J\}} \frac{p_{jh}}{m_j}\toas 0.$
Therefore
    $$g_i= \ep^2 -\frac{4\sigma}{\sqrt{2\pi}}\sqrt{h}\varepsilon  s_h [(n-N_T)I_{\{i\in\{J\}\}} +
    (n-N_T-1)I_{\{i\not\in \{J\}\}} ] - 2\sigma^2 h [N_T I_{\{i\in\{J\}\}} + (N_T+1) I_{\{i\not\in\{J\}\}} ]+\hot. $$
Further, $N_T\ll n$ and $h\ll \ep^2$, then for all $i$
     $$g_i= \ep^2 -\frac{4\sigma}{\sqrt{2\pi}}\frac{\varepsilon  s_h}{\sqrt{h}}+\hot. $$
%where $\hot= o(s_h) + o(\ep^2).$

Moreover from (27) we reach that $\sumjni a_j= \sum_{j\neq i, j\not\in \{J\}} 2\frac{s_h}{\sigma\sqrt h } +
\sum_{j\neq i, j\in \{J\}} \frac{u_{jh}}{\sigma\sqrt h }+ \hot$, and again the second sum is negligible wrt the first one,
thus, for all $i$, $$\ep\sumjni a_j=  2\frac{s_h\ep}{\sigma\sqrt h } [(n-N_T) I_{\{m_i\neq 0\}} + (n-N_T-1) I_{\{m_i=0\}} ]+
 \hot= \frac 2 \sigma \frac{s_h\ep}{h\sqrt h}+\hot.$$

Now, using (28), from\\
$\sum_{i\in \{J\}} a_i g_i = \frac{1}{\sigma}\sqrt{h}s_{h} \sum_{\ell=1}^{N_{T}}  p_{\ell h}
\Bigg[v_{h}^{2} - 2 \sigma^2N_{T}+
2 {\sigma}v_{h}s_{h} \Big(\varepsilon \sum_{k\neq{}\ell} \frac{1}{|\gamma_{k}|} p_{kh}
$ $- 2 (n-N_{T} )\Big)
\Bigg]+\hot=\\
\frac{1}{\sigma}\sqrt{h}s_{h} \sum_{\ell=1}^{N_{T}}  p_{\ell h}
\Bigg[v_{h}^{2} -
4 {\sigma}v_{h}s_{h} n
\Bigg]+\hot$\\
 we reach that\\
$\sum_{i\in \{J\}} a_i \frac{g_i|m_i|}{\sigma^2 h}= \frac{1}{\sigma}\sqrt{h}s_{h} \sum_{\ell=1}^{N_{T}}  p_{\ell h}
\Bigg[v_{h}^{2} -
4 {\sigma}v_{h}s_{h} n
\Bigg]\frac{|\gamma_\ell|}{\sigma^2 h}+\hot$\\
and from\\
$\sum_{i\not\in \{J\}} a_i g_i=(n-N_{T})\frac{2}{\sigma}\sqrt{h} s_{h}\Bigg[  v_{h}^{2} - 2 \sigma^2 (N_{T} + 1)+
2\sigma v_{h}s_{h} \left(\varepsilon \sum_{k=1}^{N_{T}} \frac{1}{|\gamma_{k}|}p_{kh}
- 2 (n-N_{T} -1)\right)
\Bigg]=$\\
 $ (n-N_{T})\frac{2}{\sigma}\sqrt{h} s_{h}\Bigg[  v_{h}^{2} -
4\sigma v_{h}s_{h}n
\Bigg]+\hot$\\
we reach that\\
$\sum_{i\not\in \{J\}} a_i \frac{g_i\ep}{\sigma^2 h}= (n-N_{T})\frac{2}{\sigma}\sqrt{h} s_{h}\Bigg[  v_{h}^{2} -
4\sigma v_{h}s_{h}n
\Bigg]\frac{\ep}{\sigma^2 h}+\hot.$

Thus $$F'\sqrt{2\pi}=v_{h}\Big[  v_{h} -4\sigma s_{h}n\Big] \Bigg[\frac{1}{\sigma}\sqrt{h} \sum_{\ell=1}^{N_{T}}
u_{\ell h}\frac{|\gamma_\ell|}{\sigma^2 h}
- (n-N_T)\frac{2}{\sigma}\sqrt{h} s_{h} \frac{\ep}{\sigma^2 h}\Bigg]$$
$$ + 2\ep \Big(1+ \frac 2  \sigma\frac{s_h\ep}{h\sqrt h}\Big)
\Big(\sum_{i\in J} \frac{u_{ih}}{\sigma\sqrt h} +  \sum_{i\not\in J} \frac{2s_h}{\sigma\sqrt h}\Big)+\hot.
$$
If now our sequence $\ep_h$ is such that $v_h= 4\sigma ns_h +\hot $, and noting that also $\sum_{j\in J} p_{jh}|\gamma_j|\toas 0$ and that
$n\ep= n\sqrt h v_h= \frac{v_h}{\sqrt h}\to +\infty$ then
$$F'(\ep_h)\sqrt{2\pi}= v_h\cdot o(ns_h)\frac{s_h}{\sigma^3 \sqrt{h}}\Bigg[ \sum_{\ell=1}^{N_{T}}
p_{\ell h}|\gamma_\ell|
- 2n \ep\Bigg]+\frac{2\ep s_h}{\sigma\sqrt h}\Big(1+ \frac 2  \sigma\frac{s_h\ep}{h\sqrt h}\Big)
%\Big(\sum_{i\in J} p_{ih} +
\cdot 2(n-N_T)+\hot
$$
$$=- 2n \ep v_h\cdot o(ns_h)\frac{s_h}{\sigma^3 \sqrt{h}}
+\frac{4n\ep s_h}{\sigma\sqrt h}\Big(1+ \frac 2  \sigma\frac{s_h\ep}{h\sqrt h}\Big)
+\hot
$$
now $v_h=4 \sigma ns_h + o (v_h)$ means also %$\ep_h=\frac{s_h}{\sqrt h} + o (\ep_h),$ i.e.
$s_h =\ep_h\sqrt h + o (\ep_h\sqrt h),$ and thus $\frac{s_h\ep_h}{h\sqrt h}= \frac{\ep^2_h}{h}+ o(\frac{\ep^2_h}{h})\to +\infty,$ therefore
$$F'(\ep_h)\sqrt{2\pi}= - 2 \ep \frac{\ep}{\sqrt h}\cdot o(ns_h)\frac{s_h}{\sigma^3h \sqrt{h}}
+\frac{4n\ep s_h}{\sigma\sqrt h}\frac 2  \sigma\frac{s_h\ep}{h\sqrt h}+\hot=$$
$$\frac{\ep}{\sqrt h}\frac{s_h\ep}{h\sqrt h} ns_h \frac{8}{\sigma^2}(1 + o(1)) +\hot
 =
%\frac{\ep}{\sqrt h}\frac{s_h\ep}{h\sqrt h} ns_h  \frac{8}{\sigma^2}=
 \frac{8}{\sigma^2}\Big(\frac{s_h\ep_h}{h\sqrt h}\Big)^2
+\hot.$$

\vspace{-0.8cm}\qed \\

\end{document}